\documentclass[12pt,a4paper]{article}
\makeatletter
\def\eqnarray{%
\stepcounter{equation}%
\let\@currentlabel=\theequation
\global\@eqnswtrue
\global\@eqcnt\z@
\tabskip\@centering
\let\\=\@eqncr
$$\halign to \displaywidth\bgroup\@eqnsel\hskip\@centering
$\displaystyle\tabskip\z@{##}$&\global\@eqcnt\@ne
\hfil$\displaystyle{{}##{}}$\hfil
&\global\@eqcnt\tw@$\displaystyle\tabskip\z@{##}$\hfil
\tabskip\@centering&\llap{##}\tabskip\z@\cr}
\makeatother
\newcommand{\kansu}[2]{{{#1}\!\left({#2}\right)}}
\newcommand{\ket}[1]{{\vert{#1}\rangle}}
\newcommand{\bra}[1]{{\langle{#1}\vert}}
\newcommand{\kett}[2]{{\vert{#1,#2}\rangle}}
\newcommand{\braa}[2]{{\langle{#1,#2}\vert}}
\newcommand{\braket}[2]{{\langle{#1}\vert{#2}\rangle}}
\newcommand{\zettai}[1]{{\vert{#1}\vert}}

\newcommand{\calh}{{\cal H}}
\newcommand{\calm}{{\cal M}}
\newcommand{\cala}{{\cal A}}
\newcommand{\calf}{{\cal F}}

\newcommand{\fukuso}{{\mathbf C}}
\newcommand{\real}{{\mathbf R}}
\newcommand{\futon}{{\bf N}}

\newcommand{\stm}{{St_m}}
\newcommand{\grm}{{Gr_m}}
\newcommand{\eem}{{E_m}}
\newcommand{\bell}[1]{{\vert \vert{#1}\rangle \rangle}}
\newcommand{\bellbraket}[2]{{\langle \langle{#1}\vert \vert{#2}\rangle 
           \rangle}}

\newcommand{\lam}{{\bf \lambda}}
\newcommand{\slam}{{\bf \lambda_0}}
\newcommand{\zetta}{{\vert z\vert}}
\newcommand{\wetta}{{\vert w\vert}}

\begin{document}

\title{\sl Introduction to Coherent States and \\
Quantum Information Theory}
\author{
  Kazuyuki FUJII
  \thanks{E-mail address : fujii@yokohama-cu.ac.jp}\  
  \thanks{Home-page : http://fujii.sci.yokohama-cu.ac.jp}\\
  Department of Mathematical Sciences\\
  Yokohama City University\\
  Yokohama 236-0027\\ 
  JAPAN
  }
\date{}
\maketitle\thispagestyle{empty}
%
%
%
%
\begin{abstract} 
  The purpose of this paper is to introduce several basic theorems of 
  coherent states and generalized coherent states based on Lie algebras 
  su(2) and su(1,1), and to give some applications of them to quantum 
  information theory for graduate students or non--experts who are 
  interested in both Geometry and Quantum Information Theory. 

  In the first half we make a general review of coherent states and 
  generalized coherent states based on Lie algebras su(2) and su(1,1) 
  from the geometric point of view 
  and, in particular, prove that each resolution of unity can be obtained 
  by the curvature form of some bundle on the parameter space.    
  
  We also make a short review of Holonomic Quantum Computation (Computer) 
  and show a geometric construction of the well--known Bell states 
  by making use of generalized coherent states. 
  
  In the latter half we apply a method of generalized coherent states 
  to some important topics in Quantum Information Theory, in particular, 
  swap of coherent states and cloning of coherent ones. 
  
  We construct the swap operator of coherent states by making use of 
  a generalized coherent operator based on su(2) and show 
  an ``imperfect cloning" of coherent states, and moreover present 
  some related problems. 
  
  We also present a problem on a possibility of calculation or 
  approximation of coherent state path integrals on 
  Holonomic Quantum Computer. 
 
  In conclusion we state our dream, namely, a construction of {\bf 
  Geometric Quantum Information Theory}. 
\end{abstract}

\newpage

%
%
%
%

\section{Introduction}
This paper is the pair to the preceding one \cite{KF1} and the aim is to 
introduce geometric aspects of coherent states and generalized coherent 
ones based on Lie algebras $su(1,1)$ and $su(2)$ and to apply them to 
quantum information theory for graduate students or 
non--experts (in this field) who are interested in both Geometry and 
Quantum Information Theory. 

Coherent states or generalized coherent states play a crucial role 
in quantum physics, in particular, quantum optics, see \cite{KS} and  
its references or \cite{MW}.  
They also play an important one in mathematical physics, see \cite{AP}. 
For example, they are very useful in performing 
stationary phase approximations to path integral, 
see \cite{FKSF1}, \cite{FKSF2} and \cite{FKS}. 

In the theory of coherent states or generalized coherent ones 
the resolution of unity is just a key concept, see \cite{KS}. Is it 
possible to understand this fact from the geometric point of view ?  \quad 
For a set of coherent or generalized coherent states we can define 
a projector from the manifold consisting of parameters of them to 
infinite--dimensional Grassmann manifold (called classifying spaces in 
K--Theory).  
Making use of this projector we can calculate several geometric quantities  
such as Chern characters, see for example \cite{MN}. In particular, 
we prove that each resolution of unity can be obtained by the curvature form 
of some bundle on the parameter space.    

Let us turn to Quantum Information Theory (QIT). 
The main subjects in QIT are 
\begin{itemize}
\item[(i)] Quantum Computation 
\item[(ii)] Quantum Cryptgraphy 
\item[(iii)] Quantum Teleportation 
\end{itemize}
As for general introduction to QIT see \cite{LPS}, \cite{AH} and 
\cite{ASt}, \cite{KF1}. \quad 
The aim of this paper is to apply geometric methods 
to QIT, or more directly 
\begin{large}
\begin{center}
  {\bf A Geometric Construction of Quantum Information Theory}.
\end{center}
\end{large}
We are developing the theory of geometric quantum computation called 
Holonomic Quantum Computation, see \cite{ZR}, \cite{PZR}, \cite{PC} and 
\cite{KF2}--\cite{KF6}, and we are also 
studying geometric construction of the Bell states or the generalized Bell 
ones, see \cite{KF12}, \cite{KF13}. 
We are interested in geometric method of Homodyne Tomography \cite{GLM}, 
\cite{MP} or geometric one of Quantum Cryptgraphy \cite{KB}, \cite{BW}. 

On the other hand, the method of path integral plays a very important role 
in Quantum Mechanics or Quantum Field Theory. However it is not easy to 
calculate complicated path integrals with classical computers. 
We are interested in it from the quantum information theory's point of view. 
That is, can we calculate or approximate some path integral in polynomial 
times with Quantum Computers (Holonomic Quantum Computer especially) ? \quad 
Unfortunately we cannot answer this question, however 
we believe this problem becomes crucial for Quantum Computers. 

By the way it seems to the author that our calculations suggest 
some profound relation to recent non--commutative differential geometry or 
non--commutative field theory, see \cite{SV} or \cite{BDLMO}. 
This is very interesting, but beyond the scope of this paper. 
We show the relation diagramatically 
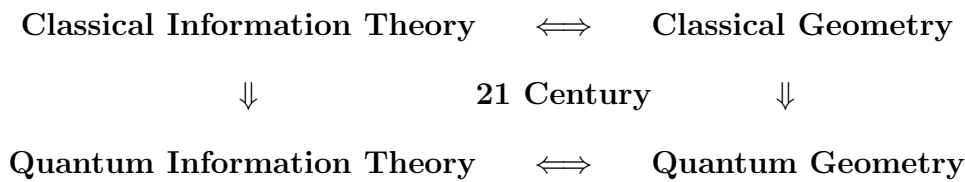
\vspace{5mm}
\begin{eqnarray}
\mbox{\bf Classical Information Theory} &\Longleftrightarrow& 
\mbox{\bf Classical Geometry}  \nonumber \\
{\Downarrow}\qquad \qquad \qquad \quad &\mbox{\bf 21 Century}& 
\qquad \qquad \Downarrow \nonumber \\
\mbox{\bf Quantum Information Theory} &\Longleftrightarrow&
\mbox{\bf Quantum Geometry}  \nonumber 
\end{eqnarray}
\par \vspace{5mm} \noindent
We expect that some readers would develop this subject. 

In the latter half of this paper 
we treat special topics in Quantum Information Theory, 
namely, swap of coherent states and cloning of coherent states. 
It is not difficult to construct a universal swap operator (see 
Appendix), however for 
coherent states we can construct a special and better one by 
making use of a generalized coherent operator based on $su(2)$. 
On the other hand, to construct a cloning operator is of course 
not easy by the no cloning theorem \cite{WZ}. 
However for coherent states we can make an approximate cloning 
(``imperfect cloning" in our terminology) by making use of the same 
coherent operator based on $su(2)$. 
This and some method in \cite{DG} may develop a better approximate cloning 
method.  We also present some related problems on these topics.

We have so many problems to be solved in the near future. 
The author expects strongly that young mathematical physicists or 
information theorists will take part in this fruitful field. 

{\bf The contents of this paper are as follows} :
\begin{itemize}
\item[1] Introduction
\item[2] Coherent States
\item[3] Generalized Coherent States Based on $su(1,1)$
 \begin{itemize}
 \item[3.1] General Theory
 \item[3.2] Some Formulas
 \item[3.3] A Supplement
 \item[3.4] Barut--Girardello Coherent States
 \end{itemize}
\item[4] Generalized Coherent States Based on $su(2)$
 \begin{itemize}
 \item[4.1] General Theory
 \item[4.2] Some Formulas
 \item[4.3] A Supplement
 \end{itemize}
\item[5] Schwinger's Boson Method
\item[6] Universal Bundles and Chern Characters
\item[7] Calculations of Curvature Forms
 \begin{itemize}
 \item[7.1] Coherent States
 \item[7.2] Generalized Coherent States Based on $su(1,1)$
 \item[7.3] Generalized Coherent States Based on $su(2)$
 \end{itemize}
\item[8] Holonomic Quantum Computation
 \begin{itemize}
 \item[8.1] One--Qubit Case
 \item[4.2] Two--Qubit Case
 \end{itemize}
\item[9] Geometric Construction of Bell States
 \begin{itemize}
 \item[9.1] Review on General Theory
 \item[9.2] Review on Projective Spaces
 \item[9.3] Bell States Revisited
 \end{itemize}
\item[10] Topics in Quantum Information Theory
 \begin{itemize}
 \item[10.1] Some Useful Formulas
 \item[10.2] Swap of Coherent States
 \item[10.3] Imperfect Cloning of Coherent States
 \item[10.4] Swap of Squeezed--like States ?
 \item[10.5] A Comment
 \end{itemize}
\item[11] Path Integral on a Quantum Computer
\item[12] Discussion and Dream
\item[] Appendix
 \begin{itemize}
 \item[A] Proof of Disentangling Formulas
 \item[B] Universal Swap Operator
 \item[C] Calculation of Path Integral
 \item[D] Representation from $SU(2)$ to $SO(3)$
 \end{itemize}
\end{itemize}

\vspace{10mm}
\section{Coherent States}

We make a review of some basic properties of displacement (coherent) 
operators within our necessity. For the proofs see \cite{AP} or 
\cite{KS}

Let $a(a^\dagger)$ be the annihilation (creation) operator of the harmonic 
oscillator.
If we set $N\equiv a^\dagger a$ (:\ number operator), then
\begin{equation}
  \label{eq:2-1}
  [N,a^\dagger]=a^\dagger\ ,\
  [N,a]=-a\ ,\
  [a^\dagger, a]=-\mathbf{1}\ .
\end{equation}
Let $\calh$ be a Fock space generated by $a$ and $a^\dagger$, and
$\{\ket{n}\vert\  n\in\futon\cup\{0\}\}$ be its basis.
The actions of $a$ and $a^\dagger$ on $\calh$ are given by
\begin{equation}
  \label{eq:2-2}
  a\ket{n} = \sqrt{n}\ket{n-1}\ ,\
  a^{\dagger}\ket{n} = \sqrt{n+1}\ket{n+1}\ ,
  N\ket{n} = n\ket{n}
\end{equation}
where $\ket{0}$ is a normalized vacuum ($a\ket{0}=0\  {\rm and}\  
\langle{0}\vert{0}\rangle = 1$). From (\ref{eq:2-2})
state $\ket{n}$ for $n \geq 1$ are given by
\begin{equation}
  \label{eq:2-3}
  \ket{n} = \frac{(a^{\dagger})^{n}}{\sqrt{n!}}\ket{0}\ .
\end{equation}
These states satisfy the orthogonality and completeness conditions
\begin{equation}
  \label{eq:2-4}
   \langle{m}\vert{n}\rangle = \delta_{mn}\ ,\quad \sum_{n=0}^{\infty}
   \ket{n}\bra{n} = \mathbf{1}\ . 
\end{equation}

 Let us state coherent states. For the normalized state $\ket{z} \in 
\calh \ {\rm for}\  z \in \fukuso$ the following three conditions are 
equivalent :
\begin{eqnarray}
  \label{eq:2-5-1}
 &&(\mbox{i})\quad a\ket{z} =  z\ket{z}\quad {\rm and}\quad 
      \langle{z}\vert{z}\rangle = 1  \\
  \label{eq:2-5-2}
 &&(\mbox{ii})\quad  \ket{z} =  \mbox{e}^{- \vert{z}\vert^{2}/2} 
          \sum_{n=0}^{\infty}\frac{z^{n}}{\sqrt{n!}}\ket{n} = 
          \mbox{e}^{- \vert{z}\vert^{2}/2}e^{za^{\dagger}}\ket{0} \\
  \label{eq:2-5-3}
 &&(\mbox{iii})\quad  \ket{z} =  \mbox{e}^{za^{\dagger}- \bar{z}a}\ket{0}. 
\end{eqnarray}
In the process from (\ref{eq:2-5-2}) to (\ref{eq:2-5-3}) 
we use the famous elementary Baker-Campbell-Hausdorff formula
\begin{equation}
  \label{eq:2-6}
 \mbox{e}^{A+B}=\mbox{e}^{-\frac1{2}[A,B]}\mbox{e}^{A}\mbox{e}^{B}
\end{equation}
whenever $[A,[A,B]] = [B,[A,B]] = 0$, see \cite{KS} or \cite{AP}.  
This is the key formula.

\noindent{\bfseries Definition}\quad The state $\ket{z}$ that 
satisfies one of (i) or (ii) or (iii) above is called the coherent state.

\noindent
The important feature of coherent states is the following resolution 
(partition) of unity.
\begin{equation}
   \label{eq:2-7}
  \int_{\fukuso} \frac{[d^{2}z]}{\pi} \ket{z}\bra{z} = 
  \sum_{n=0}^{\infty} \ket{n}\bra{n} = \mathbf{1}\ ,
\end{equation}
where we have put $[d^{2}z] = d(\mbox{Re} z)d(\mbox{Im} z)$ for simplicity. 
We note that 
\begin{equation}
   \label{eq:2-a}
  \braket{z}{w} = \mbox{e}^{-\frac{1}{2}\zetta^2 -\frac{1}{2}\wetta^2 +
                             \bar{z}w} \Longrightarrow 
  \vert{\braket{z}{w}}\vert=
  \mbox{e}^{-\frac{1}{2}\vert{z-w}\vert^2},\  
  \braket{w}{z}=\overline{\braket{z}{w}},
\end{equation}
so $\vert{\braket{z}{w}}\vert < 1$ if $z\ne w$ and 
$\vert{\braket{z}{w}}\vert \ll 1$ if $z$ and $w$ are separated enough. 
We will use this fact in the following. 

Since the operator
\begin{equation}
  \label{eq:2-8}
      D(z) = \mbox{e}^{za^{\dagger}- \bar{z}a}
      \quad \mbox{for} \quad z \in \fukuso  
\end{equation}
is unitary, we call this a displacement (coherent) operator. For these 
operators the following properties are crucial.
For  $z,\ w \in \fukuso$
\begin{eqnarray}  
  \label{eq:2-9-1} 
 && D(z)D(w) = \mbox{e}^{z\bar{w}-\bar{z}w}\ D(w)D(z), \\
  \label{eq:2-9-2}
 && D(z+w) = \mbox{e}^{-\frac{1}{2}(z\bar{w}-\bar{z}w)}\ D(z)D(w).
\end{eqnarray}
Here we list some basic properties of this operator.

\vspace{5mm}

\noindent{\bfseries (a) Matrix Elements}\quad The matrix elements of 
$D(z)$  are  
\begin{eqnarray}
   \label{eq:2-10-1}
 &&(\mbox{i})\quad n \le m \quad 
   \bra{n}D(z)\ket{m} = \mbox{e}^{-\frac{1}{2}\zetta^2}\sqrt{\frac{n!}{m!}}
                 (-\bar{z})^{m-n}{L_n}^{(m-n)}(\zetta^2), \\
   \label{eq:2-10-2}
 &&(\mbox{ii})\quad n \geq m \quad 
   \bra{n}D(z)\ket{m} = \mbox{e}^{-\frac{1}{2}\zetta^2}\sqrt{\frac{m!}{n!}}
                 z^{n-m}{L_m}^{(n-m)}(\zetta^2),
\end{eqnarray}
where ${L_n}^{(\alpha)}$ is the associated Laguerre's polynomial defined by 
\begin{equation}
   \label{eq:2-11}
 {L_n}^{(\alpha)}(x)=\sum_{j=0}^{n}(-1)^j {{n+\alpha}\choose{n-j}}
                  \frac{x^j}{j!}\ . 
\end{equation}
In particular ${L_n}^{(0)}$ is the usual Laguerre's polynomial and these 
are related to diagonal elements of $D(z)$. 
Here let us list the generating function and orthogonality condition of 
associated Laguerre's polynomials :
\begin{eqnarray}
   \label{eq:2-11-1-1}
&& \frac{\mbox{e}^{-xt/(1-t)}}{(1-t)^{\alpha +1}} = \sum_{j=0}^{\infty}
   {L_n}^{(\alpha)}(x)t^{\alpha} \quad \mbox{for}\quad 
    \vert t\vert < 1,  \\
   \label{eq:2-11-1-2}
&& \int_{0}^{\infty}\mbox{e}^{-x}x^{\alpha}{L_n}^{(\alpha)}(x)
   {L_m}^{(\alpha)}(x) dx = \frac{\Gamma(\alpha +n+1)}{n!}\delta_{nm} \quad
    \mbox{for}\quad \mbox{Re}(\alpha) > -1.
\end{eqnarray}
As an interesting application of this formula see the recent \cite{MFr}, 
or forthcoming \cite{KF15}. 

\vspace{5mm}

\noindent{\bfseries (b) Trace Formula}\quad We have 
\begin{equation}
   \label{eq:2-12}
    \mbox{Tr}U(z) = \pi{\delta^2}(z) \equiv \pi\delta(x)\delta(y) \quad 
    \mbox{if}\  z=x+iy.
\end{equation}
This is just a fundamental property. 

\vspace{5mm}

\noindent{\bfseries (c) Glauber Formula}\quad  Let $A$ be any observable. 
Then we have 
\begin{equation}
   \label{eq:2-13}
   A = \int_{\fukuso}\frac{[d^{2}z]}{\pi}\mbox{Tr}[AD^{\dagger}(z)]D(z)
\end{equation}
This formula plays an important role in the field of homodyne tomography, 
\cite{GLM} and \cite{MP}.

\vspace{5mm}

\noindent{\bfseries (d) Projection on Coherent State}\quad  
The projection on coherent state $\ket{z}$ is given by $\ket{z}\bra{z}$. 
But this projection has an interesting expression : 
\begin{equation}
   \label{eq:2-14}
          \ket{z}\bra{z} = :\mbox{e}^{-(a-z)^{\dagger}(a-z)}:
\end{equation}
where the notation $:\ :$ means normal ordering. 

\par \noindent
This formula has been used in the field of quantum cryptgraphy, \cite{KB} 
and \cite{BW}. We note that 
\[
          \ket{z}\bra{w} \ne :\mbox{e}^{-(a-z)^{\dagger}(a-w)}:
\]
for $z,\ w \in \fukuso$ with $z\ne w$.

\vspace{10mm}
\section{Generalized Coherent States Based on su(1,1)}

In this section we introduce some basic properties of generalized 
coherent operators based on $su(1,1)$, see \cite{FKSF1} or \cite{AP}.     

\subsection{General Theory}

We consider a spin $K\ (> 0)$ representation of $su(1,1) 
\subset sl(2,\fukuso)$ and set its generators 
$\{ K_+, K_-, K_3 \}\ ((K_+)^{\dagger} = K_-)$, 
\begin{equation}
  \label{eq:2-2-12}
 [K_3, K_+]=K_+, \quad [K_3, K_-]=-K_-, \quad [K_+, K_-]=-2K_3.
\end{equation}
We note that this (unitary) representation is necessarily infinite 
dimensional. 
The Fock space on which $\{ K_+, K_-, K_3 \}$ act is 
$\calh_{K} \equiv \{\kett{K}{n} \vert n\in\futon\cup\{0\} \}$ and 
whose actions are
\begin{eqnarray}
  \label{eq:2-2-13}
 K_{+} \kett{K}{n} &=& \sqrt{(n+1)(2K+n)}\kett{K}{n+1} , \nonumber \\
 K_{-} \kett{K}{n} &=& \sqrt{n(2K+n-1)}\kett{K}{n-1} ,  \nonumber   \\
 K_{3} \kett{K}{n} &=& (K+n)\kett{K}{n}, 
\end{eqnarray}
where $\kett{K}{0}$ is a normalized vacuum ($K_{-}\kett{K}{0}=0$ and 
$\langle K,0|K,0 \rangle =1$). We have written $\kett{K}{0}$ instead 
of $\ket{0}$  to emphasize the spin $K$ representation, see \cite{FKSF1}. 
We also denote by ${\bf 1}_{K}$ the unit operator on $\calh_K$. 
From (\ref{eq:2-2-13}), states $\kett{K}{n}$ are given by
\begin{equation}
  \label{eq:2-2-14}
 \kett{K}{n} =\frac{(K_+)^n}{\sqrt{n!(2K)_n}}\kett{K}{0} ,
\end{equation}
where $(a)_n$ is the Pochammer's notation 
$(a)_n \equiv  a(a+1) \cdots (a+n-1)$. 
These states satisfy the orthogonality and completeness conditions 
\begin{equation}
  \label{eq:2-2-16}
  \langle K,m \vert K,n \rangle =\delta_{mn}, 
 \quad \sum_{n=0}^{\infty}\kett{K}{n}\braa{K}{n}\ = \mathbf{1}_K.
\end{equation}
Now let us consider a generalized version of coherent states : 

\noindent{\bfseries Definition}\quad We call a state 
\begin{equation}
   \label{eq:2-2-17}
 \ket{w}  \equiv \mbox{e}^{wK_+ - \bar{w}K_-} \kett{K}{0}  
  \quad \mbox{for} \quad w \in \fukuso.
\end{equation}
the generalized coherent state based on $su(1,1)$, \cite{FS}.

\noindent  
We note that this is the extension of (\ref{eq:2-5-3}) not (\ref{eq:2-5-1}), 
see \cite{AP}. 
For this the following disentangling formula is well--known : 
\begin{eqnarray}
  \label{eq:2-2-20}
    \mbox{e}^{wK_{+} -\bar{w}K_{-}}  
  &=& \mbox{e}^{\zeta K_+}\mbox{e}^{\log (1-\vert\zeta\vert^2)K_3}
    \mbox{e}^{-\bar{\zeta}K_-} \quad \mbox{or}   \nonumber \\
  &=& \mbox{e}^{-\bar{\zeta}K_-}\mbox{e}^{-\log (1-\vert\zeta\vert^2)K_3}
    \mbox{e}^{\zeta K_+}.
\end{eqnarray}
where 
\begin{equation}
    \zeta = \zeta(w) \equiv 
    \frac{{w} \mbox{tanh}(\zettai{w})}{\zettai{w}}\ \ 
   (\Longrightarrow \zettai{\zeta} < 1). 
\end{equation}
This is the key formula for generalized coherent operators. Therefore from 
(\ref{eq:2-2-13}) 
\begin{equation}
  \label{eq:2-2-right}  
   \ket{w}=(1-\zettai{\zeta}^2)^{K}\mbox{e}^{\zeta K_{+}}\kett{K}{0}
          \equiv \ket{\zeta}.
\end{equation}
This corresponds to the right hand side of (\ref{eq:2-5-2}). Moreover 
since 
\[
    \mbox{e}^{\zeta K_{+}}\kett{K}{0}
      = \sum_{n=0}^{\infty}\frac{\zeta^n}{n!}K_{+}^{n}\kett{K}{0}
      = \sum_{n=0}^{\infty}\sqrt{\frac{(2K)_n}{n!}}
        \frac{\zeta^n K_{+}^{n}}{\sqrt{(2K)_{n}n!}}\kett{K}{0}
      = \sum_{n=0}^{\infty}\sqrt{\frac{(2K)_n}{n!}}\zeta^n \kett{K}{n}
\]
we have 
\begin{equation}
  \label{eq:2-2-left}  
   \ket{w}=(1-\zettai{\zeta}^2)^{K}
    \sum_{n=0}^{\infty}\sqrt{\frac{(2K)_n}{n!}}\zeta^n \kett{K}{n}
   \equiv \ket{\zeta}.
\end{equation}
This corresponds to the left hand side of (\ref{eq:2-5-2}).  
Then the resolution of unity corresponding to (\ref{eq:2-7}) is 
\begin{eqnarray}
  \label{eq:2-2-18}  
    &&\int_{\fukuso}\frac{2K-1}{\pi} \frac{\mbox{tanh}(\zettai{w})[d^{2}w]}
     {\left(1-\mbox{tanh}^2(\zettai{w})\right)\zettai{w}}
     \ket{w}\bra{w}
    =  
    \int_{\fukuso}\frac{2K-1}{\pi} \frac{\mbox{sinh}(2\zettai{w})[d^{2}w]}
     {2\zettai{w}}\ket{w}\bra{w}  \nonumber \\
   = &&\int_{\mbox{D}}\frac{2K-1}{\pi} \frac{[d^{2}\zeta]}{\left(1- \vert
    \zeta\vert^{2}\right)^2} \ket{\zeta}\bra{\zeta} = 
    \sum_{n=0}^{\infty}\kett{K}{n}\braa{K}{n}\ = \mathbf{1}_K,
\end{eqnarray}
where 
$\fukuso \rightarrow \mbox{D} : w \mapsto \zeta = \zeta(w) $ and 
$D$ is the Poincare disk in $\fukuso$, see \cite{KF7}. 

Here let us construct an example of spin $K$--representations.

\par \noindent 
If we set 
\begin{equation}
  \label{eq:2-2-21}
  K_+\equiv{1\over2}\left( a^{\dagger}\right)^2\ ,\
  K_-\equiv{1\over2}a^2\ ,\
  K_3\equiv{1\over2}\left( a^{\dagger}a+{1\over2}\right)\ , 
\end{equation}
then we have
\begin{equation}
  \label{eq:2-2-22}
  [K_3,K_+]=K_+\ ,\
  [K_3,K_-]=-K_-\ ,\
  [K_+,K_-]=-2K_3\ .
\end{equation}
That is, the set $\{K_+,K_-,K_3\}$ gives a unitary representation of $su(1,1)$
with spin $K = 1/4\ \mbox{and}\ 3/4$. Now we also call an operator 
\begin{equation}
  \label{eq:2-2-23}
   S(w) = \mbox{e}^{\frac{1}{2}\{w(a^{\dagger})^2 - \bar{w}a^2\}}
   \quad \mbox{for} \quad w \in \fukuso 
\end{equation}
the squeezed operator.

\subsection{Some Formulas}

We make some preliminaries for the following section. For that we list 
some useful formulas on generalized coherent states based ob $su(1,1)$. 
Since the proofs are not so difficult, we leave them to the readers. 

\noindent{\bfseries Formulas}\quad For $w_1, w_2$ we have
\begin{eqnarray}
  \label{eq:2-3-a-1}
 &&(\mbox{i})\quad \braket{w_1}{w_2}=
         \left\{\frac{(1-\zettai{\zeta_1}^2)(1-\zettai{\zeta_2}^2)}
                     {(1-{\bar \zeta_1}\zeta_2)^2} 
         \right\}^{K},    \\
&&{}  \nonumber \\
  \label{eq:2-3-b-2}
 &&(\mbox{ii})\quad 
      \bra{w_1}K_{+}\ket{w_2}=\braket{w_1}{w_2}
                   \frac{2K{\bar \zeta_1}}{1-{\bar \zeta_1}\zeta_2}\ ,  \\
&&{}  \nonumber \\
  \label{eq:2-3-c-3}
 &&(\mbox{iii})\quad  
      \bra{w_1}K_{-}\ket{w_2}=\braket{w_1}{w_2}
                   \frac{2K\zeta_2}{1-{\bar \zeta_1}\zeta_2}\ ,  \\
&&{}  \nonumber \\
  \label{eq:2-3-d-4}
 &&(\mbox{iv})\quad  
      \bra{w_1}K_{-}K_{+}\ket{w_2}=\braket{w_1}{w_2}
         \frac{2K+4K^2{\bar \zeta_1}\zeta_2}{(1-{\bar \zeta_1}\zeta_2)^2}\ .
\end{eqnarray}
where 
\begin{equation}
  \label{eq:two-zetas}
     \zeta_j=\frac{{w_j}\mbox{tanh}(\zettai{w_j})}{\zettai{w_j}}\quad 
             \mbox{for}\quad j=1,\ 2.
\end{equation}
When $w_1=w_2\equiv w$, then $\braket{w}{w}=1$, so we have 
\begin{eqnarray}
  \label{eq:kitaichi-1}
      &&\bra{w}K_{+}\ket{w}=
             \frac{2K{\bar \zeta}}{1-\zettai{\zeta}^2}\ , \quad 
      \bra{w}K_{-}\ket{w}
             \frac{2K\zeta}{1-\zettai{\zeta}^2}\ , \\
  \label{eq:kitaichi-2}
      &&\bra{w}K_{-}K_{+}\ket{w}=
         \frac{2K+4K^2\zettai{\zeta}^2}{(1-\zettai{\zeta}^2)^2}\ .
\end{eqnarray}

\par \noindent 
Here let us make a comment. From (\ref{eq:2-3-a-1}) 
\[
  \zettai{\braket{w_1}{w_2}}^2=
       \left\{\frac{(1-\zettai{\zeta_1}^2)(1-\zettai{\zeta_2}^2)}
                   {\zettai{1-{\bar \zeta_1}\zeta_2}^2} 
       \right\}^{2K}, 
\]
so we want to know the property of 
\[
   \frac{(1-\zettai{\zeta_1}^2)(1-\zettai{\zeta_2}^2)}
        {\zettai{1-{\bar \zeta_1}\zeta_2}^2}. 
\]
It is easy to see that
\begin{equation}
 \label{eq:futou-shiki-1}
 1 - \frac{(1-\zettai{\zeta_1}^2)(1-\zettai{\zeta_2}^2)}
          {\zettai{1-{\bar \zeta_1}\zeta_2}^2}
 =\frac{\zettai{\zeta_1-\zeta_2}^2}{\zettai{1-{\bar \zeta_1}\zeta_2}^2}
 \geq 0
\end{equation}
and $(\ref{eq:futou-shiki-1})=0$ if and only if (iff) $\zeta_1=\zeta_2$. 
Therefore 
\begin{equation}
 \label{eq:futou-shiki-2}
 \zettai{\braket{w_1}{w_2}}^2 = 
       \left\{\frac{(1-\zettai{\zeta_1}^2)(1-\zettai{\zeta_2}^2)}
                   {\zettai{1-{\bar \zeta_1}\zeta_2}^2} 
       \right\}^{2K} \leq 1
\end{equation}
because $2K>1$ ($2K-1 > 0$ from (\ref{eq:2-2-18})). Of course 
\begin{equation}
 \label{eq:futou-shiki-3}
  \zettai{\braket{w_1}{w_2}}=1\quad \mbox{iff}\quad \zeta_1=\zeta_2 
                              \quad \mbox{iff}\quad  w_1=w_2 .
\end{equation}
by (\ref{eq:two-zetas}).

\subsection{A Supplement}

Before ending this section let us make a brief comment on generalized 
coherent states (\ref{eq:2-2-17}). The coherent states $\ket{z}$ has been 
defined by (\ref{eq:2-5-1}) : $a\ket{z}=z\ket{z}$. Why do we define 
the generalized coherent states $\ket{w}$ as $K_{-}\ket{w}=w\ket{w}$ 
because $K_{-}$ is an annihilation operator corresponding to $a$ ? 
First let us try to calculate $K_{-}\ket{w}$ making use of 
(\ref{eq:2-2-right}). 
\[
  K_{-}\ket{w}=
     (1-\zettai{\zeta}^2)^{K}K_{-}\mbox{e}^{\zeta K_{+}}\kett{K}{0}
  =(1-\zettai{\zeta}^2)^{K}\mbox{e}^{\zeta K_{+}}\mbox{e}^{-\zeta K_{+}}
    K_{-}\mbox{e}^{\zeta K_{+}}\kett{K}{0}.
\]
Here it is easy to see 
\begin{eqnarray}
  \mbox{e}^{-\zeta K_{+}}K_{-}\mbox{e}^{\zeta K_{+}}
  &=&\sum_{n=0}^{\infty}\frac{1}{n!}
    [-\zeta K_{+},[-\zeta K_{+},[\ , \cdots, [-\zeta K_{+},K_{-}]\cdots ] ]]
     \nonumber \\
  &=&K_{-}+2\zeta K_{3}+\zeta^{2}K_{+}\ , \nonumber
\end{eqnarray}
from the relations (\ref{eq:2-2-12}), so that 
\begin{eqnarray}
   K_{-}\ket{w}&=&(1-\zettai{\zeta}^2)^{K}\mbox{e}^{\zeta K_{+}}
          (K_{-}+2\zeta K_{3}+\zeta^{2}K_{+})\kett{K}{0} \nonumber \\
   &=&2\zeta K(1-\zettai{\zeta}^2)^{K}\mbox{e}^{\zeta K_{+}}\kett{K}{0}
    + \zeta^{2}K_{+}(1-\zettai{\zeta}^2)^{K}\mbox{e}^{\zeta K_{+}}\kett{K}{0}
    \nonumber \\
   &=&(2K\zeta +\zeta^{2}K_{+})\ket{w} 
\end{eqnarray}
because $K_{-}\kett{K}{0}={\bf 0}$. Namely we have the equation
\begin{equation}
  \label{eq:perolomov-1}
 (K_{-}-\zeta^{2}K_{+})\ket{w}=2K\zeta \ket{w},\quad \mbox{where}\quad 
    \zeta=\frac{{w} \mbox{tanh}(\zettai{w})}{\zettai{w}}, 
\end{equation}
or more symmetrically 
\begin{equation}
  \label{eq:perolomov-2}
 ({\zeta}^{-1}K_{-}-\zeta K_{+})\ket{w}=2K\ket{w},\quad \mbox{where}\quad 
    \zeta=\frac{{w} \mbox{tanh}(\zettai{w})}{\zettai{w}}.  
\end{equation}
This equation is completely different from (\ref{eq:2-5-1}).

\subsection{Barut--Girardello Coherent States}

Now let us make a brief comment on Barut--Girardello coherent states, 
\cite{BGi}. 

\par \noindent 
The states $\vert\vert{w}\rangle\rangle$ ($w \in \fukuso$) defined by 
\begin{equation}
\label{eq:BG-equation}
     K_{-}\vert\vert{w}\rangle\rangle=w\vert\vert{w}\rangle\rangle 
\end{equation}
are called the Barut--Girardello coherent states. This definition is 
a natural genelization of (\ref{eq:2-5-1}) because $K_{-}$ is an 
annihilation operator. In the preceding section we denoted by a capital letter 
$K$ a level of the representation of $su(1,1)$. But to avoid some confusion 
in this subsection we use a small letter $k$ instead of $K$. 

\par \noindent
The solution is easy to find and given by 
\begin{equation}
\label{eq:BG-solution}
  \vert\vert{w}\rangle\rangle=\sum_{n=0}^{\infty}
     \frac{w^n}{\sqrt{n!(2k)_n}}\kett{k}{n}
\end{equation}
up to the normalization factor. Compare this with (\ref{eq:2-2-left}). 
Let us determine the inner product. 
\[
\langle\langle{w} \vert\vert{w'} \rangle\rangle
=\sum_{n=0}^{\infty} \frac{({\bar w}w')^{n}}{n!(2k)_{n}} 
=\sum_{n=0}^{\infty} \frac{(\sqrt{{\bar w}w'})^{2n}}{n!(2k)_{n}} 
\]
Noting that 
\[
(2k)_{n}=\frac{\Gamma(2k+n)}{\Gamma(2k)}\ \Longrightarrow \ 
\frac{1}{(2k)_{n}}=\frac{\Gamma(2k)}{\Gamma(2k+n)}
\]
we have 
\[
\langle\langle{w} \vert\vert{w'} \rangle\rangle=
\Gamma(2k)(\sqrt{{\bar w}w'})^{-2k+1}I_{2k-1}(2\sqrt{{\bar w}w'}),
\]
where $I_{\nu}(z)$ is the modified Bessel function of the first kind : 
\[
I_{\nu}(z)=\left(\frac{z}{2}\right)^{\nu}
\sum_{n=0}^{\infty} \frac{(z/2)^{2n}}{n!\Gamma(\nu+n+1)}.
\]
Therefore 
\begin{equation}
\vert\vert{w}\vert\vert=
{\langle\langle{w} \vert\vert{w'} \rangle\rangle}^{1/2}=
\left\{ \Gamma(2k)\zettai{w}^{-2k+1}I_{2k-1}(2\zettai{w}) \right\}^{1/2}.
\end{equation}
This gives the normalization factor of (\ref{eq:BG-solution}). 
Therefore the normalized solution of (\ref{eq:BG-equation}) corresponding to 
(\ref{eq:2-5-2}) is given by 
\begin{equation}
\vert\vert{w}\rangle\rangle=
\left\{ \Gamma(2k)\zettai{w}^{-2k+1}I_{2k-1}(2\zettai{w}) \right\}^{-1/2}
\sum_{n=0}^{\infty}\frac{w^n}{\sqrt{n!(2k)_n}}\kett{k}{n}.
\end{equation}

Next we show the resolution of unity. 
\begin{equation}
\label{eq:resolution-of-unity}
\int_{\fukuso} d{\mu}({\bar w},w)
\vert\vert{w}\rangle\rangle \langle\langle{w} \vert\vert
\equiv 
\int_{\fukuso} \frac{2K_{2k-1}(2\zettai{w})}{\pi\Gamma(2k)}[d^{2}w] 
\vert\vert{w}\rangle\rangle \langle\langle{w} \vert\vert
={\bf 1}_{k},
\end{equation}
where $K_{\nu}(z)$ is the modified Bessel function whose integral 
representation is given by 
\[
K_{\nu}(z)=\frac{\sqrt{\pi}}{\Gamma(\nu+1/2)}\left(\frac{z}{2}\right)^{\nu}
\int_{1}^{\infty} dy \mbox{e}^{-zy}(y^{2}-1)^{\nu-1/2}, \quad 
\nu > -\frac{1}{2}\ .
\]
The proof of (\ref{eq:resolution-of-unity}) is not so easy, so we give it. 
\begin{eqnarray}
\int_{\fukuso} d{\mu}({\bar w},w)\vert\vert{w}\rangle\rangle
\langle\langle{w}\vert\vert &=& 
\sum_{n=0}^{\infty} \sum_{m=0}^{\infty}\int_{\fukuso} d{\mu}({\bar w},w)
\frac{{\bar w}^{n}w^{m}}{\sqrt{n!(2k)_{n}m!(2k)_{m}}}
\kett{k}{n}\braa{k}{m} \\ \nonumber 
&=&\sum_{n=0}^{\infty}\int_{0}^{\infty} d{\mu}(r)
\frac{r^{2n}}{n!(2k)_{n}}\kett{k}{n}\braa{k}{n} \nonumber \\
&=&\sum_{n=0}^{\infty}\frac{1}{n!(2k)_{n}} \left\{
\int_{0}^{\infty} d{\mu}(r) r^{2n} \right\}
\kett{k}{n}\braa{k}{n} \nonumber 
\end{eqnarray}
where we have integrated on $\theta$ making use of $w=r\mbox{e}^{i\theta}$.
Here 
\begin{eqnarray}
\int_{0}^{\infty} d{\mu}(r) r^{2n} &=&
\frac{4}{\Gamma(2k)}\int_{0}^{\infty} dr r^{2k+2n}K_{2k-1}(2r)
\nonumber \\
&=&\frac{4}{\Gamma(2k)}\frac{1}{4}\Gamma(\frac{2k+2n+1+2k-1}{2})
\Gamma(\frac{2k+2n+1-2k+1}{2}) \nonumber \\
&=&\frac{\Gamma(2k+n)}{\Gamma(2k)}\Gamma(n+1)
=(2k)_{n}n!\ , \nonumber 
\end{eqnarray}
where we have used the famous formula 
\[
\int_{0}^{\infty}dx x^{\mu-1}K_{\nu}(ax)=
\frac{1}{4}\left(\frac{2}{a}\right)
\Gamma(\frac{\mu+\nu}{2})\Gamma(\frac{\mu-\nu}{2}) \qquad 
a>0, \quad  \mbox{Re}{\mu}>|\mbox{Re}{\nu}|\ .
\]
For the proof see \cite{FF1}; Appendix B. \quad Therefore we have 
\[
\int_{\fukuso} d{\mu}({\bar w},w)\vert\vert{w}\rangle\rangle
\langle\langle{w}\vert\vert 
=
\sum_{n=0}^{\infty}\frac{1}{n!(2k)_{n}}(2k)_{n}n!
\kett{k}{n}\braa{k}{n} 
=
\sum_{n=0}^{\infty}\kett{k}{n}\braa{k}{n}={\bf 1}_{k}.
\]

\par \noindent 
Their states have several interesting structures, but we don't consider 
them in this paper.  
See \cite{FF1}, \cite{FF2} and \cite{Tri} as for further developments and 
applications.  

\par \vspace{5mm} \noindent
A comment is in order. \quad Here let us compare two types of coherent 
states based on Lie algebra $su(1,1)$ $\cdots$ Perelomov type 
(section 3.1) and Barut--Girardello one (section 3.4). 
The measures satisfying resolution of unity must be positive, so we have 

\par \noindent
(1) Perelomov type \qquad \qquad $K > \frac{1}{2}$ \ ($\Longleftarrow$ 
(\ref{eq:2-2-18}))

\par \noindent
(2) Barut--Girardello type \quad  $K > 0$ \ ($\Longleftarrow$ 
(\ref{eq:resolution-of-unity}))

\vspace{10mm}
\section{Generalized Coherent States Based on su(2)}

In this section we introduce some basic properties of generalized 
coherent operators based on $su(2)$, see \cite{FKSF1} or \cite{AP}.     

\subsection{General Theory}

We consider a spin $J\ (> 0)$ representation of $su(2) 
\subset sl(2,\fukuso)$ and set its generators 
$\{J_{+}, J_{-}, J_{3}\}\ ((J_+)^{\dagger} = J_-)$, 
\begin{equation}
  \label{eq:3-2-12}
 [J_3, J_+]=J_+, \quad [J_3, J_-]=-J_-, \quad [J_+, J_-]=2J_3.
\end{equation}
We note that this (unitary) representation is necessarily finite 
dimensional. 
The Fock space on which $\{ J_+, J_-, J_3 \}$ act is 
$\calh_{J} \equiv \{\kett{J}{m} \vert 0\le m \le 2J \}$ and 
whose actions are
\begin{eqnarray}
  \label{eq:3-2-13}
 J_{+} \kett{J}{m} &=& \sqrt{(m+1)(2J-m)}\kett{J}{m+1} , \nonumber \\
 J_{-} \kett{J}{m} &=& \sqrt{m(2J-m+1)}\kett{J}{m-1} ,  \nonumber   \\
 J_{3} \kett{J}{m} &=& (-J+m)\kett{J}{m}, 
\end{eqnarray}
where $\kett{J}{0}$ is a normalized vacuum ($J_{-}\kett{J}{0}=0$ and 
$\langle J,0|J,0 \rangle =1$). We have written $\kett{J}{0}$ instead 
of $\ket{0}$  to emphasize the spin $J$ representation, see \cite{FKSF1}. 
We also denote by ${\bf 1}_{J}$ the unit operator on $\calh_{J}$. 
From (\ref{eq:2-2-13}), states $\kett{J}{m}$ are given by
\begin{equation}
  \label{eq:3-2-14}
 \kett{J}{m} =\frac{(J_+)^m}{\sqrt{m!{}_{2J}\mbox{P}_m}}\kett{J}{0} ,
\end{equation}
where ${}_{2J}\mbox{P}_m = (2J)(2J-1)\cdots(2J-m+1)$.  
These states satisfy the orthogonality and completeness conditions 
\begin{equation}
  \label{eq:3-2-16}
  \langle J,m \vert J,n \rangle =\delta_{mn}, 
 \quad \sum_{m=0}^{2J}\kett{J}{m}\braa{J}{m}\ = \mathbf{1}_J.
\end{equation}
Now let us consider a generalized version of coherent states : 

\noindent{\bfseries Definition}\quad We call a state 
\begin{equation}
   \label{eq:3-2-17}
 \ket{v}  \equiv \mbox{e}^{vJ_{+} - \bar{v}J_{-}} \kett{J}{0}  
  \quad \mbox{for} \quad v \in \fukuso.
\end{equation}
the generalized coherent state based on $su(2)$, \cite{FS}.

\noindent  
We note that this is the extension of (\ref{eq:2-5-3}) not (\ref{eq:2-5-1}), 
see \cite{AP}. 
For this the following disentangling formula is well--known : 
\begin{eqnarray}
  \label{eq:3-2-20}
    \mbox{e}^{vJ_{+} -\bar{v}J_{-}}  
  &=& \mbox{e}^{\eta J_+}\mbox{e}^{\log (1+\vert\eta\vert^2)J_3}
    \mbox{e}^{-\bar{\eta}J_-} \quad \mbox{or}   \nonumber \\
  &=& \mbox{e}^{-\bar{\eta}J_-}\mbox{e}^{-\log (1+\vert\eta\vert^2)J_3}
    \mbox{e}^{\eta J_+}.
\end{eqnarray}
where 
\begin{equation}
    \eta = \eta(v) \equiv 
    \frac{{v} \mbox{tan}(\zettai{v})}{\zettai{v}}\ . 
\end{equation}
This is the key formula for generalized coherent operators. Therefore from 
(\ref{eq:3-2-13}) 
\begin{equation}
  \label{eq:3-2-right}  
   \ket{v}=\frac{1}{(1+\zettai{\eta}^2)^{J}}\mbox{e}^{\eta J_{+}}\kett{J}{0}
          \equiv \ket{\eta}.
\end{equation}
This corresponds to the right hand side of (\ref{eq:2-5-2}). Moreover 
since 
\begin{eqnarray}
    \mbox{e}^{\eta J_{+}}\kett{J}{0}
     &=& \sum_{m=0}^{\infty}\frac{\eta^m}{m!}J_{+}^{m}\kett{J}{0}
      = \sum_{m=0}^{\infty}\sqrt{\frac{{}_{2J}\mbox{P}_m}{m!}}
        \frac{\eta^m J_{+}^{m}}{\sqrt{{}_{2J}\mbox{P}_m m!}}\kett{J}{0}
      = \sum_{m=0}^{2J}\sqrt{\frac{{}_{2J}\mbox{P}_m}{m!}}\eta^m \kett{J}{m}
          \nonumber \\
     &=& \sum_{m=0}^{2J}\sqrt{{}_{2J}\mbox{C}_m}\eta^m \kett{J}{m}
\end{eqnarray}
we have 
\begin{equation}
  \label{eq:3-2-left}  
   \ket{v}=\frac{1}{(1+\zettai{\eta}^2)^{J}}
    \sum_{m=0}^{2J}\sqrt{{}_{2J}\mbox{C}_m}\eta^m \kett{J}{m}
   \equiv \ket{\eta}.
\end{equation}
This corresponds to the left hand side of (\ref{eq:2-5-2}).  
Then the resolution of unity corresponding to (\ref{eq:2-7}) is 
\begin{eqnarray}
  \label{eq:3-2-18}  
    &&\int_{\fukuso}\frac{2J+1}{\pi} \frac{\mbox{tan}(\zettai{v})[d^{2}v]}
     {\left(1+\mbox{tan}^{2}(\zettai{v})\right)\zettai{v}}\ket{v}\bra{v} 
   = 
    \int_{\fukuso}\frac{2J+1}{\pi} \frac{\mbox{sin}(2\zettai{v})[d^{2}v]}
     {2\zettai{v}}\ket{v}\bra{v} 
 \nonumber \\
   = &&\int_{\fukuso}\frac{2J+1}{\pi} \frac{[d^{2}\eta]}{\left(1+ \vert
    \eta\vert^{2}\right)^2} \ket{\eta}\bra{\eta} = 
    \sum_{m=0}^{2J}\kett{J}{m}\braa{J}{m}\ = \mathbf{1}_J,
\end{eqnarray}
where 
$\fukuso \rightarrow \fukuso \subset {\fukuso}\mbox{P}^{1} : v \mapsto 
\eta = \eta(v) $, see \cite{KF7}.

\subsection{Some Formulas}

We make some preliminaries for the following section. For that we list 
some useful formulas on generalized coherent states based on $su(2)$. 
Since the proofs are not so difficult, we leave them to the readers. 

\noindent{\bfseries Formulas}\quad For $v_1, v_2$ we have
\begin{eqnarray}
  \label{eq:3-3-a-1}
 &&(\mbox{i})\quad \braket{v_1}{v_2}=
         \left\{\frac{(1+{\bar \eta_1}\eta_2)^2}
                     {(1+\zettai{\eta_1}^2)(1+\zettai{\eta_2}^2)}
         \right\}^{J},    \\
&&{}  \nonumber \\
  \label{eq:3-3-b-2}
 &&(\mbox{ii})\quad 
      \bra{v_1}J_{+}\ket{v_2}=\braket{v_1}{v_2}
                   \frac{2J{\bar \eta_1}}{1+{\bar \eta_1}\eta_2}\ ,  \\
&&{}  \nonumber \\
  \label{eq:3-3-c-3}
 &&(\mbox{iii})\quad  
      \bra{v_1}J_{-}\ket{v_2}=\braket{v_1}{v_2}
                   \frac{2J\eta_2}{1+{\bar \eta_1}\eta_2}\ ,  \\
&&{}  \nonumber \\
  \label{eq:3-3-d-4}
 &&(\mbox{iv})\quad  
      \bra{v_1}J_{-}J_{+}\ket{v_2}=\braket{v_1}{v_2}
         \frac{2J+4J^2{\bar \eta_1}\eta_2}{(1+{\bar \eta_1}\eta_2)^2}\ .
\end{eqnarray}
where 
\begin{equation}
  \label{eq:two-etas}
     \eta_j=\frac{{v_j}\mbox{tan}(\zettai{v_j})}{\zettai{v_j}}\quad 
             \mbox{for}\quad j=1,\ 2.
\end{equation}
When $v_1=v_2\equiv v$, then $\braket{v}{v}=1$, so we have 
\begin{eqnarray}
  \label{eq:kitaichi-1-J}
      &&\bra{v}J_{+}\ket{v}=
             \frac{2J{\bar \eta}}{1+\zettai{\eta}^2}\ , \quad 
      \bra{v}J_{-}\ket{v}
             \frac{2J\eta}{1+\zettai{\eta}^2}\ , \\
  \label{eq:kitaichi-2-J}
      &&\bra{v}J_{-}J_{+}\ket{v}=
         \frac{2J+4J^2\zettai{\eta}^2}{(1+\zettai{\eta}^2)^2}\ .
\end{eqnarray}

\par \noindent 
Here let us make a comment. From (\ref{eq:3-3-a-1}) 
\[
  \zettai{\braket{v_1}{v_2}}^2=
       \left\{\frac{\zettai{1+{\bar \eta_1}\eta_2}^2}
                   {(1+\zettai{\eta_1}^2)(1+\zettai{\eta_2}^2)}
       \right\}^{2J}, 
\]
so we want to know the property of 
\[
   \frac{\zettai{1+{\bar \eta_1}\eta_2}^2}
        {(1+\zettai{\eta_1}^2)(1+\zettai{\eta_2}^2)}. 
\]
It is easy to see that
\begin{equation}
 \label{eq:futou-shiki-1-J}
 1 - \frac{\zettai{1+{\bar \eta_1}\eta_2}^2}
          {(1+\zettai{\eta_1}^2)(1+\zettai{\eta_2}^2)}
 =\frac{\zettai{\eta_1-\eta_2}^2}
       {(1+\zettai{\eta_1}^2)(1+\zettai{\eta_2}^2)}
 \geq 0
\end{equation}
and $(\ref{eq:futou-shiki-1-J})=0$ if and only if (iff) $\eta_1=\eta_2$. 
Therefore 
\begin{equation}
 \label{eq:futou-shiki-2-J}
 \zettai{\braket{v_1}{v_2}}^2 = 
       \left\{\frac{\zettai{1+{\bar \eta_1}\eta_2}^2}
                   {(1+\zettai{\eta_1}^2)(1+\zettai{\eta_2}^2)} 
       \right\}^{2J} \leq 1
\end{equation}
because $2J>1$ (from (\ref{eq:3-2-18})). Of course 
\begin{equation}
 \label{eq:futou-shiki-3-J}
  \zettai{\braket{v_1}{v_2}}=1\quad \mbox{iff}\quad \eta_1=\eta_2 
                              \quad \mbox{iff}\quad  v_1=v_2 .
\end{equation}
by (\ref{eq:two-etas}).

\subsection{A Supplement}

Before ending this section let us make a brief comment on generalized 
coherent states (\ref{eq:3-2-17}). The coherent states $\ket{z}$ has been 
defined by (\ref{eq:2-5-1}) : $a\ket{z}=z\ket{z}$. Why do we define 
the generalized coherent states $\ket{w}$ as $J_{-}\ket{v}=v\ket{v}$ 
because $J_{-}$ is an annihilation operator corresponding to $a$ ? 
First let us try to calculate $J_{-}\ket{v}$ making use of 
(\ref{eq:3-2-right}). 
\[
  J_{-}\ket{v}=
     (1+\zettai{\eta}^2)^{-J}J_{-}\mbox{e}^{\eta J_{+}}\kett{J}{0}
  =(1+\zettai{\eta}^2)^{-J}\mbox{e}^{\eta J_{+}}\mbox{e}^{-\eta J_{+}}
    J_{-}\mbox{e}^{\eta J_{+}}\kett{J}{0}.
\]
Here it is easy to see 
\begin{eqnarray}
  \mbox{e}^{-\eta J_{+}}J_{-}\mbox{e}^{\eta J_{+}}
  &=&\sum_{m=0}^{\infty}\frac{1}{m!}
    [-\eta J_{+},[-\eta J_{+},[\ , \cdots, [-\eta J_{+},J_{-}]\cdots ] ]]
     \nonumber \\
  &=&J_{-}-2\eta J_{3}-\eta^{2}J_{+}\ , \nonumber
\end{eqnarray}
from the relations (\ref{eq:3-2-12}), so that 
\begin{eqnarray}
   J_{-}\ket{v}&=&(1+\zettai{\zeta}^2)^{-J}\mbox{e}^{\eta J_{+}}
          (J_{-}-2\eta J_{3}-\eta^{2}J_{+})\kett{J}{0} \nonumber \\
   &=&2\eta J(1+\zettai{\eta}^2)^{-J}\mbox{e}^{\eta J_{+}}\kett{J}{0}
    - \eta^{2}J_{+}(1+\zettai{\zeta}^2)^{-J}\mbox{e}^{\eta J_{+}}\kett{J}{0}
    \nonumber \\
   &=&(2J\eta -\eta^{2}J_{+})\ket{v} 
\end{eqnarray}
because $J_{-}\kett{J}{0}={\bf 0}$ and $J_{3}\kett{J}{0}=-J\kett{J}{0}$.  
Namely we have the equation
\begin{equation}
  \label{eq:perolomov-1-J}
 (J_{-}+\eta^{2}J_{+})\ket{v}=2J\eta \ket{v},\quad \mbox{where}\quad 
    \eta=\frac{{v}\mbox{tan}(\zettai{v})}{\zettai{v}}, 
\end{equation}
or more symmetrically 
\begin{equation}
  \label{eq:perolomov-2-J}
 ({\eta}^{-1}J_{-}+\eta J_{+})\ket{v}=2J\ket{v},\quad \mbox{where}\quad 
    \eta=\frac{{v}\mbox{tan}(\zettai{v})}{\zettai{v}}.  
\end{equation}
This equation is completely different from (\ref{eq:2-5-1}).

\vspace{10mm}
\section{Schwinger's Boson Method}

Here let us construct the spin $J$ and $K$--representations making 
use of Schwinger's boson method.

\par \noindent
We consider the system of two-harmonic oscillators. If we set
\begin{equation}
  \label{eq:4-2-24}
  a_1 = a \otimes 1,\  {a_1}^{\dagger} = a^{\dagger} \otimes 1;\ 
  a_2 = 1 \otimes a,\  {a_2}^{\dagger} = 1 \otimes a^{\dagger},
\end{equation}
then it is easy to see 
\begin{equation}
  \label{eq:4-2-25}
 [a_i, a_j] = [{a_i}^{\dagger}, {a_j}^{\dagger}] = 0,\ 
 [a_i, {a_j}^{\dagger}] = \delta_{ij}, \quad i, j = 1, 2. 
\end{equation}
We also denote by $N_{i} = {a_i}^{\dagger}a_i$ number operators.

Now we can construct representation of Lie algebras $su(2)$ and $su(1,1)$ by 
making use of Schwinger's boson method, see \cite{FKSF1}, \cite{FKSF2}. 
Namely if we set
\begin{eqnarray}
  \label{eq:4-2-26-1}
  su(2) &:&\quad
     J_+ = {a_1}^{\dagger}a_2,\ J_- = {a_2}^{\dagger}a_1,\ 
     J_3 = {1\over2}\left({a_1}^{\dagger}a_1 - {a_2}^{\dagger}a_2\right), \\
  \label{eq:4-2-26-2}
  su(1,1) &:&\quad
     K_+ = {a_1}^{\dagger}{a_2}^{\dagger},\ K_- = a_2 a_1,\ 
     K_3 = {1\over2}\left({a_1}^{\dagger}a_1 + {a_2}^{\dagger}a_2  + 1\right),
\end{eqnarray}
then we have
\begin{eqnarray}
  \label{eq:4-2-27-1}
  su(2) &:&\quad
     [J_3, J_+] = J_+,\ [J_3, J_-] = - J_-,\ [J_+, J_-] = 2J_3, \\
  \label{eq:4-2-27-2}
  su(1,1) &:&\quad
     [K_3, K_+] = K_+,\ [K_3, K_-] = - K_-,\ [K_+, K_-] = -2K_3.
\end{eqnarray}

In the following we define (unitary) generalized coherent operators 
based on Lie algebras $su(2)$ and $su(1,1)$. 

\noindent{\bfseries Definition}\quad We set 
\begin{eqnarray}
  \label{eq:4-2-28-1}
  su(2) &:&\quad 
U_{J}(v) = e^{vJ_{+} - \bar{v}J_{-}}\quad {\rm for}\quad v \in \fukuso , \\
  \label{eq:4-2-28-2}
  su(1,1) &:&\quad 
U_{K}(w) = e^{wK_{+} - \bar{w}K_{-}}\quad {\rm for}\quad w \in \fukuso.
\end{eqnarray}

\par \noindent 
For the latter convenience let us list well-known disentangling formulas  
once more. We have
\begin{eqnarray}
  \label{eq:j-formula}
  su(2) &:&\quad 
U_{J}(v)  = e^{\eta J_{+}}
 e^{{\rm log}\left(1 + \zettai{\eta}^{2}\right)J_{3}}
 e^{- \bar{\eta}J_{-}}, \quad 
   where\quad  \eta = \frac{{v}{\rm tan}{(\zettai{v})}}
                           {\zettai{v}},  \\
  \label{eq:k-formula}
  su(1,1) &:&\quad 
U_{K}(w) =  e^{\zeta K_{+}}
 e^{{\rm log}\left(1 - \zettai{\zeta}^{2}\right)K_{3}}
 e^{- \bar{\zeta}K_{-}},\quad 
   where\quad  \zeta = \frac{{w}{\rm tanh}{(\zettai{w})}}
                             {\zettai{w}}. 
 \end{eqnarray}
For the proof see Appendix A. 
As for a generalization of these formulas see \cite{FS}.\quad  

\par \noindent
Now let us make some mathematical preliminaries for the latter sections. 
We have easily
\begin{eqnarray}
  \label{eq:J-rotation-1}
  U_{J}(t)a_{1}U_{J}(t)^{-1}&=&
      cos(\zettai{t})a_{1}-\frac{tsin(\zettai{t})}{\zettai{t}}a_{2}, \\
  \label{eq:J-rotation-2}
  U_{J}(t)a_{2}U_{J}(t)^{-1}&=&
      cos(\zettai{t})a_{1}+\frac{{\bar t}sin(\zettai{t})}{\zettai{t}}a_{2},
\end{eqnarray}
so the map 
$
(a_{1},a_{2}) \longrightarrow 
          (U_{J}(t)a_{1}U_{J}(t)^{-1},U_{J}(t)a_{2}U_{J}(t)^{-1}) 
$
is 
\[
(U_{J}(t)a_{1}U_{J}(t)^{-1},U_{J}(t)a_{2}U_{J}(t)^{-1})=
(a_{1},a_{2})
\left(
  \begin{array}{cc}
     cos(\zettai{t})& \frac{{\bar t}sin(\zettai{t})}{\zettai{t}}\\
     -\frac{tsin(\zettai{t})}{\zettai{t}}& cos(\zettai{t})
   \end{array}
 \right).
\]
We note that 
\[
\left(
  \begin{array}{cc}
     cos(\zettai{t})& \frac{{\bar t}sin(\zettai{t})}{\zettai{t}}\\
     -\frac{tsin(\zettai{t})}{\zettai{t}}& cos(\zettai{t})
   \end{array}
 \right) \in SU(2).
\]

\par \noindent 
On the other hand we have easily
\begin{eqnarray}
  \label{eq:K-rotation-1}
  U_{K}(t)a_{1}U_{K}(t)^{-1}&=&
  cosh(\zettai{t})a_{1}-\frac{tsinh(\zettai{t})}{\zettai{t}}a_{2}^{\dagger}, \\
  \label{eq:K-rotation-2}
  U_{K}(t)a_{2}^{\dagger}U_{K}(t)^{-1}&=&
  cosh(\zettai{t})a_{2}^{\dagger}-\frac{{\bar t}sinh(\zettai{t})}{\zettai{t}}
  a_{1},
\end{eqnarray}
so the map 
$
(a_{1},a_{2}^{\dagger}) \longrightarrow 
   (U_{K}(t)a_{1}U_{K}(t)^{-1},U_{K}(t)a_{2}^{\dagger}U_{K}(t)^{-1}) 
$
is 
\[
(U_{K}(t)a_{1}U_{K}(t)^{-1},U_{K}(t)a_{2}^{\dagger}U_{K}(t)^{-1})=
(a_{1},a_{2}^{\dagger})
\left(
  \begin{array}{cc}
     cosh(\zettai{t})& -\frac{{\bar t}sinh(\zettai{t})}{\zettai{t}}\\
     -\frac{tsinh(\zettai{t})}{\zettai{t}}& cosh(\zettai{t})
   \end{array}
 \right).
\]
We note that 
\[
\left(
  \begin{array}{cc}
     cosh(\zettai{t})& -\frac{{\bar t}sinh(\zettai{t})}{\zettai{t}}\\
     -\frac{tsinh(\zettai{t})}{\zettai{t}}& cosh(\zettai{t})
   \end{array}
 \right) \in SU(1,1). 
\]

\par \vspace{10mm} \noindent 
Before ending this section let us ask a question.

What is a relation between (\ref{eq:4-2-28-2}) and (\ref{eq:2-2-23}) 
of generalized coherent operators based on $su(1.1)$ ?

\noindent
The answer is given by :

\noindent{\bfseries Formula}\quad We have 
\begin{equation}
  \label{eq:4-2-29}
  U_{J}(-\frac{\pi}{4})S_{1}(w)S_{2}(-w)U_{J}(-\frac{\pi}{4})^{-1} = U_{K}(w),
\end{equation}
where $S_{j}(w)=(\ref{eq:2-2-23})$ with $a_{j}$ instead of $a$, see 
\cite{MP}.

\noindent 
Namely, $U_{K}(w)$ is given by ``rotating'' the product $S_{1}(w)S_{2}(-w)$ 
by $U_{J}(-\frac{\pi}{4})$. 

\noindent{\bfseries Proof}\quad  It is easy to see 
\begin{equation}
 U_{J}(t)S_{1}(w)S_{2}(-w)U_{J}(t)^{-1} 
=U_{J}(t)\mbox{e}^{
   \frac{w}{2}\left\{(a_{1}^{\dagger})^{2}-(a_{2}^{\dagger})^{2}\right\} 
 - \frac{{\bar w}}{2}\left\{(a_{1})^{2}-(a_{2})^{2}\right\}
                  }
 U_{J}(t)^{-1}=\mbox{e}^{\mbox{X}}
\end{equation}  
where
\begin{eqnarray}
 \mbox{X}&=&
    \frac{w}{2}\left\{(U_{J}(t)a_{1}^{\dagger}U_{J}(t)^{-1})^{2}- 
                     (U_{J}(t)a_{2}^{\dagger}U_{J}(t)^{-1})^{2}
               \right\} \nonumber \\
    &-& \frac{{\bar w}}{2}\left\{(U_{J}(t)a_{1}U_{J}(t)^{-1})^{2}- 
                             (U_{J}(t)a_{2}U_{J}(t)^{-1})^{2} 
                         \right\}. 
\end{eqnarray}
From (\ref{eq:J-rotation-1}) and (\ref{eq:J-rotation-2}) we have 
\begin{eqnarray}
 &&\mbox{X}= \nonumber \\
 &&{}\ \ \frac{w}{2}\left\{
   \left(cos^{2}(\zettai{t})-\frac{t^2sin^{2}(\zettai{t})}{\zettai{t}^{2}}
   \right)(a_{1}^{\dagger})^2
 - \left(cos^{2}(\zettai{t})-\frac{{\bar t}^2sin^{2}(\zettai{t})}
         {\zettai{t}^{2}}
   \right)(a_{2}^{\dagger})^2
 - \frac{(t+{\bar t})sin(2\zettai{t})}{\zettai{t}}
         a_{1}^{\dagger}a_{2}^{\dagger}
        \right\}      \nonumber \\
 &&-\frac{{\bar w}}{2}\left\{
   \left(cos^{2}(\zettai{t})-\frac{{\bar t}^2sin^{2}(\zettai{t})}
   {\zettai{t}^{2}}
   \right)a_{1}^2
 - \left(cos^{2}(\zettai{t})-\frac{t^2sin^{2}(\zettai{t})}{\zettai{t}^{2}}
   \right)a_{2}^2
 - \frac{(t+{\bar t})sin(2\zettai{t})}{\zettai{t}}a_{1}a_{2}
        \right\}. \nonumber \\
 &&{} 
\end{eqnarray}
Here we set $t=\frac{-\pi}{4}$, then 
\[
 \mbox{X}= \frac{w}{2}(2a_{1}^{\dagger}a_{2}^{\dagger})
           -\frac{{\bar w}}{2}(2a_{1}a_{2})
         = {w}a_{1}^{\dagger}a_{2}^{\dagger}-{\bar w}a_{1}a_{2} 
         \ \Longrightarrow\  \mbox{e}^{X}=U_{K}(w). 
\]
Namely, we obtain the formula. 

\vspace{5mm}
Next let us prove the following 

\noindent{\bfseries Formula}\quad 
\begin{equation}
\label{eq:squeezed-adjoint-formula}
 U_{J}(t)S_{1}(\alpha)S_{2}(\beta)U_{J}(t)^{-1} 
=U_{J}(t)\mbox{e}^{\left\{
   \frac{\alpha}{2}(a_{1}^{\dagger})^{2}-
   \frac{{\bar \alpha}}{2}(a_{1})^{2}
 + \frac{\beta}{2}(a_{2}^{\dagger})^{2}-
   \frac{{\bar \beta}}{2}(a_{2})^{2}
                   \right\}
                  }U_{J}(t)^{-1}=\mbox{e}^{\mbox{X}}
\end{equation}  
where
\begin{eqnarray}
 \mbox{X}&=&
           \frac{\alpha}{2}(U_{J}(t)a_{1}^{\dagger}U_{J}(t)^{-1})^{2}- 
           \frac{{\bar \alpha}}{2}(U_{J}(t)a_{1}U_{J}(t)^{-1})^{2}
     \nonumber \\
 &+&       \frac{\beta}{2}(U_{J}(t)a_{2}^{\dagger}U_{J}(t)^{-1})^{2}- 
           \frac{{\bar \beta}}{2}(U_{J}(t)a_{2}U_{J}(t)^{-1})^{2}. 
     \nonumber 
\end{eqnarray}
From (\ref{eq:J-rotation-1}) and (\ref{eq:J-rotation-2}) we have 
\begin{eqnarray}
\mbox{X}&=&
\frac{1}{2}\left\{cos^{2}(\zettai{t})\alpha + 
\frac{t^2 sin^{2}(\zettai{t})}{\zettai{t}^2}\beta\right\}
    (a_{1}^{\dagger})^2
-\frac{1}{2}\left\{cos^{2}(\zettai{t}){\bar \alpha} + 
\frac{{\bar t}^2 sin^{2}(\zettai{t})}{\zettai{t}^2}{\bar \beta}\right\}
a_{1}^2  \nonumber \\
&+&\frac{1}{2}\left\{cos^{2}(\zettai{t})\beta + 
\frac{{\bar t}^2 sin^{2}(\zettai{t})}{\zettai{t}^2}\alpha\right\}
    (a_{2}^{\dagger})^2
-\frac{1}{2}\left\{cos^{2}(\zettai{t}){\bar \beta} + 
\frac{t^2 sin^{2}(\zettai{t})}{\zettai{t}^2}{\bar \alpha}\right\}a_{2}^2
\nonumber \\
&+&(\beta t-\alpha {\bar t})\frac{sin(2\zettai{t})}{2\zettai{t}}
      a_{1}^{\dagger}a_{2}^{\dagger}
-({\bar \beta}{\bar t}-{\bar \alpha}t)\frac{sin(2\zettai{t})}{2\zettai{t}}
      a_{1}a_{2}. 
\end{eqnarray}
If we set 
\begin{equation}
\label{eq:strong-condition}
\beta t-\alpha {\bar t}=0 \Longleftrightarrow \beta t=\alpha {\bar t}, 
\end{equation}
then it is easy to check 
\[
cos^{2}(\zettai{t})\alpha + 
\frac{t^2 sin^{2}(\zettai{t})}{\zettai{t}^2}\beta=\alpha, 
\quad 
cos^{2}(\zettai{t})\beta + 
\frac{{\bar t}^2 sin^{2}(\zettai{t})}{\zettai{t}^2}\alpha=\beta, 
\]
so, in this case, 
\[
X=\frac{1}{2}\alpha(a_{1}^{\dagger})^2-\frac{1}{2}{\bar \alpha}a_{1}^2 + 
  \frac{1}{2}\beta(a_{2}^{\dagger})^2-\frac{1}{2}{\bar \beta}a_{2}^2\ . 
\]
Therefore 
\begin{equation}
\label{eq:2-invariant-property}
U_{J}(t)S_{1}(\alpha)S_{2}(\beta)U_{J}(t)^{-1}=
S_{1}(\alpha)S_{2}(\beta).
\end{equation}
That is, $S_{1}(\alpha)S_{2}(\beta)$ commutes with $U_{J}(t)$ under the 
condition (\ref{eq:strong-condition}). 
We use this formula in the following.

\vspace{10mm}
\section{Universal Bundles and Chern Characters}

In this section we introduce some basic properties of pull--backed ones 
of universal bundles over the infinite--dimensional Grassmann manifolds 
and Chern characters, see \cite{MN}.

Let $\calh$ be a separable Hilbert space over $\fukuso$.
For $m\in{\bf N}$, we set
\begin{equation}
  \label{eq:stmh}
  \kansu{\stm}{\cal H}
  \equiv
  \left\{
    V=\left(v_1,\cdots,v_m\right)
    \in
    \calh\times\cdots\times\calh\ \vert\  V^\dagger V \in GL(m;\fukuso)
  \right\}\ .
\end{equation}
This is called a (universal) Stiefel manifold.
Note that the unitary group $U(m)$ acts on $\kansu{\stm}{\calh}$
from the right :
\begin{equation}
  \label{eq:stmsha}
  \kansu{\stm}{\calh}\times\kansu{U}{m}
  \longrightarrow
  \kansu{\stm}{\calh}\  :\  \left( V,a\right)\longmapsto Va\ .
\end{equation}
Next we define a (universal) Grassmann manifold
\begin{equation}
  \kansu{\grm}{\calh}
  \equiv
  \left\{
    X\in\kansu{M}{\calh}\ \vert\ 
    X^2=X, X^\dagger=X\  \mathrm{and}\  \mathrm{tr}X=m\right\}\ ,
\end{equation}
where $M(\calh)$ denotes a space of all bounded linear operators on $\calh$.
Then we have a projection
\begin{equation}
  \label{eq:piteigi}
  \pi : \kansu{\stm}{\calh}\longrightarrow\kansu{\grm}{\calh}\ ,
  \quad \kansu{\pi}{V}\equiv V(V^{\dagger}V)^{-1}V^\dagger\ ,
\end{equation}
compatible with the action (\ref{eq:stmsha}) 
($\kansu{\pi}{Va}=Va\{a^{-1}(V^{\dagger}V)^{-1}a\}(Va)^\dagger
=\kansu{\pi}{V}$).

Now the set
\begin{equation}
  \label{eq:principal}
  \left\{
    \kansu{U}{m}, \kansu{\stm}{\calh}, \pi, \kansu{\grm}{\calh}
  \right\}\ ,
\end{equation}
is called a (universal) principal $U(m)$ bundle, 
see \cite{MN} and \cite{KF1}. \quad We set
\begin{equation}
  \label{eq:emh}
  \kansu{\eem}{\cal H}
  \equiv
  \left\{
    \left(X,v\right)
    \in
    \kansu{\grm}{\calh}\times\calh\ \vert\  Xv=v \right\}\ .
\end{equation}
Then we have also a projection 
\begin{equation}
  \label{eq:piemgrm}
  \pi : \kansu{\eem}{\calh}\longrightarrow\kansu{\grm}{\calh}\ ,
  \quad \kansu{\pi}{\left(X,v\right)}\equiv X\ .
\end{equation}
The set
\begin{equation}
  \label{eq:universal}
  \left\{
    \fukuso^m, \kansu{\eem}{\calh}, \pi, \kansu{\grm}{\calh}
  \right\}\ ,
\end{equation}
is called a (universal) $m$--th vector bundle. This vector bundle is 
one associated with the principal $U(m)$ bundle (\ref{eq:principal}). 

Next let ${\calm}$ be a finite or infinite dimensional differentiable manifold 
and the map 
\begin{equation}
  \label{eq:abstract-projector}
   P : {\calm} \longrightarrow \kansu{\grm}{\calh}
\end{equation} 
be given (called a projector). Using this $P$ we can make the bundles 
(\ref{eq:principal}) and (\ref{eq:universal}) pullback over ${\calm}$ :
\begin{eqnarray}
  \label{eq:hikimodoshi1}
  &&\left\{\kansu{U}{m},\widetilde{St}, \pi_{\widetilde{St}}, {\calm}\right\}
  \equiv
  P^*\left\{\kansu{U}{m}, \kansu{\stm}{\calh}, \pi, 
  \kansu{\grm}{\calh}\right\}
  \ , \\
  \label{eq:hikimodoshi2}
  &&\left\{\fukuso^m,\widetilde{E}, \pi_{\widetilde{E}}, {\calm}\right\}
  \equiv
  P^*\left\{\fukuso^m, \kansu{\eem}{\calh}, \pi, \kansu{\grm}{\calh}\right\}
  \ , 
\end{eqnarray}

\[    
   \matrix{
    \kansu{U}{m}&&\kansu{U}{m}\cr
    \Big\downarrow&&\Big\downarrow\cr
    \widetilde{St}&\longrightarrow&\kansu{\stm}{\calh}\cr
    \Big\downarrow&&\Big\downarrow\cr
    {\calm}&\stackrel{P}{\longrightarrow}&\kansu{\grm}{\calh}\cr
           } \qquad \qquad  
   \matrix{
    \fukuso^m&&\fukuso^m\cr
    \Big\downarrow&&\Big\downarrow\cr
    \widetilde{E}&\longrightarrow&\kansu{E_m}{\calh}\cr
    \Big\downarrow&&\Big\downarrow\cr
    {\calm}&\stackrel{P}{\longrightarrow}&\kansu{\grm}{\calh}\cr
           } 
\]
\par \vspace{5mm} \noindent
see \cite{MN}. (\ref{eq:hikimodoshi2}) is of course a vector bundle 
associated with (\ref{eq:hikimodoshi1}).

For this bundle the (global) curvature ($2$--) form $\bf\Omega$ is given by 
\begin{equation}
  \label{eq:P-curvature}
  {\bf\Omega}=PdP\wedge dP 
\end{equation}
making use of (\ref{eq:abstract-projector}), where $d$ is the usual 
differential form on $\bf\Omega$. 
For the bundles Chern characters play an essential role in several geometric 
properties. In this case Chern characters are given by 
\begin{equation}
  \label{eq:Chern--classes}
  {\bf\Omega},\ {\bf\Omega}^2,\ \cdots,\ {\bf\Omega}^{m/2}; \quad 
  {\bf\Omega}^2={\bf\Omega}\wedge{\bf\Omega},\ \mbox{etc}, 
\end{equation}
where we have assumed that $m=\mbox{dim}{\calm}$ is even. 
In this paper we don't take the trace of (\ref{eq:Chern--classes}), so 
it may be better to call them densities for Chern characters.  

To calculate these quantities in infinite--dimensional cases is not so 
easy. In the next section let us calculate these ones in the special cases.

Let us now define our projectors for the latter aim. In the following, for 
$\calh$ we treat $\calh=\calh$ in section 2, $\calh=\calh_{K}$ in section 3 
and $\calh=\calh_{J}$ in section 4 at the same time. 
For $u_1, u_2, \cdots, u_m \in \fukuso$ we consider a set of coherent or 
generalized coherent states 
$\{\ket{u_1}, \ket{u_2}, \cdots,\ket{u_m}\}$ and set 
\begin{equation}
  \label{eq:a set of coherent states}   
   V_{m}({\bf u})=(\ket{u_1},\ket{u_2}, \cdots, \ket{u_m})\equiv V_{m}
\end{equation}
where ${\bf u}=(u_1, u_2, \cdots, u_m)$. 
Since ${V_{m}}^{\dagger}V_{m}=(\braket{u_i}{u_j}) \in M(m,\fukuso)$,  
we define 
\begin{equation}
  \label{eq:domain}
    {\cal D}_m \equiv \{{\bf u}\in \fukuso^{m}\ \vert\ 
               \mbox{det}({V_{m}}^{\dagger}V_{m}) \ne 0 \}. 
\end{equation}
We note that ${\cal D}_m$ is an open set in $\fukuso^{m}$. 
For example, for $m=1$ and $m=2$ 
\begin{eqnarray}
   {V_{1}}^{\dagger}V_{1}&=&1,    \nonumber \\
   \mbox{det}({V_{2}}^{\dagger}V_{2})&=& 
  \left|
    \begin{array}{cc}
        1 & a \\
        \bar{a} & 1 
    \end{array}
  \right|
   = 1-\vert{a}\vert^{2} \geq 0\ ,  \nonumber 
\end{eqnarray}
where $a=\braket{u_1}{u_2}$. So from (\ref{eq:2-a}), 
(\ref{eq:futou-shiki-3}) and (\ref{eq:futou-shiki-3-J}) we have 
\begin{eqnarray}
  \label{eq:condition-1}   
   {\cal D}_1 &=&\{u \in \fukuso \ \vert\ \mbox{no conditions}\} 
               = \fukuso ,  \\
  \label{eq:condition-2}   
   {\cal D}_2 &=&\{(u_1,u_2) \in \fukuso^2 \ \vert\ u_1\ne u_2 \}.
\end{eqnarray}
For ${\cal D}_m\ (m\geq 3)$ it is not easy for us to give a simple 
condition like (\ref{eq:condition-2}). 
\begin{flushleft}
{\bf Problem}\quad For the case $m=3$ make the condition (\ref{eq:domain}) 
clear like (\ref{eq:condition-2}). 
\end{flushleft}
At any rate 
$
V_{m} \in \kansu{\stm}{\calh} \ \mbox{for}\ {\bf u} \in {\cal D}_m 
$ .
Now let us define our projector $P$ as follows : 
\begin{eqnarray}
  \label{eq:ultimate-projector}
  P : {\cal D}_m \longrightarrow \kansu{\grm}{\calh}\ , \quad 
    P({\bf u})= V_{m}(V_{m}^{\dagger}V_{m})^{-1}V_{m}^{\dagger}\ .
\end{eqnarray}
In the following we set $V=V_{m}$ for simplicity.  Let us calculate 
(\ref{eq:P-curvature}). Since 
\begin{equation}
    dP=V(V^{\dagger}V)^{-1}dV^{\dagger}
     \{
        {\bf 1}-V(V^{\dagger}V)^{-1}V^{\dagger}
     \}
     + \{{\bf 1}-V(V^{\dagger}V)^{-1}V^{\dagger}\}dV(V^{\dagger}V)^{-1}
      V^{\dagger} 
\end{equation}
where $d=\sum_{j=1}^{m}\left(du_{j}\frac{\partial}{\partial u_{j}}+ 
d{\bar u_{j}}\frac{\partial}{\partial {\bar u_{j}}}\right)$, we have 
\[
  PdP=V(V^{\dagger}V)^{-1}dV^{\dagger}
      \{{\bf 1}-V(V^{\dagger}V)^{-1}V^{\dagger}\} 
\]
after some calculation. Therefore we obtain  
\begin{equation}
  \label{eq:curvature-local}
  PdP\wedge dP=V(V^{\dagger}V)^{-1}[dV^{\dagger}\{
               {\bf 1}-V(V^{\dagger}V)^{-1}V^{\dagger}
               \}dV](V^{\dagger}V)^{-1}V^{\dagger}\ .
\end{equation}
Our main calculation is $dV^{\dagger}
\{{\bf 1}-V(V^{\dagger}V)^{-1}V^{\dagger}\}dV$, which is rewritten as 
\begin{equation}
  \label{eq:curvature-decomposition}
  dV^{\dagger}\{{\bf 1}-V(V^{\dagger}V)^{-1}V^{\dagger}\}dV=
  {[\{{\bf 1}-V(V^{\dagger}V)^{-1}V^{\dagger}\}dV]}^{\dagger}\ 
  [\{{\bf 1}-V(V^{\dagger}V)^{-1}V^{\dagger}\}dV] 
\end{equation}
since $Q \equiv {\bf 1}-V(V^{\dagger}V)^{-1}V^{\dagger}$ is also a projector
($Q^2=Q$ and $Q^{\dagger}=Q$). Therefore the first step for us is to 
calculate the term 
\begin{equation}
  \label{eq:main-term}
  \{{\bf 1}-V(V^{\dagger}V)^{-1}V^{\dagger}\}dV\ .
\end{equation}
\par \noindent 
Let us summarize {\bf our process of calculations} : 
\begin{eqnarray}
  &&\mbox{1--st step}\qquad  \{{\bf 1}-V(V^{\dagger}V)^{-1}V^{\dagger}\}dV\ 
     \cdots (\ref {eq:main-term}),  \nonumber \\
  &&\mbox{2--nd step}\qquad 
    dV^{\dagger}\{{\bf 1}-V(V^{\dagger}V)^{-1}V^{\dagger}\}dV\  \cdots 
    (\ref{eq:curvature-decomposition}),  \nonumber \\
  &&\mbox{3--rd step}\qquad 
    V(V^{\dagger}V)^{-1}
    [dV^{\dagger}\{{\bf 1}-V(V^{\dagger}V)^{-1}V^{\dagger}\}dV]   
    (V^{\dagger}V)^{-1}V^{\dagger}\  \cdots 
    (\ref{eq:curvature-local}).  \nonumber 
\end{eqnarray}

\vspace{1cm}
\section{Calculations of Curvature Forms}

In this section we only calculate the curvature forms ($m=1$). 
The calculations even for the case $m=2$ are complicated enough, 
see \cite{KF9} and \cite{KF10}.  For $m\geq 3$ calculations may become 
miserable. 

\subsection{Coherent States}

In this case $\braket{z}{z}=1$, so our projector is very simple to be 
\begin{equation}
  \label{eq:s-1-projector} 
     P(z)=\ket{z}\bra{z}. 
\end{equation}
In this case the calculation of curvature is relatively simple. 
From (\ref{eq:curvature-local}) we have 
\begin{equation}
  \label{eq:s-1-curvature} 
  PdP\wedge dP=\ket{z}\{d\bra{z}({\bf 1}-\ket{z}\bra{z})d\ket{z}\}\bra{z}
         =\ket{z}\bra{z}\{d\bra{z}({\bf 1}-\ket{z}\bra{z})d\ket{z}\}.
\end{equation}
Since $\ket{z}=\mbox{exp}(-\frac{1}{2}\zetta^2)\mbox{exp}(za^{\dagger})
\ket{0}$ by (\ref{eq:2-5-2}), 
\[
 d\ket{z}=\left\{
                 (a^{\dagger}-\frac{{\bar z}}{2})dz-\frac{z}{2}d{\bar z}
          \right\}\ket{z}
         =\left\{
                 a^{\dagger}dz-\frac{1}{2}({\bar z}dz+zd{\bar z})
          \right\}\ket{z}
         =\left\{
                 a^{\dagger}dz-\frac{1}{2}d(\zetta^2)
          \right\}\ket{z},
\]
so that 
\[
  ({\bf 1}-\ket{z}\bra{z})d\ket{z}=
     ({\bf 1}-\ket{z}\bra{z})a^{\dagger}\ket{z}dz =
     (a^{\dagger}-\bra{z}a^{\dagger}\ket{z})\ket{z}dz =
     (a-z)^{\dagger}dz\ket{z}
\]
because $({\bf 1}-\ket{z}\bra{z})\ket{z}={\bf 0}$. Similarly 
$d\bra{z}({\bf 1}-\ket{z}\bra{z})=\bra{z}(a-z) d{\bar z}$. 
\par \noindent 
Let us summarize : 
\begin{equation}
  \label{eq:s-ralations}
   ({\bf 1}-\ket{z}\bra{z})d\ket{z}=(a-z)^{\dagger}dz\ket{z}, \quad 
   d\bra{z}({\bf 1}-\ket{z}\bra{z})=\bra{z}(a-z) d{\bar z}\ .
\end{equation}
Now we are in a position to determine the curvature form 
(\ref{eq:s-1-curvature}).
\[
  d\bra{z}({\bf 1}-\ket{z}\bra{z})d\ket{z}=
  \bra{z}(a-z)(a-z)^{\dagger}\ket{z}d{\bar z}\wedge dz=
  d{\bar z}\wedge dz
\]
after some algebra. Therefore
\begin{equation}
  \label{eq:t-result-1}
 {\bf\Omega}= PdP\wedge dP=\ket{z}\bra{z}d{\bar z}\wedge dz\ .
\end{equation}
From this result we know 
\[
  \frac{{\bf\Omega}}{2\pi i}=\ket{z}\bra{z}\frac{dx\wedge dy}{\pi}
\]
when $z=x+iy$. This just gives the resolution of unity in (\ref{eq:2-7}).

\subsection{Generalized Coherent States Based on $su(1,1)$}

In this case $\braket{w}{w}=1$, so our projector is very simple to be 
\begin{equation}
  \label{eq:w-1-projector} 
     P(w)=\ket{w}\bra{w}. 
\end{equation}
In this case the calculation of curvature is relatively simple. 
From (\ref{eq:curvature-local}) we have 
\begin{equation}
  \label{eq:w-1-curvature} 
  PdP\wedge dP=\ket{w}\{d\bra{w}({\bf 1}_{K}-\ket{w}\bra{w})d\ket{w}\}\bra{w}
         =\ket{w}\bra{w}\{d\bra{w}({\bf 1}_{K}-\ket{w}\bra{w})d\ket{w}\}, 
\end{equation}
where $d=dw\frac{\partial}{\partial w}+d{\bar w}\frac{\partial}
{\partial {\bar w}}$. 
Since $\ket{w}=(1-\zettai{\zeta}^2)^{K}\mbox{exp}(\zeta K_{+})\kett{K}{0}$
by (\ref{eq:2-2-right}), 
\begin{equation}
  \label{eq:bibun-koushiki} 
 d\ket{w}=\left\{d\zeta K_{+}+Kd\mbox{log}(1-\zettai{\zeta}^2)\right\}\ket{w}
\end{equation}
by some calculation, so that 
\begin{eqnarray}
  ({\bf 1}_{K}-\ket{w}\bra{w})d\ket{w}&=&
    ({\bf 1}_{K}-\ket{w}\bra{w})K_{+}\ket{w}d\zeta =
    (K_{+}-\bra{w}K_{+}\ket{w})\ket{w}d\zeta  \nonumber \\
  &=&\left(K_{+}-\frac{2K{\bar \zeta}}{1-\zettai{\zeta}^2}\right)d\zeta\ket{w}
\end{eqnarray}
because $({\bf 1}_{K}-\ket{w}\bra{w})\ket{w}={\bf 0}$. Similarly we have 
\begin{equation}
 d\bra{w}({\bf 1}_{K}-\ket{w}\bra{w})=\bra{w}
 \left(K_{-}-\frac{2K{\zeta}}{1-\zettai{\zeta}^2}\right)d{\bar \zeta}
\end{equation} 
Now we are in a position to determine the curvature form 
(\ref{eq:w-1-curvature}).
\begin{eqnarray}
  &&d\bra{w}({\bf 1}_{K}-\ket{w}\bra{w})d\ket{w}    \nonumber \\
  &&=\bra{w}
     \left(K_{-}-\frac{2K{\zeta}}{1-\zettai{\zeta}^2}\right)
     \left(K_{+}-\frac{2K{\bar \zeta}}{1-\zettai{\zeta}^2}\right)
     \ket{w} d{\bar \zeta}\wedge d\zeta   \nonumber \\
 &&=\left\{\bra{w}K_{-}K_{+}\ket{w}
           -\frac{2K{\bar \zeta}}{1-\zettai{\zeta}^2}\bra{w}K_{-}\ket{w}
           -\frac{2K{\zeta}}{1-\zettai{\zeta}^2}\bra{w}K_{+}\ket{w}
           +\frac{4K^{2}\zettai{\zeta}^2}{(1-\zettai{\zeta}^2)^2}
    \right\}d{\bar \zeta}\wedge d\zeta  \nonumber \\
 &&=\frac{2K}{(1-\zettai{\zeta}^2)^2}d{\bar \zeta}\wedge d\zeta
\end{eqnarray}
after some algebra with (\ref{eq:kitaichi-1}) and (\ref{eq:kitaichi-2}). 
Therefore
\begin{equation}
  \label{eq:w-result-1}
 {\bf\Omega}= PdP\wedge dP=\ket{w}\bra{w}
              \frac{2Kd{\bar \zeta}\wedge d\zeta}
                   {(1-\zettai{\zeta}^2)^2}.
\end{equation}
From this result we know 
\[
  \frac{{\bf\Omega}}{2\pi i}
  =\frac{2K}{\pi}
   \frac{d{\zeta_1}\wedge d{\zeta_2}}{(1-\zettai{\zeta}^2)^2}\ket{w}\bra{w}
  =\frac{2K}{\pi}\frac{d{\zeta_1}\wedge d{\zeta_2}}{(1-\zettai{\zeta}^2)^2}
   \ket{\zeta}\bra{\zeta}
\]
by (\ref{eq:2-2-right}) 
when $\zeta=\zeta_{1}+\sqrt{-1}\zeta_{2}$. If we define a constant  
\begin{equation}
     C_{K}=\frac{2K-1}{2K},
\end{equation}
then we have 
\begin{equation}
     C_{K}\frac{{\bf\Omega}}{2\pi i}=\frac{2K-1}{\pi}
  \frac{d{\zeta_1}\wedge d{\zeta_2}}{(1-\zettai{\zeta}^2)^2}
  \ket{\zeta}\bra{\zeta}.  
\end{equation}
This gives the resolution of unity in (\ref{eq:2-2-18}). But the situation 
is a bit different from \cite{KF9} in which the constant corresponding to 
$C_{K}$ is just one.  
\begin{flushleft}
{\bf Problem}\quad What is a (deep) meaning of $C_{K}$  ? 
\end{flushleft}

\subsection{Generalized Coherent States Based on $su(2)$}

In this case $\braket{v}{v}=1$, so our projector is very simple to be 
\begin{equation}
  \label{eq:v-1-projector} 
     P(v)=\ket{v}\bra{v}. 
\end{equation}
In this case the calculation of curvature is relatively simple. 
From (\ref{eq:curvature-local}) we have 
\begin{equation}
  \label{eq:v-1-curvature} 
  PdP\wedge dP=\ket{v}\{d\bra{v}({\bf 1}_{J}-\ket{v}\bra{v})d\ket{v}\}\bra{v}
         =\ket{v}\bra{v}\{d\bra{v}({\bf 1}_{J}-\ket{v}\bra{v})d\ket{v}\}, 
\end{equation}
where $d=dv\frac{\partial}{\partial v}+d{\bar v}\frac{\partial}
{\partial {\bar v}}$. 
Since $\ket{v}=(1+\zettai{\eta}^2)^{-J}\mbox{exp}(\eta J_{+})\kett{J}{0}$
by (\ref{eq:3-2-right}), 
\begin{equation}
  \label{eq:bibun-koushiki-J} 
 d\ket{v}=\left\{d\eta J_{+}-Jd\mbox{log}(1+\zettai{\eta}^2)\right\}\ket{v}
\end{equation}
by some calculation, so that 
\begin{eqnarray}
  ({\bf 1}_{J}-\ket{v}\bra{v})d\ket{v}&=&
    ({\bf 1}_{J}-\ket{v}\bra{v})J_{+}\ket{v}d\eta =
    (J_{+}-\bra{v}J_{+}\ket{v})\ket{v}d\eta  \nonumber \\
  &=&\left(J_{+}-\frac{2J{\bar \eta}}{1+\zettai{\eta}^2}\right)d\eta\ket{v}
\end{eqnarray}
because $({\bf 1}_{J}-\ket{v}\bra{v})\ket{v}={\bf 0}$. Similarly we have 
\begin{equation}
 d\bra{v}({\bf 1}_{J}-\ket{v}\bra{v})=\bra{v}
 \left(J_{-}-\frac{2J{\eta}}{1+\zettai{\eta}^2}\right)d{\bar \eta}
\end{equation} 
Now we are in a position to determine the curvature form 
(\ref{eq:v-1-curvature}).
\begin{eqnarray}
  &&d\bra{v}({\bf 1}_{J}-\ket{v}\bra{v})d\ket{v}    \nonumber \\
  &&=\bra{v}
     \left(J_{-}-\frac{2J{\eta}}{1+\zettai{\eta}^2}\right)
     \left(J_{+}-\frac{2J{\bar \eta}}{1+\zettai{\eta}^2}\right)
     \ket{v} d{\bar \eta}\wedge d\eta   \nonumber \\
 &&=\left\{\bra{v}J_{-}J_{+}\ket{v}
           -\frac{2J{\bar \eta}}{1+\zettai{\eta}^2}\bra{v}J_{-}\ket{v}
           -\frac{2J{\eta}}{1+\zettai{\eta}^2}\bra{v}J_{+}\ket{v}
           +\frac{4J^{2}\zettai{\eta}^2}{(1+\zettai{\eta}^2)^2}
    \right\}d{\bar \eta}\wedge d\eta  \nonumber \\
 &&=\frac{2J}{(1+\zettai{\eta}^2)^2}d{\bar \eta}\wedge d\eta
\end{eqnarray}
after some algebra with (\ref{eq:kitaichi-1-J}) and (\ref{eq:kitaichi-2-J}). 
Therefore
\begin{equation}
  \label{eq:v-result-1}
 {\bf\Omega}= PdP\wedge dP=\ket{v}\bra{v}
              \frac{2Jd{\bar \eta}\wedge d\eta}
                   {(1+\zettai{\eta}^2)^2}.
\end{equation}
From this result we know 
\[
  \frac{{\bf\Omega}}{2\pi i}=\frac{2J}{\pi}
    \frac{d{\eta_1}\wedge d{\eta_2}}{(1+\zettai{\eta}^2)^2}\ket{v}\bra{v}
  =\frac{2J}{\pi}\frac{d{\eta_1}\wedge d{\eta_2}}{(1+\zettai{\eta}^2)^2}
   \ket{\eta}\bra{\eta}
\]
by (\ref{eq:3-2-right}) 
when $\eta=\eta_{1}+\sqrt{-1}\eta_{2}$. If we define a constant  
\begin{equation}
     C_{J}=\frac{2J+1}{2J},
\end{equation}
then we have 
\begin{equation}
     C_{J}\frac{{\bf\Omega}}{2\pi i}=\frac{2J+1}{\pi}
  \frac{d{\eta_1}\wedge d{\eta_2}}{(1+\zettai{\eta}^2)^2}
  \ket{\eta}\bra{\eta}.  
\end{equation}
This gives the resolution of unity in (\ref{eq:3-2-18}). But the situation 
is a bit different from \cite{KF9} in which the constant corresponding to 
$C_{J}$ is just one.  
\begin{flushleft}
{\bf Problem}\quad What is a (deep) meaning of $C_{J}$  ? 
\end{flushleft}

\vspace{10mm}
\section{Holonomic Quantum Computation}

In this section we introduce the concept of Holonomic Quantum Computation, 
see \cite{KF6}. 

Let $\calm$ be a parameter space and we denote by $\lam$ its element. 
Let $\slam$ be a fixed reference point of $\calm$. Let $H_\lam$ be 
a family of Hamiltonians parameterized by $\calm$ which act on a Fock space 
$\calh$. We set $H_0$ = $H_\slam$ for simplicity and assume that this has 
a $m$-fold degenerate vacuum :
\begin{equation}
  H_{0}v_j = \mathbf{0},\quad j = 1 \sim m. 
\end{equation}
These $v_j$'s form a $m$-dimensional vector space. We may assume that 
$\langle v_{i}\vert v_{j}\rangle = \delta_{ij}$. Then $\left(v_1,\cdots,v_m
\right) \in \kansu{\stm}{\calh}$ and 
\[
  F_0 \equiv \left\{\sum_{j=1}^{m}x_{j}v_{j}\vert x_j \in \fukuso \right\} 
  \cong \fukuso^m.
\]
Namely, $F_0$ is a vector space associated with o.n.basis 
$\left(v_1,\cdots,v_m\right)$.

Next we assume for simplicity 
that a family of unitary operators parameterized by $\calm$
\begin{equation}
  \label{eq:ufamily} 
  W : \calm \rightarrow U(\calh),\quad W(\slam) = {\rm id}.
\end{equation}
is given and $H_{\lam}$ above is given by the following isospectral family
\begin{equation}
 H_{\lam} \equiv W(\lam)H_0 W(\lam)^{-1}.
\end{equation}
In this case there is no level crossing of eigenvalues. Making use of 
$W(\lam)$ we can define a projector
\begin{equation}
  \label{eq:pfamily}
 P : \calm \rightarrow \kansu{\grm}{\calh}, \quad 
 P(\lam) \equiv
  W(\lam) \left(\sum^{m}_{j=1}v_{j}{v_{j}}^{\dagger}\right)W(\lam)^{-1}
\end{equation}
and have the pullback bundles over $\calm$ from (\ref{eq:hikimodoshi1}) and 
(\ref{eq:hikimodoshi2})
\begin{equation}
  \label{eq:target}
 \left\{\kansu{U}{m},\widetilde{St}, \pi_{\widetilde{St}}, \calm\right\},\quad 
 \left\{\fukuso^m,\widetilde{E}, \pi_{\widetilde{E}}, \calm\right\}.
\end{equation}

For the latter we set
\begin{equation}
  \label{eq:vacuum}
 \ket{vac} = \left(v_1,\cdots,v_m\right).
\end{equation}
In this case a canonical connection form $\cala$ of 
$\left\{\kansu{U}{m},\widetilde{St}, \pi_{\widetilde{St}}, \calm\right\}$ is 
given by 
\begin{equation}
  \label{eq:cform}
 \cala = \bra{vac}W(\lam)^{-1}d W(\lam)\ket{vac},
\end{equation}
where $d$ is a usual differential form on $\calm$, and its curvature form by
\begin{equation}
  \label{eq:curvature}
  \calf \equiv d\cala+\cala\wedge\cala,
\end{equation}
see \cite{SW} and \cite{MN}.

Let $\gamma$ be a loop in $\calm$ at $\slam$., 
\[
\gamma : [0,1] \longrightarrow \calm, \quad \gamma(0) = \gamma(1). 
\]
For this $\gamma$ a holonomy operator $\Gamma_{\cala}$ is defined as 
the path--ordered integral of $\cala$ along $\gamma$ : 
\begin{equation}
  \label{eq:holonomy}
  \Gamma_{\cala}(\gamma)={\cal P}\mbox{exp}\left\{\oint_{\gamma}\cala\right\} 
\ \in \ \kansu{U}{m},
\end{equation}
where ${\cal P}$ means path-ordered. See \cite{MN}.  

This acts on the fiber $F_0$ at $\slam$ of the vector bundle 
$\left\{\fukuso^m,\widetilde{E}, \pi_{\widetilde{E}}, M\right\}$ as follows :
${\textbf x} \rightarrow \Gamma_{\cala}(\gamma){\textbf x}$.\quad 
The holonomy group $Hol(\cala)$ is in general subgroup of $\kansu{U}{m}$ 
. In the case of $Hol(\cala) = \kansu{U}{m}$, $\cala$ is called irreducible, 
see \cite{MN}.

\par \noindent
In the Holonomic Quantum Computation we take 
\begin{eqnarray}
  \label{eq:information}
  &&{\rm Encoding\ of\ Information} \Longrightarrow {\textbf x} \in F_0 , 
  \nonumber \\
  &&{\rm Processing\ of\ Information} \Longrightarrow \Gamma_{\cala}(\gamma) : 
  {\textbf x} \rightarrow \Gamma_{\cala}(\gamma){\textbf x}.
\end{eqnarray}
%
\begin{center}
\setlength{\unitlength}{1mm}
\begin{picture}(160,130)
\put(80,100){\circle*{4}}
\put(80,60){\circle*{4}}
\put(80,20){\circle*{4}}
\put(80,102){\line(0,1){15}}
\put(80,62){\line(0,1){56}}
\put(80,32){\line(0,1){26}}
\put(79.5,24){$\cdot$}
\put(79.5,26){$\cdot$}
\put(79.5,28){$\cdot$}
\qbezier(80,60)(140,80)(80,100)
\qbezier(80,20)(100,5)(110,20)
\qbezier(80,20)(100,35)(110,20)
\put(72,59){{\bf X}}
\put(69,99){A{\bf X}}
\put(72,18){$\lambda_{0}$}
\put(77,121){${\bf F_{0}}$}
\qbezier(50,0)(90,10)(150,0)
\qbezier(50,0)(20,20)(30,30)
\qbezier(150,0)(140,20)(120,30)
\qbezier(30,30)(80,40)(120,30)
\put(120,10){${\cal M}$}
\end{picture}
\end{center}
\begin{center}
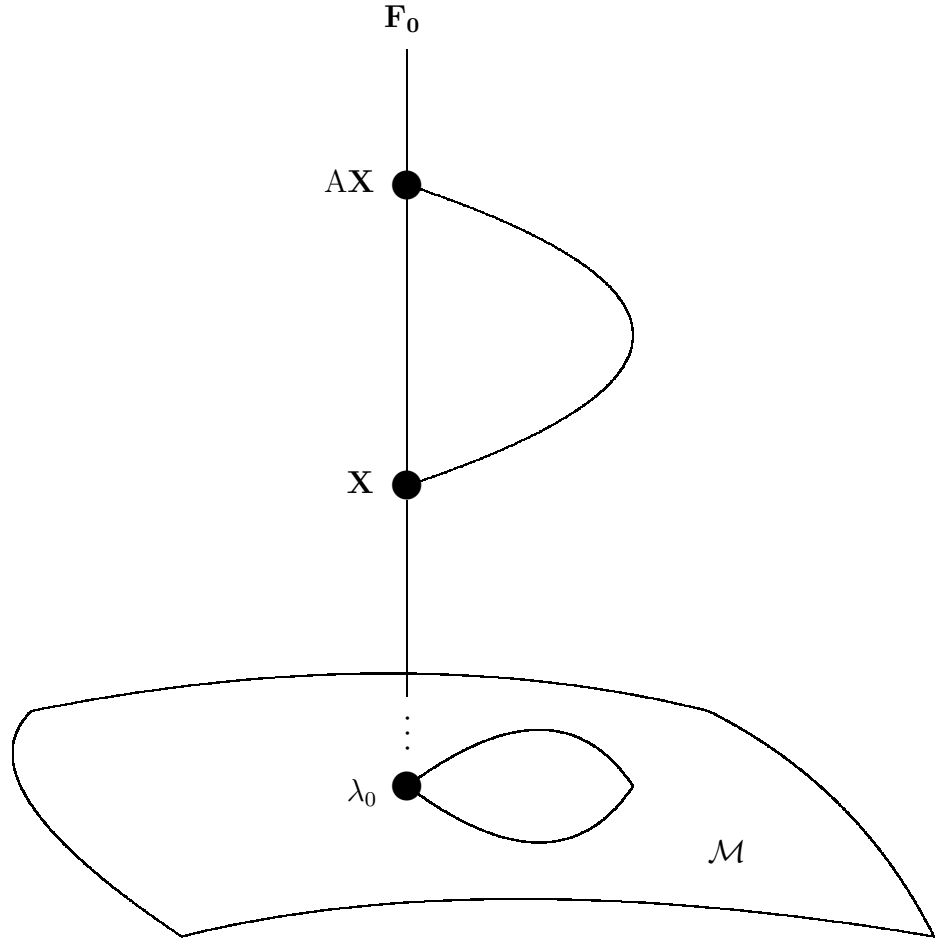
 \textbf{Quantum Computational Bundle}
\end{center}

\subsection{One--Qubit Case}

Let $H_0$ be a Hamiltonian with nonlinear interaction produced by 
a Kerr medium., that is  
\begin{equation}
H_{0} = \hbar {\rm X} N(N-1), 
\end{equation}
where X is a certain constant, see \cite{MW} and \cite{PC}. 
The eigenvectors of $H_0$ corresponding to $0$ is 
$\left\{\ket{0},\ket{1}\right\}$, so its eigenspace is 
${\rm Vect}\left\{\ket{0},\ket{1}\right\} \cong \fukuso^2$. 
The vector space ${\rm Vect}\left\{\ket{0},\ket{1}\right\}$ is called 
1-qubit ({\bf qu}antum {\bf bit}) space and we set 
\[
F_{0}={\rm Vect}\left\{\ket{0},\ket{1}\right\}\quad \mbox{and}\quad 
\ket{vac} = (\ket{0},\ket{1}).
\]

Now we consider the following isospectral family of $H_0$ : 
\begin{eqnarray}
  \label{eq:onefamily}
   &&H_{(\alpha,\beta)}
  = W(\alpha,\beta)H_{0} W(\alpha,\beta)^{-1},\\
  \label{eq:onedouble}
  &&W(\alpha,\beta) = D(\alpha)S(\beta).
\end{eqnarray}
In this case 
\begin{equation}
 {\cal M}=\left\{(\alpha,\beta) \in {\mathbf C}^{2} \right\}
\end{equation}
and we want calculate 
\begin{equation}
  \label{eq:1-connection}
  {\cal A}=\bra{vac}W^{-1}dW\ket{vac} 
\end{equation}
where 
\begin{equation}
    d=d\alpha\frac{\partial}{\partial \alpha}+
      d\bar{\alpha}\frac{\partial}{\partial \bar{\alpha}}+
      d\beta\frac{\partial}{\partial \beta}+
      d\bar{\beta}\frac{\partial}{\partial \bar{\beta}}.
\end{equation}
Since ${\cal A}$ is anti--hermitian (${\cal A}^{\dagger}=-{\cal A}$), 
we can write 
\begin{equation}
   {\cal A}=
            A_{\alpha}d\alpha + A_{\beta}d\beta - 
            {A_{\alpha}}^{\dagger}d{\bar \alpha} - 
            {A_{\beta}}^{\dagger}d{\bar \beta}  
\end{equation}
where
\[
   A_{\alpha}=\bra{vac}W^{-1}\frac{\partial W}{\partial \alpha}\ket{vac} 
     \qquad 
   A_{\beta}=\bra{vac}W^{-1}\frac{\partial W}{\partial \beta}\ket{vac} .
\]
The calculation of $A_{\alpha}$ and $A_{\beta}$ is as follows (\cite{KF2}) :
\begin{eqnarray}
A_{\alpha}&=&\frac{{\bar \alpha}}{2}L+cosh(\zettai{\beta})F+
\frac{{\bar \beta}sinh(\zettai{\beta})}{\zettai{\beta}}E,   \\
A_{\beta}&=&\frac{ {\bar \beta}(-1+cosh(2\zettai{\beta})) }
{4\zettai{\beta}^{2}}\left(K+\frac{1}{2}L\right),
\end{eqnarray}
where 
\[
  E=\pmatrix{&1\cr0&\cr} ,\ 
  F=\pmatrix{&0\cr1&\cr} ,\ 
  K=\pmatrix{0&\cr&1\cr} ,\ 
  L=\pmatrix{1&\cr&1\cr} .
\]
\par \vspace{5mm} \noindent
Then making use of these ones 
we can show that the holonomy group generated by (\ref{eq:1-connection}) 
is irreducible in $U(2)$, namely just $U(2)$, see \cite{PZR} and \cite{KF2}. 
This is very crucial fact to Holonomic Quantum Computation.

\vspace{10mm}
\subsection{Two--Qubit Case}

We consider the system of two particles, so the Hamiltonian that 
we treat in the following is 
\begin{equation}
  \label{eq:hamiltonian}
  H_0 = \hbar {\rm X} N_{1}(N_{1}-1) + \hbar {\rm X} N_{2}(N_{2}-1).
\end{equation}
The eigenspace of $0$ of this Hamiltonian becomes therefore
\begin{equation}
  \label{eq:2-eigenspace}
   F_0 = {\rm Vect}\left\{\ket{0},\ket{1}\right\}\otimes 
         {\rm Vect}\left\{\ket{0},\ket{1}\right\} 
    = {\rm Vect}\left\{\kett{0}{0},\kett{0}{1}, \kett{1}{0},
                       \kett{1}{1}\right\}
    \cong \fukuso^{4}. 
\end{equation}
We set $\ket{vac} = 
\left(\kett{0}{0}, \kett{0}{1}, \kett{1}{0}, \kett{1}{1} \right)$.

\par \noindent
Next we consider the following isospectral family of $H_0$ : 
\begin{eqnarray}
  \label{eq:twofamily}
   &&H_{(\alpha_{1},\beta_{1},\lambda,\mu,\alpha_{2},\beta_{2})}
  = W(\alpha_{1},\beta_{1},\lambda,\mu,\alpha_{2},\beta_{2})H_0 
     W(\alpha_{1},\beta_{1},\lambda,\mu,\alpha_{2},\beta_{2})^{-1},\\
  \label{eq:double}
  &&W(\alpha_{1},\beta_{1},\lambda,\mu,\alpha_{2},\beta_{2}) 
  = W_{1}(\alpha_{1},\beta_{1})O_{12}(\lambda,\mu)
      W_{2}(\alpha_{2},\beta_{2}). 
\end{eqnarray}
where 
\begin{equation}
   O_{12}(\lambda, \mu) = U_{J}(\lambda)U_{K}(\mu), \quad 
   W_{j}(\alpha_{j},\beta_{j}) = D_{j}(\alpha_{j})S_{j}(\beta_{j}) \quad 
     \mbox{for}\quad j = 1, 2.   
\end{equation}
%
%
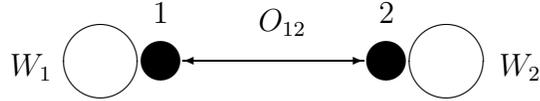
\begin{center}
\setlength{\unitlength}{1mm}
\begin{picture}(80,22)
  \put(30,8){\circle*{6}}
  \put(60,8){\circle*{6}}
  \put(22,8){\circle{10}}
  \put(68,8){\circle{10}}
  \put(45,8){\vector(-1,0){12}}
  \put(45,8){\vector(1,0){12}}
  \put(29,13){1}
  \put(59,13){2}
  \put(10,6){$W_{1}$}
  \put(75,6){$W_{2}$}
  \put(43,12){$O_{12}$}
\end{picture}
\end{center}
%
%
In this case
\begin{equation}
 {\cal M}=\left\{(\alpha_{1},\beta_{1},\lambda,\mu,\alpha_{2},\beta_{2}) 
 \in {\mathbf C}^{6} \right\}
\end{equation}
and we have only to calculate the following 
\begin{equation}
 \label{eq:2-connection}
  {\cal A}=\bra{vac}W^{-1}dW\ket{vac},
\end{equation}
where 
\begin{eqnarray}
    d&=&d\alpha_{1}\frac{\partial}{\partial \alpha_{1}}+
      d\bar{\alpha_{1}}\frac{\partial}{\partial \bar{\alpha_{1}}}+
      d\beta_{1}\frac{\partial}{\partial \beta_{1}}+
      d\bar{\beta_{1}}\frac{\partial}{\partial \bar{\beta_{1}}}+ 
      d\lambda\frac{\partial}{\partial \lambda}+
      d\bar{\lambda}\frac{\partial}{\partial \bar{\lambda}}+
      d\mu\frac{\partial}{\partial \mu}+
      d\bar{\mu}\frac{\partial}{\partial \bar{\mu}}  \nonumber \\
     &+&
      d\alpha_{2}\frac{\partial}{\partial \alpha_{2}}+
      d\bar{\alpha_{2}}\frac{\partial}{\partial \bar{\alpha_{2}}}+
      d\beta_{2}\frac{\partial}{\partial \beta_{2}}+
      d\bar{\beta_{2}}\frac{\partial}{\partial \bar{\beta_{2}}}\ . 
\end{eqnarray}
The calculation of (\ref{eq:2-connection}) is not easy, but we can 
determine it, see \cite{KF2}, \cite{KF3} and \cite{KF5} for the details. 
But we cannot determine its curvature form which is necessary to 
look for the holonomy group (Ambrose--Singer theorem) due to too
complication.

\par \noindent
Then the essential point is 

\par \noindent
\textbf{Problem}\quad Is the connection form (\ref{eq:2-connection}) 
irreducible in $U(4)$ ?
\par \noindent

\par \noindent
Our analysis in \cite{KF5} shows that the holonomy group generated 
by ${\cal A}$ may be $SU(4)$ not $U(4)$. To obtain $U(4)$ a sophisticated 
trick $\cdots$ higher dimensional holonomies \cite{AFG} $\cdots$ may be 
necessary. See also \cite{DL}. 

\vspace{10mm}
\section{Geometric Construction of Bell States}

In this section we introduce the geometric constraction of Bell states by 
making use of coherent states based on $su(2)$, \cite{KF12}. 
One of purpose of Quantum Information Theory is to clarify a role of 
{\bf entanglement} of states, so that we would like to look for geometric 
meaning of entanglement. 

\par \noindent
The famous Bell states (\cite{JB}, \cite{BMR}) given by 
\begin{eqnarray}
  \label{eq:Bell-states-1}
 && \frac{1}{\sqrt{2}}
      (\ket{0}\otimes \ket{0} + \ket{1}\otimes \ket{1}), \\
  \label{eq:Bell-states-2}
 && \frac{1}{\sqrt{2}}
      (\ket{0}\otimes \ket{0} - \ket{1}\otimes \ket{1}),  \\
  \label{eq:Bell-states-3}
 && \frac{1}{\sqrt{2}}
      (\ket{0}\otimes \ket{1} + \ket{1}\otimes \ket{0}), \\
  \label{eq:Bell-states-4}
 &&\frac{1}{\sqrt{2}}
      (\ket{0}\otimes \ket{1} - \ket{1}\otimes \ket{0}) 
\end {eqnarray}
are typical examples of entanglement. It is very interesting that these 
play an essential role in Quantum Teleportation, see \cite{LPS}. 
We would like to reconstruct these states in a geometric manner.

\subsection{Review on General Theory}
Let us make a review of \cite{DF} and rewrite the result with our method. 
Let $G$ be a compact 
linear Lie group (for example $G = U(n)$) and consider a coherent state 
representation of $G$ whose parameter space is a compact complex manifold 
$S = G/H$, where $H$ is a subgroup of $G$. For example $G = U(n)$ and 
$H = U(k) \times U(n-k)$, then $S$ = $U(n)/ U(k) \times U(n-k) \cong 
G_{k}({\fukuso}^{n})$, which is called a complex Grassmann manifold, 
\cite{AP}, \cite{FKS}. 
Let $Z$ be a local coordinate on $S$ and $\ket{Z}$ a generalized 
coherent state in some representation space $V$ ($\cong {\fukuso}^{K}$ for 
some high $K \in \futon$). Then we have, by definition,  
the measure $d\mu(Z,Z^{\dagger})$ that satisfies the resolution 
of unity 
\begin{equation}
 \label{eq:the resolution-of-unity}
  \int_{S} d\mu(Z,Z^{\dagger}) \ket{Z}\bra{Z} = {\bf 1}_{V}\quad  
  \mbox{and}\quad \int_{S} d\mu(Z,Z^{\dagger}) = \mbox{dim}V\ .
\end{equation}

Next we define an anti-automorphism $\flat : S \longrightarrow S$. We call 
$Z \longrightarrow {Z}^{\flat}$ an anti-automorphism if and only if
\begin{eqnarray}
 \label{eq:automorphism} 
 &&(\mbox{i})\ \ Z \longrightarrow {Z}^{\flat}\  \mbox{induces an 
   automorphism of}\ S, \\
 \label{eq:anti-map} 
 &&(\mbox{ii})\ \ {\flat}\  \mbox{is an anti-map, namely}\quad
   \braket{Z^{\flat}}{W^{\flat}}=\braket{W}{Z}. 
\end{eqnarray}
Now let us redefine the generalized Bell state in \cite{DF} as follows :
\par \noindent
 \textbf{Definition}\  The generalized Bell state is defined as 
\begin{equation}
 \label{eq:generalized-Bell-state}
  \bell{B}= \frac{1}{\sqrt{\mbox{dim}V}} \int_{S} d\mu(Z,Z^{\dagger}) 
            \ket{Z}\otimes \ket{Z^{\flat}}.
\end{equation}
Then we have 
\begin{eqnarray}
  \bellbraket{B}{B}&=&\frac{1}{\mbox{dim}V} \int_{S} \int_{S} 
     d\mu(Z,Z^{\dagger})d\mu(W,W^{\dagger})  
     (\bra{Z}\otimes \bra{Z^{\flat}})( \ket{W}\otimes \ket{W^{\flat}})
  \nonumber \\
  &=&\frac{1}{\mbox{dim}V} \int_{S} \int_{S}
     d\mu(Z,Z^{\dagger})d\mu(W,W^{\dagger})
    \braket{Z}{W}\braket{Z^{\flat}}{W^{\flat}} 
  \nonumber \\
  &=&\frac{1}{\mbox{dim}V} \int_{S} \int_{S}
     d\mu(Z,Z^{\dagger})d\mu(W,W^{\dagger})
    \braket{Z}{W}\braket{W}{Z} 
  \nonumber \\
  &=&\frac{1}{\mbox{dim}V} \int_{S}
     d\mu(Z,Z^{\dagger})\braket{Z}{Z} 
  \nonumber \\
  &=& \frac{1}{\mbox{dim}V} \int_{S} d\mu(Z,Z^{\dagger}) 
   = 1, \nonumber
\end{eqnarray}
where we have used (\ref{eq:the resolution-of-unity}) and 
(\ref{eq:anti-map}).

\subsection{Review on Projective Space}
We make a review of complex projective spaces, \cite{MN}, \cite{FKSF2} 
and \cite{KF5}. 
For $N \in \futon$ the complex projective space ${\fukuso}P^{N}$ is defined 
as follows : For \mbox{\boldmath $\zeta$}, \mbox{\boldmath $\mu$} $\in 
{\fukuso}^{N+1}-\{{\bf 0}\}$\   \mbox{\boldmath $\zeta$} is equivalent to 
\mbox{\boldmath $\mu$} (\mbox{\boldmath $\zeta$} $\sim$ 
\mbox{\boldmath $\mu$}) if and only if \mbox{\boldmath $\zeta$} = $\lambda$
\mbox{\boldmath $\mu$} for $\lambda \in \fukuso - \{0 \}$. We show 
the equivalent relation class as [\mbox{\boldmath $\zeta$}] and set 
${\fukuso}P^{N} \equiv {\fukuso}^{N+1}-\{{\bf 0}\} / \sim $. When  
\mbox{\boldmath $\zeta$} = $({\zeta}_{0}, {\zeta}_{1}, \cdots, {\zeta}_{N})$ 
we write usually as [\mbox{\boldmath $\zeta$}] = $[{\zeta}_{0}: {\zeta}_{1}:  
\cdots : {\zeta}_{N}]$. Then it is well--known that ${\fukuso}P^{N}$ has 
$N+1$ local charts, namely
\begin{equation}
  {\fukuso}P^{N} = \bigcup_{j=0}^{N} U_{j}\ ,  \quad 
    U_{j} = \{ [{\zeta}_{0}: \cdots : {\zeta}_{j}: \cdots : {\zeta}_{N}]\ |\  
          {\zeta}_{j} \ne 0 \}.
\end{equation} 
Since
\[
  ({\zeta}_{0}, \cdots , {\zeta}_{j}, \cdots , {\zeta}_{N}) =  
  {\zeta}_{j}\left(\frac{{\zeta}_{0}}{{\zeta}_{j}}, \cdots, 
  \frac{{\zeta}_{j-1}}{{\zeta}_{j}}, 1, \frac{{\zeta}_{j+1}}{{\zeta}_{j}}, 
  \cdots, \frac{{\zeta}_{N}}{{\zeta}_{j}}\right),
\]
we have the local coordinate on $U_{j}$ 
\begin{equation}
  \left(\frac{{\zeta}_{0}}{{\zeta}_{j}}, \cdots, 
  \frac{{\zeta}_{j-1}}{{\zeta}_{j}}, \frac{{\zeta}_{j+1}}{{\zeta}_{j}}, 
  \cdots, \frac{{\zeta}_{N}}{{\zeta}_{j}}\right). 
\end{equation}

However the above definition of ${\fukuso}P^{N}$ is not tractable, so we use 
the well--known expression by projections
\begin{equation}
 {\fukuso}P^{N} \cong G_{1}({\fukuso}^{N+1}) = 
     \{P \in M(N+1; \fukuso)\ |\ P^{2} = P,\ P^{\dagger} = P \ \mbox{and}\ 
       \mbox{tr}P = 1 \}
\end{equation}
and the correspondence 
\begin{equation}
 \label{eq:correspondence}
  [{\zeta}_{0}: {\zeta}_{1}: \cdots : {\zeta}_{N}] \Longleftrightarrow 
  \frac{1}{\zettai{{\zeta}_{0}}^2 + \zettai{{\zeta}_{1}}^2 + \cdots + 
          \zettai{{\zeta}_{N}}^2 }
  \left(
     \begin{array}{ccccc} 
         \zettai{{\zeta}_{0}}^2& {\zeta}_{0}{\bar {\zeta}_{1}}& 
         \cdot& \cdot& {\zeta}_{0}{\bar {\zeta}_{N}}  \\
         {\zeta}_{1}{\bar {\zeta}_{0}} & \zettai{{\zeta}_{1}}^2&  
         \cdot& \cdot& {\zeta}_{1}{\bar {\zeta}_{N}}  \\
         \cdot& \cdot& & & \cdot  \\
         \cdot& \cdot& & & \cdot  \\
         {\zeta}_{N}{\bar {\zeta}_{0}}& {\zeta}_{N}{\bar {\zeta}_{1}}& 
         \cdot& \cdot& \zettai{{\zeta}_{N}}^2     
     \end{array}
  \right) \equiv P\ .
\end{equation}
If we set 
\begin{equation}
  \ket{\mbox{\boldmath $\zeta$}}=
 \frac{1}{\sqrt{\sum_{j=0}^{N} \zettai{\zeta_{j}}^2} }
  \left(
     \begin{array}{c}
        {\zeta}_{0} \\
        {\zeta}_{1} \\
         \cdot  \\
         \cdot  \\
        {\zeta}_{N} 
     \end{array} 
  \right)\ , 
\end{equation}
then we can write the right hand side of (\ref{eq:correspondence}) as 
\begin{equation}
 \label{eq:real-projector}
  P = \ket{\mbox{\boldmath $\zeta$}}\bra{\mbox{\boldmath $\zeta$}} \quad 
  \mbox{and} \quad 
    \braket{\mbox{\boldmath $\zeta$}}{\mbox{\boldmath $\zeta$}} = 1.
\end{equation}
For example on $U_{1}$ 
\[
  \left(z_{1}, z_{2}, \cdots, z_{N} \right) = 
  \left(\frac{{\zeta}_{1}}{{\zeta}_{0}},\frac{{\zeta}_{2}}{{\zeta}_{0}}, 
  \cdots, 
  \frac{{\zeta}_{N}}{{\zeta}_{0}}\right) ,  
\]
we have 
\begin{eqnarray}
  P(z_{1}, \cdots, z_{N}) &=& 
  \frac{1}{1 + \sum_{j=1}^{N} \zettai{z_{j}}^2}
     \left(
         \begin{array}{ccccc}
             1& {\bar z_{1}}& \cdot& \cdot& {\bar z_{N}}  \\
             z_{1}& \zettai{z_{1}}^2& \cdot& \cdot& z_{1}{\bar z_{N}} \\
             \cdot& \cdot& & & \cdot \\
             \cdot& \cdot& & & \cdot \\
             z_{N}& z_{N}{\bar z_{1}}& \cdot& \cdot& \zettai{z_{N}}^2
         \end{array}
     \right)  \nonumber \\
   &=& \ket{\left(z_{1}, z_{2}, \cdots, z_{N}\right)}
       \bra{\left(z_{1}, z_{2}, \cdots, z_{N}\right)}\ ,
\end{eqnarray}
where 
\begin{equation}
  \ket{\left(z_{1}, z_{2}, \cdots, z_{N} \right)} = 
  \frac{1}{\sqrt{1 + \sum_{j=1}^{N} \zettai{z_{j}}^2}}
     \left(
         \begin{array}{c}
             1 \\
             z_{1} \\
             \cdot \\
             \cdot \\
             z_{N} 
         \end{array}
     \right).  \nonumber \\
\end{equation}

Let us give a more detail description for the cases $N$ = $1$ and $2$.

\par \noindent
 (a) {\bf $N = 1$} : 
\begin{eqnarray}
 \label{eq:cp1-1}
  P(z)&=&\frac{1}{1+\zettai{z}^2}
     \left(
         \begin{array}{cc}
             1& {\bar z} \\
             z& \zettai{z}^2 
         \end{array}
     \right)  
   = \ket{z}\bra{z}, \nonumber  \\
  &&\mbox{where}\ \ket{z}=\frac{1}{\sqrt{1+\zettai{z}^2}}
     \left(
         \begin{array}{c}
             1 \\
             z 
         \end{array}
     \right), 
  \quad z=\frac{\zeta_{1}}{\zeta_{0}}, \quad \mbox{on}\ U_{1}\ ,  \\
 \label{eq:cp1-2}
  P(w)&=&\frac{1}{\zettai{w}^2+1}
     \left(
         \begin{array}{cc}
             \zettai{w}^2 & w \\
             {\bar w}& 1
         \end{array}
     \right)  
   = \ket{w}\bra{w},  \nonumber  \\
  &&\mbox{where}\ \ket{w}=\frac{1}{\sqrt{\zettai{w}^2+1}}
     \left(
         \begin{array}{c}
             w \\
             1
         \end{array}
     \right), 
  \quad w=\frac{\zeta_{0}}{\zeta_{1}}, \quad \mbox{on}\ U_{2}\ . 
\end{eqnarray}

\vspace{5mm}
\par \noindent
 (b) {\bf $N = 2$} : 
\begin{eqnarray}
 \label{eq:cp2-1}
  P(z_{1},z_{2})&=&\frac{1}{1+\zettai{z_{1}}^2+\zettai{z_{2}}^2}
     \left(
         \begin{array}{ccc}
             1& {\bar z_{1}}& {\bar z_{2}} \\
             z_{1}& \zettai{z_{1}}^2& z_{1}{\bar z_{2}} \\
             z_{2}& z_{2}{\bar z_{1}}& \zettai{z_{2}}^2 
         \end{array}
     \right)  
   = \ket{(z_{1},z_{2})}\bra{(z_{1},z_{2})}, \nonumber  \\
  \mbox{where}\quad  
  &\ket{(z_{1},z_{2})}&=\frac{1}{\sqrt{1+\zettai{z_{1}}^2+\zettai{z_{2}}^2}}
     \left(
         \begin{array}{c}
             1 \\
             z_{1} \\
             z_{2} 
         \end{array}
     \right), 
\quad (z_{1},z_{2})=\left(\frac{\zeta_{1}}{\zeta_{0}},
   \frac{\zeta_{2}}{\zeta_{0}} \right)  \quad \mbox{on}\ U_{1}\ ,  \\
 \label{eq:cp2-2}
  P(w_{1},w_{2})&=&\frac{1}{\zettai{w_{1}}^2+1+\zettai{w_{2}}^2}
     \left(
         \begin{array}{ccc}
             \zettai{w_{1}}^2& w_{1}& w_{1}{\bar w_{2}} \\
             {\bar w_{1}}& 1& {\bar w_{2}} \\
             w_{2}{\bar w_{1}}& w_{2}& \zettai{w_{2}}^2 
         \end{array}
     \right)  
   = \ket{(w_{1},w_{2})}\bra{(w_{1},w_{2})}, \nonumber \\
  \mbox{where}\quad 
  &\ket{(w_{1},w_{2})}&=\frac{1}{\sqrt{\zettai{w_{1}}^2+1+\zettai{w_{2}}^2}}
     \left(
         \begin{array}{c}
             w_{1} \\
              1  \\
             w_{2} 
         \end{array}
     \right), 
\quad  (w_{1},w_{2})=\left(\frac{\zeta_{0}}{\zeta_{1}},
   \frac{\zeta_{2}}{\zeta_{1}} \right)\ \  \mbox{on}\ U_{2}\ , \\
 \label{eq:cp2-3}
  P(v_{1},v_{2})&=&\frac{1}{\zettai{v_{1}}^2+\zettai{v_{2}}^2+1}
     \left(
         \begin{array}{ccc}
             \zettai{v_{1}}^2& v_{1}{\bar v_{2}}& v_{1} \\
             v_{2}{\bar v_{1}}& \zettai{v_{2}}^2& v_{2} \\
             {\bar v_{1}}& {\bar v_{2}}& 1 
         \end{array}
     \right)  
   = \ket{(v_{1},v_{2})}\bra{(v_{1},v_{2})}, \nonumber \\
  \mbox{where}\quad 
  &\ket{(v_{1},v_{2})}&=\frac{1}{\sqrt{\zettai{v_{1}}^2+\zettai{v_{2}}^2+1}}
     \left(
         \begin{array}{c}
             v_{1} \\
             v_{2}  \\
              1 
         \end{array}
     \right), 
\quad (v_{1},v_{2})=\left(\frac{\zeta_{0}}{\zeta_{2}},
   \frac{\zeta_{1}}{\zeta_{2}} \right)  \quad \mbox{on}\ U_{3}\ . 
\end{eqnarray}

\subsection{Bell States Revisited}

In this subsection we show that (\ref{eq:generalized-Bell-state}) coinsides 
with the Bell states (\ref{eq:Bell-states-1})--(\ref{eq:Bell-states-4}) 
by choosing anti-automorphism $\flat$ suitably.

We treat first of all the case of spin $\frac{1}{2}$. From here we 
identify 
\[
\ket{0}= {1\choose 0}\quad \mbox{and}\quad \ket{1}={0\choose 1},  
\]
so we have 
\begin{equation}
\ket{\eta}=\frac{1}{\sqrt{1+\zettai{\eta}^2}}(\ket{0}+\eta\ket{1})
=\frac{1}{\sqrt{1+\zettai{\eta}^2}}
\left(
  \begin{array}{c}
  1 \\
  \eta
  \end{array}
\right).
\end{equation}
In this case we consider the following four anti-automorphisms 
(\ref{eq:automorphism}) and (\ref{eq:anti-map}) :
\begin{equation}
 \label{eq:4-anti-automorphisms}
  \mbox{(1)}\ \  \eta^{\flat}={\bar \eta}\qquad  
  \mbox{(2)}\ \  \eta^{\flat}=-{\bar \eta}\qquad  
  \mbox{(3)}\ \  \eta^{\flat}=\frac{1}{\bar \eta}\qquad  
  \mbox{(4)}\ \  \eta^{\flat}=\frac{-1}{\bar \eta}\ .
\end{equation}
Now by making use of these we define 

\par \noindent 
\textbf{Definition}
\begin{eqnarray}
&&\mbox{(1)}\ \ \bell{B}=\frac{1}{\sqrt{2}}\int_{\fukuso}d\mu(\eta,{\bar \eta})
  \ket{\eta}\otimes \ket{{\bar \eta}}, \\
&&\mbox{(2)}\ \ \bell{B}=\frac{1}{\sqrt{2}}\int_{\fukuso}d\mu(\eta,{\bar \eta})
  \ket{\eta}\otimes \ket{-{\bar \eta}}, \\
&&\mbox{(3)}\ \ \bell{B}=\frac{1}{\sqrt{2}}\int_{\fukuso}d\mu(\eta,{\bar \eta})
  \ket{\eta}\otimes \ket{1/{\bar \eta}}, \\
&&\mbox{(4)}\ \ \bell{B}=\frac{1}{\sqrt{2}}\int_{\fukuso}d\mu(\eta,{\bar \eta})
  \ket{\eta}\otimes \ket{-1/{\bar \eta}}, 
\end{eqnarray}
where we have put for simplicity 
\[
 d\mu(\eta,{\bar \eta})=\frac{2}{\pi}
 \frac{[d^{2}\eta]}{(1+\zettai{\eta}^2)^2}\ . 
\]
It is easy to see from (\ref{eq:cp1-1}) and (\ref{eq:cp1-2}) 
\begin{eqnarray}
&&\mbox{(1)}\ \ \ket{\eta^{\flat}}=\ket{{\bar \eta}}=
      \frac{1}{\sqrt{1+\zettai{\eta}^2}}(\ket{0}+{\bar \eta}\ket{1}), \\
&&\mbox{(2)}\ \ \ket{\eta^{\flat}}=\ket{-{\bar \eta}}=
      \frac{1}{\sqrt{1+\zettai{\eta}^2}}(\ket{0}-{\bar \eta}\ket{1}), \\
&&\mbox{(3)}\ \ \ket{\eta^{\flat}}=\ket{1/{\bar \eta}}=
      \frac{1}{\sqrt{1+\zettai{\eta}^2}}({\bar \eta}\ket{0}+\ket{1}), \\
&&\mbox{(4)}\ \ \ket{\eta^{\flat}}=\ket{-1/{\bar \eta}}=
      \frac{1}{\sqrt{1+\zettai{\eta}^2}}(-{\bar \eta}\ket{0}+\ket{1}), 
\end{eqnarray}
Then making use of elementary facts  
\begin{eqnarray}
&&\frac{2}{\pi}\int_{\fukuso} \frac{[d^{2}\eta]}{(1+\zettai{\eta}^2)^2}
   \frac{1}{1+\zettai{\eta}^2}=
 \frac{2}{\pi}\int_{\fukuso} \frac{[d^{2}\eta]}{(1+\zettai{\eta}^2)^2}
   \frac{\zettai{\eta}^2}{1+\zettai{\eta}^2}=1, \nonumber \\
&&\frac{2}{\pi}\int_{\fukuso} \frac{[d^{2}\eta]}{(1+\zettai{\eta}^2)^2}
   \frac{\eta}{1+\zettai{\eta}^2}=
 \frac{2}{\pi}\int_{\fukuso} \frac{[d^{2}\eta]}{(1+\zettai{\eta}^2)^2}
   \frac{{\bar \eta}}{1+\zettai{\eta}^2}=0, \nonumber 
\end{eqnarray}
we obtain easily 
\begin{eqnarray}
&&\mbox{(1)}\ \ \bell{B}
 = \frac{1}{\sqrt{2}}(\ket{0}\otimes \ket{0} + \ket{1}\otimes \ket{1}), \\
&&\mbox{(2)}\ \ \bell{B}
 = \frac{1}{\sqrt{2}}(\ket{0}\otimes \ket{0} - \ket{1}\otimes \ket{1}), \\
&&\mbox{(3)}\ \ \bell{B}
 = \frac{1}{\sqrt{2}}(\ket{0}\otimes \ket{1} + \ket{1}\otimes \ket{0}), \\
&&\mbox{(4)}\ \ \bell{B}
 = \frac{1}{\sqrt{2}}(\ket{0}\otimes \ket{1} - \ket{1}\otimes \ket{0}). 
\end{eqnarray}
We just recovered the Bell states (\ref{eq:Bell-states-1})--
(\ref{eq:Bell-states-4}) !! \quad 
We can say that four $\bell{B}$ in {\bf Definition} are overcomplete 
expression (making use of generalized coherent states) of the Bell states. 
This is an important point of view. 

\par \noindent 
Since we consider the case of higher spin $J$, we write $\ket{\eta}$ as 
\begin{equation}
  {\ket{\eta}}_{J}=\frac{1}{\left(1+\zettai{\eta}^2 \right)^J}
    \sum_{k=0}^{2J}\sqrt{{}_{2J}C_k}\ \eta^k \ket{k}  
\end{equation}
to emphasize the dependence of spin $J$. Here we have set $\ket{k}=
\kett{J}{k}$ for simplicity. 
From the above result it is very 
natural to define {\bf Bell states with spin $J$} as follows because the 
parameter space is the same ${\fukuso}P^1$ : 
\par \noindent 
\textbf{Definition}
\begin{eqnarray}
&&\mbox{(1)}\ \ \bell{B}=\frac{1}{\sqrt{2J+1}}\int_{\fukuso} 
d\mu(\eta,{\bar \eta})
  {\ket{\eta}}_{J}\otimes {\ket{{\bar \eta}}}_{J},  \\
&&\mbox{(2)}\ \ \bell{B}=\frac{1}{\sqrt{2J+1}}\int_{\fukuso} 
d\mu(\eta,{\bar \eta})
  {\ket{\eta}}_{J}\otimes {\ket{-{\bar \eta}}}_{J}, \\
&&\mbox{(3)}\ \ \bell{B}=\frac{1}{\sqrt{2J+1}}\int_{\fukuso} 
d\mu(\eta,{\bar \eta})
  {\ket{\eta}}_{J}\otimes {\ket{1/{\bar \eta}}}_{J}, \\
&&\mbox{(4)}\ \ \bell{B}=\frac{1}{\sqrt{2J+1}}\int_{\fukuso} 
d\mu(\eta,{\bar \eta})
  {\ket{\eta}}_{J}\otimes {\ket{-1/{\bar \eta}}}_{J}, 
\end{eqnarray}
where 
\[
d\mu(\eta,{\bar \eta})=\frac{2J+1}{\pi}
\frac{[d^{2}\eta]}{(1+\zettai{\eta}^2)^2}\ .
\]

\par \noindent
Let us calculate ${\ket{{\bar \eta}}}_{J}$, ${\ket{-{\bar \eta}}}_{J}$, 
${\ket{1/{\bar \eta}}}_{J}$ and ${\ket{-1/{\bar \eta}}}_{J}$. 
It is easy to see 
\begin{eqnarray}
&&\mbox{(1)}\ \ {\ket{{\bar \eta}}}_{J}=
\frac{1}{\left(1+\zettai{\eta}^2 \right)^J}
    \sum_{k=0}^{2J}\sqrt{{}_{2J}C_k}\ {\bar \eta}^k \ket{k}, \\  
&&\mbox{(2)}\ \ {\ket{-{\bar \eta}}}_{J}
=\frac{1}{\left(1+\zettai{\eta}^2 \right)^J}
    \sum_{k=0}^{2J}\sqrt{{}_{2J}C_k}\ (-1)^k{\bar \eta}^k \ket{k}, \\  
&&\mbox{(3)}\ \ {\ket{1/{\bar \eta}}}_{J}=
    \frac{1}{\left(1+\zettai{\eta}^2 \right)^J}
    \sum_{k=0}^{2J}\sqrt{{}_{2J}C_k}\ {\bar \eta}^k \ket{2J-k}, \\  
&&\mbox{(4)}\ \ {\ket{-1/{\bar \eta}}}_{J}=
    \frac{1}{\left(1+\zettai{\eta}^2 \right)^J}
    \sum_{k=0}^{2J}\sqrt{{}_{2J}C_k}\ (-1)^k{\bar \eta}^k \ket{2J-k}. 
\end{eqnarray}
From this lemma and the elementary facts  
\[
  \frac{2J+1}{\pi}\int_{\fukuso} \frac{[d^{2}\eta]}{(1+\zettai{\eta}^2)^2}
   \frac{\zettai{\eta}^{2k}}{(1+\zettai{\eta}^2)^{2J}}=\frac{1}{{}_{2J}C_k}
  \quad \mbox{for}\quad 0 \le k \le 2J\ ,
\]
we can give explicit forms to the Bell states with spin $J$ : 
\begin{eqnarray}
&&\mbox{(1)}\ \ \bell{B}=\frac{1}{\sqrt{2J+1}} \sum_{k=0}^{2J}
   \ket{k}\otimes \ket{k},  \\
&&\mbox{(2)}\ \ \bell{B}=\frac{1}{\sqrt{2J+1}} \sum_{k=0}^{2J}(-1)^{k}
   \ket{k}\otimes \ket{k}, \\
&&\mbox{(3)}\ \ \bell{B}=\frac{1}{\sqrt{2J+1}} \sum_{k=0}^{2J}
  \ket{k}\otimes \ket{2J-k}, \\
&&\mbox{(4)}\ \ \bell{B}=\frac{1}{\sqrt{2J+1}} \sum_{k=0}^{2J}(-1)^{k}
  \ket{k}\otimes \ket{2J-k}.
\end{eqnarray}
We obtained the Bell states with spin $J$ which are a natural extension of 
usual ones ($J=1/2$). 

\par \noindent 
A comment is in order.  For the case $J=1$ :
\begin{eqnarray}
&&\mbox{(1)}\ \ \frac{1}{\sqrt{3}} 
   (\ket{0}\otimes \ket{0}+\ket{1}\otimes \ket{1}+\ket{2}\otimes \ket{2}), 
   \nonumber \\
&&\mbox{(2)}\ \ \frac{1}{\sqrt{3}} 
   (\ket{0}\otimes \ket{0}-\ket{1}\otimes \ket{1}+\ket{2}\otimes \ket{2}), 
  \nonumber \\
&&\mbox{(3)}\ \ \frac{1}{\sqrt{3}} 
   (\ket{0}\otimes \ket{2}+\ket{1}\otimes \ket{1}+\ket{2}\otimes \ket{0}), 
   \nonumber \\
&&\mbox{(4)}\ \ \frac{1}{\sqrt{3}} 
   (\ket{0}\otimes \ket{2}-\ket{1}\otimes \ket{1}+\ket{2}\otimes \ket{0}). 
    \nonumber 
\end{eqnarray}
It is easy to see that they are not linearly independent, so that only this 
case is very special (peculiar).

\par \vspace{5mm} \noindent
{\bf Comment} \quad We cannot give a geometric construction of Bell states 
by making use of generalized coherent states based on $su(1,1)$ (Lie 
algebra of non--compact Lie group). Because the parameter space in this 
case is a Poincare disk $D=\{\zeta \in \fukuso\ |\ \zettai{\zeta} < 1\}$ and 
the measure on it is given by 
\[
d\mu(\zeta, {\bf \zeta})=\frac{2K-1}{\pi}
\frac{[d^{2}\zeta]}{(1-\zettai{\zeta}^{2})^{2}}\ , 
\]
see (\ref{eq:2-2-18}). Therefore we have 
\[
\int_{D}d\mu(\zeta, {\bf \zeta}) = (2K-1)\int_{0}^{1}dr 
\frac{1}{(1-r)^{2}} = \infty \ ! 
\]
Compare this with (\ref{eq:the resolution-of-unity}). This is a reason why 
we cannot determine a normalization.

\vspace{10mm}
\section{Topics in Quantum Information Theory}

In this section we don't introduce a general theory of quantum information 
theory (see for example \cite{LPS}), but focus our attension to special topics 
of it, that is, 
\begin{itemize}
   \item swap of coherent states 
   \item cloning of coherent states 
\end{itemize}
Because this is just a good one as examples
of applications of coherent and generalized coherent states and 
our method developed in the following may open a new possibility. 

\par \noindent 
First let us define a swap operator : 
\begin{equation}
S : {\calh}\otimes {\calh} \longrightarrow {\calh}\otimes {\calh}, \quad 
  S(a\otimes b)=b\otimes a \quad \mbox{for any}\ a, b \in {\calh} 
\end{equation}
where ${\calh}$ is the Fock space in Section 2.

\par \noindent 
It is not difficult to construct this operator in a universal manner, 
see Appendix B. But 
for coherent states we can construct a better one by making use of 
generalized coherent operators in the preceding section. 

\par \vspace{5mm} \noindent 
Next let us introduce no cloning theorem, \cite{WZ}.  For that 
we define a cloning (copying) operator C which is unitary 
\begin{equation}
 C : {\calh}\otimes {\calh} \longrightarrow {\calh}\otimes {\calh}, \quad 
    C(h\otimes \ket{0})=h\otimes h \quad \mbox{for any}\ h \in {\calh}\ .
\end{equation}
It is very known that there is no cloning theorem 

\noindent{\bfseries ``No Cloning Theorem''}\quad We have no $C$ above. 

The proof is very easy (almost trivial). Because $2h=h+h \in {\calh}$ and $C$ 
is a linear operator, so 
\begin{equation}
 \label{eq:C-equality}
   C(2h\otimes \ket{0})=2C(h\otimes \ket{0}). 
\end{equation}
The LHS of (\ref{eq:C-equality}) is 
\[
   C(2h\otimes \ket{0})= 2h\otimes 2h= 4(h\otimes h),
\]
while the RHS of (\ref{eq:C-equality}) 
\[
   2C(h\otimes \ket{0})= 2(h\otimes h).
\]
This is a contradiction.\quad  This is called no cloning theorem. 

\par \noindent 
Let us return to the case of coherent states. 
For coherent states $\ket{\alpha}$ and $\ket{\beta}$\ \ the superposition 
$\ket{\alpha}+\ket{\beta}$ is no longer a coherent state, so that coherent 
states may not suffer from the theorem above. 

\begin{flushleft}
 {\bf Problem}\ \ Is it possible to clone coherent states ? 
\end{flushleft}

\par \noindent
At this stage it is not easy, so we will make do with 
approximating it (imperfect cloning in our terminology) 
instead of making a perfect cloning. 

\par \noindent
We write notations once more. 
\begin{eqnarray}
   \mbox{Coherent States}\quad &&\ket{\alpha}=D(\alpha)\ket{0}
   \quad \mbox{for} \quad \alpha \in \fukuso   \nonumber \\
   \mbox{Squeezed--like States}\quad &&\ket{\beta}=S(\beta)\ket{0}
   \ \quad \mbox{for} \quad \beta \in \fukuso    \nonumber 
\end{eqnarray}

\vspace{10mm}
\subsection{Some Useful Formulas}

We list and prove some useful formulas in the following. 
Now we prepare some parameters $\alpha,\ \epsilon,\ \kappa$ in which 
$\epsilon, \kappa$ are free ones, while $\alpha$ is unknown one in the 
cloning case. 
Let us unify the notations as follows. 
\begin{eqnarray}
 &&\alpha : \mbox{(unknown)}\quad  \alpha=\zettai{\alpha}\mbox{e}^{i\chi},  \\
 &&\epsilon : \mbox{known}\quad  \quad \quad \ 
              \epsilon=\zettai{\epsilon}\mbox{e}^{i\phi}, \\
 &&\kappa :  \mbox{known}\quad  \quad \ \ \ 
              \kappa=\zettai{\kappa}\mbox{e}^{i\delta},
\end{eqnarray}

\par \noindent 
Let us start. 

\par \noindent
\mbox{(i)}\ First let us calculate
\begin{equation}
S(\epsilon)D(a)S(\epsilon)^{-1}.
\end{equation}
For that we show 
\begin{equation}
\label{eq:squeezed-rotation}
S(\epsilon)aS(\epsilon)^{-1}=cosh(\zettai{\epsilon})a
        -\mbox{e}^{i\phi}sinh(\zettai{\epsilon})a^{\dagger}.
\end{equation}
Proof is as follows.\quad For $X=(1/2)\{\epsilon(a^{\dagger})^{2}-
{\bar \epsilon}a^{2}\}$ we have easily $[X,a]=-{\epsilon}a^{\dagger}$ and 
$[X,a^{\dagger}]=-{\bar \epsilon}a$, so 
\begin{eqnarray}
S(\epsilon)aS(\epsilon)^{-1}&=&\mbox{e}^{X}a\mbox{e}^{-X}
=a+[X,a]+\frac{1}{2!}[X,[X,a]]+\frac{1}{3!}[X,[X,[X,a]]]+\cdots
   \nonumber \\
&=&a-{\epsilon}a^{\dagger}+\frac{\zettai{\epsilon}^2}{2!}a-
\frac{\epsilon\zettai{\epsilon}^2}{3!}a^{\dagger}+\cdots
   \nonumber \\
&=&\left\{1+\frac{\zettai{\epsilon}^2}{2!}+\cdots \right\}a-
   \frac{\epsilon}{\zettai{\epsilon}}\left\{\zettai{\epsilon}+
   \frac{\zettai{\epsilon}^3}{3!}+\cdots \right\}a^{\dagger}
   \nonumber \\
&=&cosh(\zettai{\epsilon})a-\frac{\epsilon sinh(\zettai{\epsilon})}{
\zettai{\epsilon}}a^{\dagger}
=cosh(\zettai{\epsilon})a-\mbox{e}^{i\phi}sinh(\zettai{\epsilon})a^{\dagger}. 
   \nonumber
\end{eqnarray}

\par \noindent
From this it is easy to check 
\begin{eqnarray}
\label{eq:adjoint-alpha}
 S(\epsilon)D(\alpha)S(\epsilon)^{-1}&=&
 D\left(\alpha S(\epsilon)a^{\dagger}S(\epsilon)^{-1}-
 {\bar \alpha}S(\epsilon)aS(\epsilon)^{-1}\right)
   \nonumber \\
 &=&D\left(cosh(\zettai{\epsilon})\alpha
    +\mbox{e}^{i\phi}sinh(\zettai{\epsilon}){\bar \alpha}\right).  
\end{eqnarray}
Therefore 
\begin{equation}
\label{eq:adjoint-alpha-choice}
S(\epsilon)D(\alpha)S(\epsilon)^{-1}=\left\{
   \begin{array}{ll}
     \displaystyle{D(\mbox{e}^{\zettai{\epsilon}}\alpha)}\ 
       \qquad  \mbox{if}\quad \phi=2\chi   \\
     \displaystyle{D(\mbox{e}^{-\zettai{\epsilon}}\alpha)}
       \qquad  \mbox{if}\quad \phi=2\chi+\pi
   \end{array}
 \right.
\end{equation}
By making use of this formula we can change a scale of $\alpha$.

\par \vspace{5mm} \noindent
\mbox{(ii)}\ Next le us calculate 
\begin{equation}
S(\epsilon)S(\alpha)S(\epsilon)^{-1}.
\end{equation}
From the definition 
\[
 S(\epsilon)S(\alpha)S(\epsilon)^{-1}
 =S(\epsilon)\mbox{exp}\left\{ \frac{1}{2}\left(\alpha (a^{\dagger})^{2}-
 {\bar \alpha}a^{2}\right) \right\}S(\epsilon)^{-1}
 \equiv \mbox{e}^{Y/2}
\]
where 
\[
Y=\alpha\left(S(\epsilon)a^{\dagger}S(\epsilon)^{-1}\right)^{2}
  -{\bar \alpha}\left(S(\epsilon)aS(\epsilon)^{-1}\right)^{2}.
\]
From (\ref{eq:squeezed-rotation}) and after some calculations we have 
\begin{eqnarray}
Y&=&\left\{ cosh^{2}(\zettai{\epsilon})\alpha
         -\mbox{e}^{2i\phi}sinh^{2}(\zettai{\epsilon}){\bar \alpha}
  \right\}(a^{\dagger})^2-
  \left\{ cosh^{2}(\zettai{\epsilon}){\bar \alpha}
         -\mbox{e}^{-2i\phi}sinh^{2}(\zettai{\epsilon})\alpha
  \right\}a^2      \nonumber \\
  &+&\frac{(-\mbox{e}^{-i\phi}\alpha+\mbox{e}^{i\phi}{\bar \alpha})}{2}
       sinh(2\zettai{\epsilon})(a^{\dagger}a+aa^{\dagger}) \nonumber \\
 &=&\left\{ cosh^{2}(\zettai{\epsilon})\alpha
         -\mbox{e}^{2i\phi}sinh^{2}(\zettai{\epsilon}){\bar \alpha}
  \right\}(a^{\dagger})^2-
  \left\{ cosh^{2}(\zettai{\epsilon}){\bar \alpha}
         -\mbox{e}^{-2i\phi}sinh^{2}(\zettai{\epsilon})\alpha
  \right\}a^2     \nonumber \\
  &+&(-\mbox{e}^{-i\phi}\alpha+\mbox{e}^{i\phi}{\bar \alpha})
    sinh(2\zettai{\epsilon})(a^{\dagger}a+\frac{1}{2}) \quad 
    (\Longleftarrow  [a, a^{\dagger}]=1),   \nonumber 
\end{eqnarray}
or 
\begin{eqnarray}
 \frac{1}{2}Y&=&\left\{ cosh^{2}(\zettai{\epsilon})\alpha
         -\mbox{e}^{2i\phi}sinh^{2}(\zettai{\epsilon}){\bar \alpha}
  \right\}K_{+}-
  \left\{ cosh^{2}(\zettai{\epsilon}){\bar \alpha}
         -\mbox{e}^{-2i\phi}sinh^{2}(\zettai{\epsilon})\alpha
  \right\}K_{-}    \nonumber \\
  &+&(-\mbox{e}^{-i\phi}\alpha+\mbox{e}^{i\phi}{\bar \alpha})
    sinh(2\zettai{\epsilon})K_{3}
\end{eqnarray}
with $\{K_{+},K_{-},K_{3}\}$ in (\ref{eq:2-2-21}). This is our formula. 

\par \noindent 
Now
\[
  -\mbox{e}^{-i\phi}\alpha+\mbox{e}^{i\phi}{\bar \alpha}
  =\zettai{\alpha}(-\mbox{e}^{-i(\phi-\chi)}+\mbox{e}^{i(\phi-\chi)})
  =2i\zettai{\alpha}sin(\phi-\chi),
\]
so if we choose $\phi=\chi$, then $\mbox{e}^{2i\phi}{\bar \alpha}=
\mbox{e}^{2i\chi}\mbox{e}^{-i\chi}\zettai{\alpha}=
\alpha$ and 
\[
   cosh^{2}(\zettai{\epsilon}){\alpha}
         -\mbox{e}^{2i\phi}sinh^{2}(\zettai{\epsilon}){\bar \alpha} 
  =\left(cosh^{2}(\zettai{\epsilon})-
         sinh^{2}(\zettai{\epsilon})\right)\alpha 
  =\alpha
\]
, and finally
\[
Y=\alpha(a^{\dagger})^2-{\bar \alpha}a^{2}.
\]
That is, 
\[
  S(\epsilon)S(\alpha)S(\epsilon)^{-1}=S(\alpha) \Longleftrightarrow 
  S(\epsilon)S(\alpha)=S(\alpha)S(\epsilon).
\]
The operators $S(\epsilon)$ and $S(\alpha)$ commute if the phases of 
$\epsilon$ and $\alpha$ coincide.

\par \vspace{5mm} \noindent
\mbox{(iii)}\ Third formula is : \quad For $V(t)=\mbox{e}^{itN}$\ where 
$N=a^{\dagger}a$ (a number operator) 
\begin{equation}
\label{eq:phase-factor}
V(t)D(\alpha)V(t)^{-1}=D(\mbox{e}^{it}\alpha).
\end{equation}
{\noindent}The proof is as follows. \quad 
\[
V(t)D(\alpha)V(t)^{-1}=\mbox{exp}\left(\alpha V(t)a^{\dagger}V(t)^{-1}-
  {\bar \alpha}V(t)aV(t)^{-1}\right).
\]
It is easy to see 
\begin{eqnarray}
V(t)aV(t)^{-1}&=&\mbox{e}^{itN}a\mbox{e}^{-itN}
              =a + [itN,a] + \frac{1}{2!}[itN,[itN,a]]+ \cdots \nonumber \\
              &=&a+(-it)a+\frac{(-it)^2}{2!}a+\cdots \nonumber \\
              &=&\mbox{e}^{-it}a. \nonumber
\end{eqnarray}
Therefore we obtain 
\[
V(t)D(\alpha)V(t)^{-1}=\mbox{exp}\left(\alpha \mbox{e}^{it}a^{\dagger}
                    -{\bar \alpha}\mbox{e}^{-it}a^{\dagger}\right)
                    =D(\mbox{e}^{it}\alpha).
\]
This formula is often used as follows. 
\begin{equation}
\label{eq:phase-operator}
\ket{\alpha}\ \longrightarrow \ 
V(t)\ket{\alpha}=V(t)D(\alpha)V(t)^{-1}V(t)\ket{0}
=D(\mbox{e}^{it}\alpha)\ket{0}
=\ket{\mbox{e}^{it}\alpha},
\end{equation}
where we have used 
\[
V(t)\ket{0}=\ket{0}
\]
becase $N\ket{0}=0$. 
That is, we can add a phase to $\alpha$ by making use of this formula.

\subsection{Swap of Coherent States}
The purpose of this section is to construct a swap operator 
satifying 
\begin{equation}
     \ket{\alpha_{1}}\otimes \ket{\alpha_{2}} \longrightarrow 
     \ket{\alpha_{2}}\otimes \ket{\alpha_{1}}.
\end{equation}
Let us remember $U_{J}(\kappa)$ once more 
\[
U_{J}(\kappa)=\mbox{e}^{\kappa a_{1}^{\dagger}a_{2}-{\bar \kappa}
a_{1}a_{2}^{\dagger}} \quad \mbox{for} \quad \kappa \in \fukuso .
\]
We note an important property of this operator : 
\begin{equation}
\label{eq:invariant-property}
U_{J}(\kappa)\ket{0}\otimes \ket{0}=\ket{0}\otimes \ket{0}.
\end{equation}
The construction is as follows. 
\begin{eqnarray}
\label{eq:swap-operator}
U_{J}(\kappa)\ket{\alpha_{1}}\otimes \ket{\alpha_{2}}
&=&U_{J}(\kappa)D(\alpha_{1})\otimes D(\alpha_{2})\ket{0}\otimes \ket{0}
=U_{J}(\kappa)D_{1}(\alpha_{1})D_{2}(\alpha_{2})\ket{0}\otimes \ket{0}
      \nonumber \\
&=&U_{J}(\kappa)D_{1}(\alpha_{1})D_{2}(\alpha_{2})U_{J}(\kappa)^{-1}
U_{J}(\kappa)\ket{0}\otimes \ket{0}
      \nonumber \\
&=&U_{J}(\kappa)D_{1}(\alpha_{1})D_{2}(\alpha_{2})U_{J}(\kappa)^{-1}
\ket{0}\otimes \ket{0} \quad \mbox{by}\quad (\ref{eq:invariant-property}), 
\end{eqnarray}
and 
\begin{eqnarray}
&&U_{J}(\kappa)D_{1}(\alpha_{1})D_{2}(\alpha_{2})U_{J}(\kappa)^{-1}=
U_{J}(\kappa)
\mbox{exp}\left\{\alpha_{1}a_{1}^{\dagger}-{\bar \alpha_{1}}a_{1} + 
          \alpha_{2}a_{2}^{\dagger}-{\bar \alpha_{2}}a_{2}
          \right\}
U_{J}(\kappa)^{-1}  \nonumber \\
&&=\mbox{exp}
\left\{
\alpha_{1}(U_{J}(\kappa)a_{1}U_{J}(\kappa)^{-1})^{\dagger}-
{\bar \alpha_{1}}U_{J}(\kappa)a_{1}U_{J}(\kappa)^{-1} 
\right.  \nonumber \\
&&\left. \qquad \ + 
\alpha_{2}(U_{J}(\kappa)a_{2}U_{J}(\kappa)^{-1})^{\dagger}-
{\bar \alpha_{2}}U_{J}(\kappa)a_{2}U_{J}(\kappa)^{-1}
\right\}  \nonumber \\
     \label{eq:tensor-product}
&&\equiv \mbox{exp}(X).
\end{eqnarray}
From (\ref{eq:J-rotation-1}) and (\ref{eq:J-rotation-2}) we have 
\begin{eqnarray}
X
&=&
\left\{
cos(\zettai{\kappa})\alpha_{1}+
\frac{\kappa sin(\zettai{\kappa})}{\zettai{\kappa}}\alpha_{2}
\right\}a_{1}^{\dagger}
-
\left\{
cos(\zettai{\kappa}){\bar \alpha_{1}}+
\frac{{\bar \kappa}sin(\zettai{\kappa})}{\zettai{\kappa}}{\bar \alpha_{2}}
\right\}a_{1}  \nonumber \\
&+&
\left\{
cos(\zettai{\kappa})\alpha_{2}-
\frac{{\bar \kappa}sin(\zettai{\kappa})}{\zettai{\kappa}}\alpha_{1}
\right\}a_{2}^{\dagger}
-
\left\{
cos(\zettai{\kappa}){\bar \alpha_{2}}-
\frac{\kappa sin(\zettai{\kappa})}{\zettai{\kappa}}{\bar \alpha_{1}}
\right\}a_{2},  \nonumber
\end{eqnarray}
so
\begin{eqnarray}
\mbox{exp}(X)
&=&
D_{1}
\left(
cos(\zettai{\kappa})\alpha_{1}+
\frac{\kappa sin(\zettai{\kappa})}{\zettai{\kappa}}\alpha_{2}
\right)\ 
D_{2}
\left(
cos(\zettai{\kappa})\alpha_{2}-
\frac{{\bar \kappa}sin(\zettai{\kappa})}{\zettai{\kappa}}\alpha_{1}
\right)  \nonumber \\
&=&
D
\left(
cos(\zettai{\kappa})\alpha_{1}+
\frac{\kappa sin(\zettai{\kappa})}{\zettai{\kappa}}\alpha_{2}
\right)
\otimes
D
\left(
cos(\zettai{\kappa})\alpha_{2}-
\frac{{\bar \kappa}sin(\zettai{\kappa})}{\zettai{\kappa}}\alpha_{1}
\right).  \nonumber 
\end{eqnarray}
Therefore we have from (\ref{eq:tensor-product})
\[
\ket{\alpha_{1}}\otimes \ket{\alpha_{2}}\  \longrightarrow \ 
\ket{
cos(\zettai{\kappa})\alpha_{1}+
\frac{\kappa sin(\zettai{\kappa})}{\zettai{\kappa}}\alpha_{2}
} 
\otimes 
\ket{
cos(\zettai{\kappa})\alpha_{2}-
\frac{{\bar \kappa}sin(\zettai{\kappa})}{\zettai{\kappa}}\alpha_{1}
}.
\]
If we write $\kappa$ as $\zettai{\kappa}\mbox{e}^{i\delta}$, then 
the above formula reduces to
\[
\ket{\alpha_{1}}\otimes \ket{\alpha_{2}}\  \longrightarrow \ 
\ket{
cos(\zettai{\kappa})\alpha_{1}+
\mbox{e}^{i\delta}sin(\zettai{\kappa})\alpha_{2}
}
\otimes
\ket{
cos(\zettai{\kappa})\alpha_{2}-
\mbox{e}^{-i\delta}sin(\zettai{\kappa})\alpha_{1}
}.
\]
Here if we choose $sin(\zettai{\kappa})=1$, then 
\[
\ket{\alpha_{1}}\otimes \ket{\alpha_{2}}\  \longrightarrow \ 
\ket{\mbox{e}^{i\delta}\alpha_{2}}
\otimes
\ket{-\mbox{e}^{-i\delta}\alpha_{1}}
=
\ket{\mbox{e}^{i\delta}\alpha_{2}}
\otimes
\ket{\mbox{e}^{-i(\delta+\pi)}\alpha_{1}}.
\]
Now by operating the operator $V=\mbox{e}^{-i\delta N}\otimes 
\mbox{e}^{i(\delta+\pi) N}$ where $N=a^{\dagger}a$ 
from the left (see (\ref{eq:phase-operator})) 
we obtain the swap 
\[
\ket{\alpha_{1}}\otimes \ket{\alpha_{2}}\  \longrightarrow \ 
\ket{\alpha_{2}}\otimes \ket{\alpha_{1}}.
\]
A comment is in order.\ \ In the formula we set $\alpha_{1}=\alpha$ and 
$\alpha_{2}=0$, then the formula reduces to 
\begin{equation}
\label{eq:adjoint-form}
U_{J}(\kappa)D_{1}(\alpha)U_{J}(\kappa)^{-1}
=D_{1}(cos(\zettai{\kappa})\alpha)
 D_{2}(-\mbox{e}^{-i\delta}sin(\zettai{\kappa})\alpha).
\end{equation}

\subsection{Imperfect Cloning of Coherent States}
We cannot clone coherent states in a perfect manner likely 
\begin{equation}
   \ket{\alpha}\otimes \ket{0} \longrightarrow 
   \ket{\alpha}\otimes \ket{\alpha} \quad \mbox{for}\ \alpha \in \fukuso .
\end{equation}
Then our question is : is it possible to approximate ?  
We show that we can at least make an ``imperfect cloning" in our terminology 
against the statement of \cite{DG}. 

\par \noindent 
Let us start. The method is almost same with one in the preceding 
subsection, but we repeat it once more. 
Operating the operator $U_{J}(\kappa)$ on $\ket{\alpha}\otimes \ket{0}$ 
\begin{eqnarray}
&&U_{J}(\kappa)\ket{\alpha}\otimes \ket{0}=
U_{J}(\kappa)\left\{ D(\alpha)\otimes {\bf 1} \right\}\ket{0}\otimes \ket{0}
=U_{J}(\kappa)D_{1}(\alpha)\ket{0}\otimes \ket{0}  \nonumber \\
=&& 
U_{J}(\kappa)D_{1}(\alpha)U_{J}(\kappa)^{-1}U_{J}(\kappa)
\ket{0}\otimes \ket{0}
=U_{J}(\kappa)D_{1}(\alpha)U_{J}(\kappa)^{-1}\ket{0}\otimes \ket{0}   
\quad \mbox{by (\ref{eq:invariant-property})}   \nonumber \\
=&&D_{1}(cos(\zettai{\kappa})\alpha)
   D_{2}(-\mbox{e}^{-i\delta}sin(\zettai{\kappa})\alpha) 
   \ket{0}\otimes \ket{0} \quad \mbox{by (\ref{eq:adjoint-form})}
   \nonumber \\
=&&D_{1}(cos(\zettai{\kappa})\alpha)
 D_{2}(\mbox{e}^{-i(\delta+\pi)}sin(\zettai{\kappa})\alpha)
 \ket{0}\otimes \ket{0}  \nonumber \\
=&&\left\{
 D(cos(\zettai{\kappa})\alpha)\otimes 
 D(\mbox{e}^{-i(\delta+\pi)}sin(\zettai{\kappa})\alpha)
 \right\}\ket{0}\otimes \ket{0}. \nonumber 
\end{eqnarray}
Operating the operator ${\bf 1}\otimes \mbox{e}^{i(\delta+\pi)N}$ on 
the last equation 
\begin{eqnarray}
 &&D(cos(\zettai{\kappa})\alpha)\otimes 
 \mbox{e}^{i(\delta+\pi)N}
 D(\mbox{e}^{-i(\delta+\pi)}sin(\zettai{\kappa})\alpha)
 \ket{0}\otimes \ket{0} \nonumber \\
=&&
 D(cos(\zettai{\kappa})\alpha)\otimes 
 \mbox{e}^{i(\delta+\pi)N}
 D(\mbox{e}^{-i(\delta+\pi)}sin(\zettai{\kappa})\alpha)
 \mbox{e}^{-i(\delta+\pi)N}
 \mbox{e}^{i(\delta+\pi)N}\ket{0}\otimes \ket{0} \nonumber \\
=&&
 D(cos(\zettai{\kappa})\alpha)\otimes 
 \mbox{e}^{i(\delta+\pi)N}
 D(\mbox{e}^{-i(\delta+\pi)}sin(\zettai{\kappa})\alpha)
 \mbox{e}^{-i(\delta+\pi)N}
 \ket{0}\otimes \ket{0} 
 \nonumber \\
=&&D(cos(\zettai{\kappa})\alpha)\otimes 
 D(\mbox{e}^{-i(\delta+\pi)}sin(\zettai{\kappa})\alpha
 \mbox{e}^{i(\delta+\pi)})
 \ket{0}\otimes \ket{0} \quad \mbox{by (\ref{eq:phase-factor})}
 \nonumber \\
=&&D(cos(\zettai{\kappa})\alpha)\otimes 
 D(sin(\zettai{\kappa})\alpha)
 \ket{0}\otimes \ket{0} \nonumber \\
=&&\ket{cos(\zettai{\kappa})\alpha}\otimes 
 \ket{sin(\zettai{\kappa})\alpha}. \nonumber 
\end{eqnarray}
Namely we have constructed 
\begin{equation}
  \ket{\alpha}\otimes \ket{0} \longrightarrow 
 \ket{cos(\zettai{\kappa})\alpha}\otimes 
 \ket{sin(\zettai{\kappa})\alpha}. 
\end{equation}
This is an ``imperfect cloning" what we have called. 

{\bf A comment is in order.} \quad 
The authors in \cite{DG} state that the ``perfect cloning" (in their 
terminology) for coherent states is possible. But it is not correct as 
shown below. 
Their method is very interesting, so let us introduce it. 

\par \noindent 
Before starting let us prepare a notation for simplicity 
(\ref{eq:adjoint-alpha}) : 
\[
S(\epsilon)D(\alpha)S(\epsilon)^{-1}=D({\tilde \alpha}), \quad 
{\tilde \alpha}\equiv cosh(\zettai{\epsilon})\alpha+
\mbox{e}^{i\phi}sinh(\zettai{\epsilon}){\bar \alpha}.
\]
Operating the operator $
S(\epsilon)\otimes S(\mbox{e}^{-2i\delta})$ from the left 
\begin{eqnarray}
S(\epsilon)\otimes S(\mbox{e}^{-2i\delta})\ket{\alpha}\otimes \ket{0}
&=&\left\{ S(\epsilon)\otimes S(\mbox{e}^{-2i\delta}) \right\}
   \left\{ D(\alpha)\otimes {\bf 1} \right\}\ket{0}\otimes \ket{0}
   \nonumber \\
&=&S(\epsilon)D(\alpha)\otimes S(\mbox{e}^{-2i\delta}) \ket{0}\otimes \ket{0}
   \nonumber \\
&=&S(\epsilon)D(\alpha)S(\epsilon)^{-1}S(\epsilon)\otimes 
   S(\mbox{e}^{-2i\delta}) \ket{0}\otimes \ket{0}
   \nonumber \\
&=&D({\tilde \alpha})S(\epsilon)\otimes 
   S(\mbox{e}^{-2i\delta}) \ket{0}\otimes \ket{0}
   \nonumber \\
&=&\left\{D({\tilde \alpha})\otimes {\bf 1}\right\} \left\{
S(\epsilon)\otimes S(\mbox{e}^{-2i\delta})\right\}
\ket{0}\otimes \ket{0}
   \nonumber \\
&=&D_{1}({\tilde \alpha}) \left\{
S(\epsilon)\otimes S(\mbox{e}^{-2i\delta}) \right\}\ket{0}\otimes \ket{0}.
   \nonumber 
\end{eqnarray}
Operating the operator $U_{J}(\kappa)$ (remember that $\kappa=\zettai{
\kappa}\mbox{e}^{i\delta}$) from the left 
\begin{eqnarray}
&&{}U_{J}(\kappa)D_{1}({\tilde \alpha}) \left\{
S(\epsilon)\otimes S(\mbox{e}^{-2i\delta}) \right\}\ket{0}\otimes \ket{0}
  \nonumber \\
=&&{}U_{J}(\kappa)D_{1}({\tilde \alpha})\left\{
S(\epsilon)\otimes S(\mbox{e}^{-2i\delta}) \right\}U_{J}(\kappa)^{-1}
U_{J}(\kappa)\ket{0}\otimes \ket{0}
  \nonumber \\
=&&{}U_{J}(\kappa)D_{1}({\tilde \alpha})\left\{
S(\epsilon)\otimes S(\mbox{e}^{-2i\delta}) \right\}U_{J}(\kappa)^{-1}
\ket{0}\otimes \ket{0}\qquad \mbox{by}\ (\ref{eq:invariant-property})
  \nonumber \\
=&&{}U_{J}(\kappa)D_{1}({\tilde \alpha})U_{J}(\kappa)^{-1} 
U_{J}(\kappa)\left\{
S(\epsilon)\otimes S(\mbox{e}^{-2i\delta}\epsilon) \right\}U_{J}(\kappa)^{-1}
\ket{0}\otimes \ket{0}
  \nonumber \\
=&&{}D_{1}(cos(\zettai{\kappa}){\tilde \alpha})
     D_{2}(-\mbox{e}^{-i\delta}sin(\zettai{\kappa}){\tilde \alpha})
\left\{S(\epsilon)\otimes S(\mbox{e}^{-2i\delta}\epsilon)\right\}
\ket{0}\otimes \ket{0}  \quad \mbox{by}\ (\ref{eq:2-invariant-property}) 
\ \mbox{and}\ (\ref{eq:adjoint-form})  \nonumber \\
=&&{}\left\{D(cos(\zettai{\kappa}){\tilde \alpha})\otimes 
D(-\mbox{e}^{-i\delta}sin(\zettai{\kappa}){\tilde \alpha})\right\}
\left\{S(\epsilon)\otimes S(\mbox{e}^{-2i\delta}\epsilon)\right\}
\ket{0}\otimes \ket{0}  \nonumber \\
=&&{}D(cos(\zettai{\kappa}){\tilde \alpha})S(\epsilon)\otimes 
D(-\mbox{e}^{-i\delta}sin(\zettai{\kappa}){\tilde \alpha})
S(\mbox{e}^{-2i\delta}\epsilon)\ket{0}\otimes \ket{0}.
  \nonumber \\
=&&{}D(cos(\zettai{\kappa}){\tilde \alpha})S(\epsilon)\otimes
D(-isin(\zettai{\kappa}){\tilde \alpha})S(-\epsilon)
\ket{0}\otimes \ket{0}, \nonumber 
\end{eqnarray}
where we have chosen in the last step 
\begin{equation}
  \mbox{e}^{-i\delta}=i\quad (\Longleftarrow \ 
  \mbox{for example}\quad \delta=-\frac{\pi}{2}).
\end{equation}
Operating the operator $S(-\epsilon)\otimes S(\epsilon)$ from the left
\begin{eqnarray}
&&{}\left\{ S(-\epsilon)\otimes S(\epsilon) \right\}
D(cos(\zettai{\kappa}){\tilde \alpha})S(\epsilon)\otimes
D(-isin(\zettai{\kappa}){\tilde \alpha})S(-\epsilon)
\ket{0}\otimes \ket{0}, \nonumber \\
=&&{}S(-\epsilon)D(cos(\zettai{\kappa}){\tilde \alpha})S(\epsilon)\otimes 
S(\epsilon)D({-i}sin(\zettai{\kappa}){\tilde \alpha})S(\epsilon)^{-1}
\ket{0}\otimes \ket{0}. \nonumber 
\end{eqnarray}
Here let us calculate the last term : 
\begin{equation}
S(-\epsilon)D(cos(\zettai{\kappa}){\tilde \alpha})S(\epsilon)
=D(cos(\zettai{\kappa}){\alpha})
\end{equation}
and we obtain 
\begin{equation}
S(\epsilon)D({-i}sin(\zettai{\kappa}){\tilde \alpha})S(\epsilon)^{-1}
=D({-i}sin(\zettai{\kappa})\alpha)
\end{equation}
against the equation (38) in \cite{DG} 
\begin{equation}
S(\epsilon)D({-i} sin(\zettai{\kappa}){\tilde \alpha})S(\epsilon)^{-1}
=D(-i sin(\zettai{\kappa}){\tilde {\tilde \alpha}}) 
\end{equation}
where 
\[
{\tilde {\tilde \alpha}}=
cosh(2\zettai{\epsilon})\alpha+
\mbox{e}^{i\phi}sinh(2\zettai{\epsilon}){\bar \alpha}.
\]
Therefore one cannot follow their method from this stage. 

\par \noindent 
But as stated above their method is simple and very interesting, so it may 
be possible to modify that more subtly by making use of 
(\ref{eq:adjoint-alpha-choice}). 

\par \noindent
{\bf Problem} Is it possible to make a ``perfect cloning" in the sense of 
\cite{DG} ?

\vspace{10mm}
\subsection{Swap of Squeezed--like States ?}
We would like to construct an operator like
\begin{equation}
     \ket{\beta_{1}}\otimes \ket{\beta_{2}} \longrightarrow 
     \ket{\beta_{2}}\otimes \ket{\beta_{1}}. 
\end{equation}
In this case we cannot use an operator $U_{J}(\kappa)$. 
Let us explain the reason. 

\par \nonumber 
Similar to (\ref{eq:swap-operator}) 
\begin{eqnarray}
U_{J}(\kappa)\ket{\beta_{1}}\otimes \ket{\beta_{2}}
&=&U_{J}(\kappa)S(\beta_{1})\otimes S(\beta_{2})\ket{0}\otimes \ket{0}
      \nonumber \\
&=&U_{J}(\kappa)S_{1}(\beta_{1})S_{2}(\beta_{2})\ket{0}\otimes \ket{0}
      \nonumber \\
&=&U_{J}(\kappa)S_{1}(\beta_{1})S_{2}(\beta_{2})U_{J}(\kappa)^{-1}
\ket{0}\otimes \ket{0}.
\end{eqnarray}
On the other hand by (\ref{eq:squeezed-adjoint-formula}) 
\[
U_{J}(\kappa)S_{1}(\beta_{1})S_{2}(\beta_{2})U_{J}(\kappa)^{-1}
=\mbox{e}^{X},
\]
where 
\begin{eqnarray}
\mbox{X}&=&
\frac{1}{2}\left\{cos^{2}(\zettai{\kappa})\beta_{1} + 
\frac{\kappa^2 sin^{2}(\zettai{\kappa})}{\zettai{\kappa}^2}\beta_{2}\right\}
    (a_{1}^{\dagger})^2
-\frac{1}{2}\left\{cos^{2}(\zettai{\kappa}){\bar \beta_{1}} + 
\frac{{\bar \kappa}^2 sin^{2}(\zettai{\kappa})}{\zettai{\kappa}^2}
{\bar \beta_{2}}\right\}a_{1}^2       \nonumber \\
&+&\frac{1}{2}\left\{cos^{2}(\zettai{\kappa})\beta_{2} + 
\frac{{\bar \kappa}^2 sin^{2}(\zettai{\kappa})}{\zettai{\kappa}^2}
\beta_{1}\right\}(a_{2}^{\dagger})^2
-\frac{1}{2}\left\{cos^{2}(\zettai{\kappa}){\bar \beta_{2}} + 
\frac{\kappa^2 sin^{2}(\zettai{\kappa})}{\zettai{\kappa}^2}
{\bar \beta_{1}}\right\}a_{2}^2        \nonumber \\
&+&(\beta_{2} \kappa-\beta_{1} {\bar \kappa})
\frac{sin(2\zettai{\kappa})}{2\zettai{\kappa}}a_{1}^{\dagger}a_{2}^{\dagger}
-({\bar \beta_{2}}{\bar \kappa}-{\bar \beta_{1}}\kappa)
\frac{sin(2\zettai{\kappa})}{2\zettai{\kappa}}a_{1}a_{2}\ .   \nonumber 
\end{eqnarray}
Here an extra term containing $a_{1}^{\dagger}a_{2}^{\dagger}$ appeared. 
To remove this we must set $\beta_{2} \kappa-\beta_{1} {\bar \kappa}=0$, 
but in this case we meet 
\[
U_{J}(\kappa)S_{1}(\beta_{1})S_{2}(\beta_{2})U_{J}(\kappa)^{-1}
=S_{1}(\beta_{1})S_{2}(\beta_{2})
\]
by (\ref{eq:2-invariant-property}). That is, there is no change. 

\par \noindent 
We could not construct an operator likely in the subsection 8.1 in spite of 
very our efforts , so we present 

\par \noindent
\begin{flushleft}
{\bf Problem}\quad 
Is it possible to find an operator such as $U_{J}(\kappa)$ in the preceding 
subsection for performing the swap ? 
\end{flushleft}

\vspace{10mm}
\subsection{A Comment}
We have used in the process of proofs 
both a displacement operator $D(\alpha)$ and a squeezed one 
$S(\epsilon)$ 
\[
D(\alpha)=\mbox{exp}(\alpha a^{\dagger}-{\bar \alpha}a), \quad 
S(\epsilon)=\mbox{exp}\frac{1}{2}\left(\epsilon (a^{\dagger})^{2}-
{\bar \epsilon}a^{2}\right)
\]
as a product operator
\begin{equation}
S(\epsilon)D(\alpha).
\end{equation}
We note that this product operator with the parameter space 
$\{(\alpha, \epsilon) \in \fukuso^{2}\}$ plays a crucial role in our 
Holonomic Quantum Computation (Computer), see section 8.1. 

\par \noindent
Similarly, the product operator 
\begin{equation}
U_{K}(w)U_{J}(v)
\end{equation}
with the parameter space $\{(v, w) \in \fukuso^{2}\}$ also plays a crucial 
role in it, see section 8.2. 

\par \noindent
We believe that Holonomic Quantum Computation and our geometric method (
involving swap or imperfect cloning) in Quantum Information Theory
are well--matched.

\vspace{20mm}
\section{Path Integral on A Quantum Computer}

In this section we present a very important problem (at least to the 
author) about the possibility of calculation of path integral on a 
Quantum Computer. 

\par \nonumber 
The path integral method plays an essential role in Quantum Mechanics or 
Quantum Field Theory. But it is, in general, not easy to calculate it 
except for Gaussian cases.
Some specialists must, in a perturbation theory, calculate many Feynman's 
graphs by making use of a classical computer(s). This is a hard and painful 
task. 

\par \noindent
Now let us present our general problem. 

\par \noindent
\begin{flushleft}
{\bf Problem}\quad 
Is it possible to calculate a path integral in polynomial times 
by making use of a quantum computer ?
\end{flushleft}

For this subject refer \cite{TW} and its references.  
But our method or interest is a bit different from \cite{TW}. 
To match our method with path integrals we should use {\bf coherent 
state path integral method}, see \cite{FKSF1}, \cite{FKSF2}, \cite{FKS}. 
\cite{IKJ} is also recommended. 

\par \noindent
To calculate a physical quantity such as a trace formula of the Hamiltonian 
we, for example, give a coherent state path integral expression to it. We 
want to calculate it, but it is usally not easy to do so. Therefore we have 
to make do with some approximations (WKB approximation, etc). 
Then our next problem is 

\par \noindent
\begin{flushleft}
{\bf Problem}\quad 
Is it possible to give it in polynomial times with 
Holonomic Quantum Computer ?
\end{flushleft}

\par \noindent
For the readers who are not familiar with coherent state path integral  
method let us show a simple, but very instructive example, \cite{KF11}. 

\par \noindent
Let us consider the Hamiltonian of harmonic oscillator 
\begin{equation}
H=\omega N =\omega a^{\dagger}a ,
\end{equation}
where we have omitted the constant term for simplicity. The eigenvalues of 
$H$ are well--known to be $\{n\omega \ |\ n=0,1,\cdots\}$ and its trace 
formula is given as 
\begin{equation}
\label{eq:Abel sum}
\mbox{tr}\ \mbox{e}^{-iTH}=\sum_{n=0}^{\infty}\mbox{e}^{-in\omega T}
=\frac{1}{1-\mbox{e}^{-i\omega T}}\quad (\mbox{Abel sum}).
\end{equation}

\par \noindent
Let us give a coherent state path integral expression to this 
trace formula. Making use of the resolution of unity (\ref{eq:2-7}) we 
obtain 
\begin{equation}
\label{eq:trace-formula}
\mbox{tr}\ \mbox{e}^{-iTH}=\mbox{tr}\ {\bf 1} \mbox{e}^{-iTH}=
\mbox{tr}\ \int_{\fukuso} \frac{[d^{2}z]}{\pi} \ket{z}\bra{z} \mbox{e}^{-iTH}
= 
\int_{\fukuso} \frac{[d^{2}z]}{\pi} \bra{z}\mbox{e}^{-iTH}\ket{z} \ .
\end{equation}
This is just an analytical expression of the trace formula. It is easy to 
calculate this directly, but we give this a path integral expression. 
Noting 
\[
\mbox{e}^{X}=\lim_{N \to \infty}\left(1+\frac{X}{N}\right)^{N},
\]
we have 
\begin{equation}
\mbox{RHS of (\ref{eq:trace-formula})}
=\lim_{N \to \infty}\int_{\fukuso} \frac{[d^{2}z]}{\pi} 
\bra{z}\left(1-i{\Delta}t H\right)^{N}\ket{z},
\end{equation}
where we have set ${\Delta}t=T/N$.

\par \noindent
By inserting the resolution of unity (\ref{eq:2-7}) at each step 
likely 
\begin{eqnarray}
\left(1-i{\Delta}t H\right)^{N} 
&=&\left(1-i{\Delta}t H\right){\bf 1}\left(1-i{\Delta}t H\right){\bf 1}
\cdots \left(1-i{\Delta}t H\right){\bf 1}\left(1-i{\Delta}t H\right)
 \nonumber \\
{\bf 1}&=&\int_{\fukuso} \frac{[d^{2}z_{j}]}{\pi}
   \ket{z_{j}}\bra{z_{j}} \quad \mbox{for any}\ \ 1\leq j \leq N-1\ ,
\nonumber 
\end{eqnarray}
we have
\begin{equation}
\mbox{RHS of (\ref{eq:trace-formula})}
=
\lim_{N \to \infty} 
\int_{PBC} \prod_{j=1}^{N} \frac{[d^{2}z_{j}]}{\pi} \prod_{j=1}^{N}
\bra{z_{j}}1-i{\Delta}tH\ket{z_{j-1}}, \nonumber 
\end{equation}
where PBC (periodic boundary condition) means $z_{N}=z_{0}=z$. We note that 
the choice of $\{z_{1}, z_{2}, \cdots, z_{N-1}\}$ is random. 

\par \noindent
Let us calculate the term $\bra{z_{j}}1-i{\Delta}tH\ket{z_{j-1}}$ : 
\begin{eqnarray}
\bra{z_{j}}1-i{\Delta}tH\ket{z_{j-1}}
&=&\braket{z_{j}}{z_{j-1}}-i{\Delta}t\bra{z_{j}}H\ket{z_{j-1}}
 \nonumber \\
&=&\braket{z_{j}}{z_{j-1}}
\left\{1-i{\Delta}t\frac{\bra{z_{j}}H\ket{z_{j-1}}}{\braket{z_{j}}{z_{j-1}}}
\right\}
 \nonumber \\
&=&\braket{z_{j}}{z_{j-1}}
\mbox{exp}\left\{
-i{\Delta}t\frac{\bra{z_{j}}H\ket{z_{j-1}}}{\braket{z_{j}}{z_{j-1}}}
\right\}\quad \mbox{up to}\ \mbox{O}(({\Delta}t)^2).
  \nonumber
\end{eqnarray}
On the other hand from (\ref{eq:2-a}) and (\ref{eq:2-5-1}) 
\begin{eqnarray}
\braket{z_{j}}{z_{j-1}}&=&\mbox{exp}\left(-\frac{1}{2}|z_{j}|^2
-\frac{1}{2}|z_{j-1}|^2+{\bar z_{j}}z_{j-1}\right), \nonumber \\
\bra{z_{j}}H\ket{z_{j-1}}&=&\omega \bra{z_{j}}a^{\dagger}a\ket{z_{j-1}}
=\omega {\bar z_{j}}z_{j-1}\braket{z_{j}}{z_{j-1}}, \quad 
\frac{\bra{z_{j}}H\ket{z_{j-1}}}{\braket{z_{j}}{z_{j-1}}}=
\omega {\bar z_{j}}z_{j-1}, \nonumber 
\end{eqnarray}
so we have 
\[
\bra{z_{j}}1-i{\Delta}tH\ket{z_{j-1}}
=\mbox{exp}\left\{ -\frac{1}{2}|z_{j}|^2-\frac{1}{2}|z_{j-1}|^2+
{\bar z_{j}}z_{j-1}\ -i\omega{\Delta}t{\bar z_{j}}z_{j-1} \right\}
\]
after some algebra. Here taking the periodic bountary condition $z_{N}=
z_{0}$ it is easy to see 
\[
\prod_{j=1}^{N}\bra{z_{j}}1-i{\Delta}tH\ket{z_{j-1}}
=\mbox{exp}\left\{ -\sum_{j=1}^{N} \left\{ {\bar z_{j}}(z_{j}-z_{j-1}) + 
i\omega{\Delta}t{\bar z_{j}}z_{j-1} \right\} 
\right\}.
\]
Therefore from this we reach 
\begin{equation}
\label{eq:path-integral-formula}
\mbox{RHS of (\ref{eq:trace-formula})}
=
\lim_{N \to \infty} 
\int_{PBC} \prod_{j=1}^{N} \frac{[d^{2}z_{j}]}{\pi} 
\mbox{exp}\left\{ -\sum_{j=1}^{N} \left\{ {\bar z_{j}}(z_{j}-z_{j-1}) +
i\omega{\Delta}t{\bar z_{j}}z_{j-1} \right\} 
\right\}.
\end{equation}
This is just the coherent state path integral expression of trace formula 
of the harmonic oscillator. 
As for calculation of (\ref{eq:path-integral-formula}) see Appendix C.

\vspace{20mm}
\section{Discussion and Dream}

Here we state our dream once more.  
The main subjects of Quantum Information Theory are 
\begin{itemize}
\item Quantum Computer (Computation)
\item Quantum Cryptgraphy
\item Quantum Teleportation
\end{itemize}
The purpose is to understand Quantum Information Theory from the geometric 
point of view or, more clearly, to construct 
\begin{Large}
\begin{center}
{\bf Geometric Quantum Information Theory}.
\end{center}
\end{Large}
For example, 
\begin{itemize}
\item {\bf Geometric} Quantum Computer (Computation)
\item {\bf Geometric} Quantum Cryptgraphy
\item {\bf Geometric} Quantum Teleportation
\end{itemize}
Geometric understanding of several concepts is very important because 
we can view them from global point of view. 
We believe that we have taken a first step towards this dream.

\vspace{5mm}
\noindent
{\it Acknowledgment.}\quad 
The author wishes to thank Kunio Funahashi for useful comments.

\par \vspace{20mm} \noindent
\begin{center}
\begin{Large}
  {\bf Appendix}
\end{Large}
\end{center}

\par \vspace{10mm} \noindent
\begin{Large}
{\bf Appendix A\quad Proof of Disentangling Formulas}
\end{Large}

\par \noindent
Here we prove the disentangling formulas (\ref{eq:k-formula}) and 
(\ref{eq:j-formula}) for generalized coherent operators based on 
Lie algebras $su(1,1)$ and $su(2)$. 

In general a representation of Lie algebra cannot be lifted to 
the representation of its Lie group if a Lie group is not simply 
connected. We note that $SU(1,1)$ is not simply connected because 
$\pi_{1}(SU(1,1))=\pi_{1}(U(1))={\cal Z}$. 

\par \noindent 
First we start under the assumption that there is a representation of 
Lie group $SU(1,1)$. Namely, 
let $\rho$ be a representation of Lie group $SU(1,1) \subset SL(2,\fukuso)$ 
\begin{equation}
      \rho : SL(2,\fukuso) \longrightarrow U(\calh \otimes \calh) 
\end{equation}
and
\begin{equation}
    K_{+}=d\rho(k_{+}),\quad K_{-}=d\rho(k_{-}), \quad K_{3}=d\rho(k_{-})
\end{equation}
where
\begin{equation}
     k_{+}=
   \left(
     \begin{array}{cc}
        0& 1\\
        0& 0
     \end{array}
   \right), \quad
     k_{-}=
   \left(
     \begin{array}{cc}
        0& 0\\
       -1& 0
     \end{array}
   \right), \quad
     k_{3}=\frac{1}{2}
   \left(
     \begin{array}{cc}
        1& 0\\
        0& -1
     \end{array}
   \right).
\end{equation}
It is easy to see 
\[
[k_{3},k_{+}]=k_{+}, \quad [k_{3},k_{-}]=-k_{-}, 
\quad [k_{+},k_{-}]=-2k_{3} \quad \mbox{but} \quad {k_{+}}^{\dagger}=-k_{-}.
\]
In this case 
\begin{eqnarray}
   &&\mbox{exp}\left(wK_{+} - \bar{w}K_{-}\right)
     = \mbox{exp}\left(d\rho(wk_{+}-\bar{w}k_{-})\right)
         \nonumber \\
   &&= \mbox{exp}\left(d\rho 
        \left(
           \begin{array}{cc}
                 0       & w \\
                \bar{w}  & 0
           \end{array}
        \right)
                \right)
    =\rho \left(
         \mbox{exp} 
        \left(
           \begin{array}{cc}
                0       & w \\
                \bar{w} & 0
           \end{array}
        \right)
          \right)
    \equiv \rho(\mbox{e}^{A}) .       
\end{eqnarray}
From 
\[
  A^2 = |w|^2 E  
\]
we have 
\begin{eqnarray}
  \mbox{e}^{A}=\cosh(|w|) E + \frac{\sinh(|w|)}{|w|}A
             = 
        \left(
           \begin{array}{cc}
               \cosh(|w|)&   \frac{\sinh(|w|)}{|w|}w \\
               \frac{\sinh(|w|)}{|w|}\bar{w}& \cosh(|w|)
           \end{array}
        \right).
\end{eqnarray}
For 
$\mbox{e}^{A}=
        \left(
           \begin{array}{cc}
             a & b \\
             c & d
           \end{array}
        \right)\ 
(ad - bc = 1)$, 
the Gauss decomposition of this matrix is given by
\begin{equation}
        \left(
           \begin{array}{cc}
             a & b \\
             c & d
           \end{array}
        \right)
=
        \left(
           \begin{array}{cc}
             1 & \frac{b}{d} \\
             0 & 1
           \end{array}
        \right)\ 
        \left(
           \begin{array}{cc}
             \frac{1}{d} & 0 \\
             0 & d
           \end{array}
        \right)\ 
        \left(
           \begin{array}{cc}
             1 & 0 \\
             \frac{c}{d} & 1
           \end{array}
        \right).
\end{equation}
Since $\rho$ is a representation of Lie group (not Lie algebra !) we have 
\begin{eqnarray}
 &&\rho\left(
        \left(
           \begin{array}{cc}
             1 & \frac{b}{d} \\
             0 & 1
           \end{array}
        \right)\ 
        \left(
           \begin{array}{cc}
             \frac{1}{d} & 0 \\
             0 & d
           \end{array}
        \right)\ 
        \left(
           \begin{array}{cc}
             1 & 0 \\
             \frac{c}{d} & 1
           \end{array}
        \right)
   \right)  \nonumber \\
 &&=
 \rho\left(
        \left(
           \begin{array}{cc}
             1 & \frac{b}{d} \\
             0 & 1
           \end{array}
        \right)
   \right)\
 \rho\left(
        \left(
           \begin{array}{cc}
             \frac{1}{d} & 0 \\
             0 & d
           \end{array}
        \right)
   \right)\
 \rho\left(
        \left(
           \begin{array}{cc}
             1 & 0 \\
             \frac{c}{d} & 1
           \end{array}
        \right)
   \right)   \nonumber \\
&&=
 \rho\left(\mbox{exp}{
        \left(
           \begin{array}{cc}
             0 & \frac{b}{d} \\
             0 & 0
           \end{array}
        \right)}
   \right)\
 \rho\left(\mbox{exp}{
        \left(
           \begin{array}{cc}
             -\mbox{log}d& 0 \\
             0 & \mbox{log}d
           \end{array}
        \right)}
   \right)\
 \rho\left(\mbox{exp}{
        \left(
           \begin{array}{cc}
             0 & 0 \\
             \frac{c}{d} & 0
           \end{array}
        \right)}
   \right) \nonumber \\
&&=
\mbox{exp}{\left(
    d\rho
        \left(
           \begin{array}{cc}
             0 & \frac{b}{d} \\
             0 & 0
           \end{array}
        \right)
      \right)}\ 
\mbox{exp}{\left(
    d\rho
        \left(
           \begin{array}{cc}
             -\mbox{log}d& 0 \\
             0 & \mbox{log}d
           \end{array}
        \right)
      \right)}\ 
\mbox{exp}{\left(
    d\rho
        \left(
           \begin{array}{cc}
             0 & 0 \\
             \frac{c}{d} & 0
           \end{array}
        \right)
      \right)} \nonumber \\
&&=
\mbox{exp}{\left(\frac{b}{d}
    d\rho(k_{+})\right)}\ 
\mbox{exp}{\left(-2\mbox{log}d\ 
    d\rho(k_{3})\right)}\ 
\mbox{exp}{\left(-\frac{c}{d}
    d\rho(k_{-})\right)} \nonumber \\
&&=
\mbox{exp}\left(\frac{b}{d}K_{+}\right)\
\mbox{exp}\left(-2\mbox{log}dK_{3}\right)\
\mbox{exp}\left(-\frac{c}{d}K_{-}\right) \nonumber \\
&&=
\mbox{exp}\left(\frac{b}{d}K_{+}\right)\
\mbox{exp}\left(\mbox{log}\left(\frac{1}{d^2}\right)K_{3}\right)\
\mbox{exp}\left(-\frac{c}{d}K_{-}\right) 
\end{eqnarray}
where 
\begin{eqnarray}
 &&\frac{b}{d}=\frac{\frac{\sinh(|w|)}{|w|}w}{\cosh(|w|)}
              = \frac{\tanh(|w|)w}{|w|}  \nonumber \\
 &&\frac{c}{d}=\frac{\tanh(|w|)\bar{w}}{|w|} \nonumber \\
 &&\frac{1}{d^2}=\frac{1}{\cosh^{2}(|w|)}=1-\tanh^{2}(|w|).
\end{eqnarray}
If we set
\[
   \zeta=\frac{\tanh(|w|)w}{|w|} \quad \Longrightarrow \quad 
   |\zeta|=\tanh(|w|)
\]
then we have (\ref{eq:k-formula}).  That is, we could prove 
(\ref{eq:k-formula}) under the assumption. 
To remove this we needs some tricks. 
We define 
\begin{eqnarray}
f(t)&=&\exp\left\{t(wK_{+}-\bar{w}K_{-})\right\} \\
g(t)&=&\exp\{\zeta(t)K_{+}\}\exp\{\log(1-|\zeta(t)|^{2})K_{3}\}
       \exp\{-{\bar \zeta(t)}K_{-}\},  \nonumber \\ 
    &&\ \mbox{where}\quad \zeta(t)=\frac{w\tanh(t|w|)}{|w|}. 
\end{eqnarray}
Then 
\begin{equation}
 \label{eq:f-equation}
  f(0)={\bf 1}, \quad \frac{d}{dt}f(t)=(wK_{+}-\bar{w}K_{-})f(t).
\end{equation}
On the other hand
\begin{eqnarray}
  g(0)&=&{\bf 1}, \nonumber \\
  \frac{d}{dt}g(t)&=&\zeta^{\prime}(t)K_{+}g(t)+
    \frac{d}{dt}\log(1-|\zeta(t)|^{2})e^{\zeta(t)K_{+}} K_{3}
    e^{\log(1-|\zeta(t)|^{2})K_{3}}e^{-{\bar \zeta(t)}K_{-}}
    -{\bar \zeta}^{\prime}(t)g(t)K_{-}  \nonumber \\
   &=&\zeta^{\prime}(t)K_{+}g(t) 
   +\frac{d}{dt}\log(1-|\zeta(t)|^{2})e^{\zeta(t)K_{+}} K_{3}
      e^{-\zeta(t)K_{+}}g(t)  
   -{\bar \zeta}^{\prime}(t)g(t)K_{-}g(t)^{-1}g(t) \nonumber \\
   &=&\left\{ \zeta^{\prime}(t)K_{+}+ 
   \frac{d}{dt}\log(1-|\zeta(t)|^{2})
        e^{\zeta(t)K_{+}}K_{3}e^{-\zeta(t)K_{+}}
   -{\bar \zeta}^{\prime}(t)g(t)K_{-}g(t)^{-1} \right\}g(t).
   \nonumber \\
   &{}& 
\end{eqnarray}
Then it is not difficult to see
\begin{eqnarray}
e^{\zeta(t)K_{+}}K_{3}e^{-\zeta(t)K_{+}}&=&K_{3}-\zeta(t)K_{+}\ ,
    \nonumber \\
g(t)K_{-}g(t)^{-1}&=&e^{\zeta(t)K_{+}}e^{\log(1-|\zeta(t)|^{2})K_{3}}
  K_{-}e^{-\log(1-|\zeta(t)|^{2})K_{3}}e^{-\zeta(t)K_{+}} \nonumber \\
 &=&e^{-\log(1-|\zeta(t)|^{2})}e^{\zeta(t)K_{+}}K_{-}e^{-\zeta(t)K_{+}}
                                                          \nonumber \\
 &=&\frac{1}{1-|\zeta(t)|^{2}}
       \left\{K_{-}-2\zeta(t)K_{3}+\zeta(t)^{2}K_{+}\right\}\ ,  \nonumber 
\end{eqnarray}
so that 
\begin{eqnarray}
&&\frac{d}{dt}g(t)  \nonumber \\
&&=
   \left\{ \zeta^{\prime}(t)K_{+} 
  -\frac{\zeta^{\prime}(t){\bar \zeta(t)}+\zeta(t){\bar \zeta}^{\prime}(t)}
   {1-|\zeta(t)|^{2}}\left(K_{3}-\zeta(t)K_{+}\right) 
  -\frac{{\bar \zeta}^{\prime}(t)}{1-|\zeta(t)|^{2}}
   \left(K_{-}-2\zeta(t)K_{3}+\zeta(t)^{2}K_{+}\right) 
   \right\}g(t).   \nonumber 
\end{eqnarray}
If we notice
\begin{eqnarray}
  &&|\zeta(t)|=\tanh(t|w|),  \nonumber \\ 
  &&\zeta^{\prime}(t){\bar \zeta}(t)=|w|(1-\tanh^{2}(t|w|))\tanh(t|w|)=
    \zeta(t){\bar \zeta}^{\prime}(t),  \nonumber \\
  &&\frac{\zeta^{\prime}(t)}{1-|\zeta(t)|^{2}}=w, \ 
    \frac{{\bar \zeta}^{\prime}(t)}{1-|\zeta(t)|^{2}}=\bar{w},  \nonumber 
\end{eqnarray}
then we reach after some algebra
\begin{equation}
 \label{eq:g-equation}
 \frac{d}{dt}g(t)=(wK_{+}-\bar{w}K_{-})g(t). 
\end{equation}
Comparing (\ref{eq:g-equation}) with (\ref{eq:f-equation}) we obtain 
(\ref{eq:k-formula}). 

\par \vspace{5mm} \noindent
Similar method is still valid for a representation of Lie group $SU(2)$ 
to prove (\ref{eq:j-formula}). Since we don't repeat here, so we leave it 
to the readers.

\par \vspace{10mm} \noindent
\begin{Large}
{\bf Appendix B\quad Universal Swap Operator}
\end{Large}

\par \noindent
Let us construct the swap operator in a universal manner 
\[
U : \calh\otimes \calh \longrightarrow \calh\otimes \calh \ ,\quad 
    U(a\otimes b)=b\otimes a \quad \mbox{for\ any}\ a,\ b \in \calh 
\]
where $\calh$ is an infinite--dimensional Hilbert space. Before constructing 
it we show in the finite--dimensional case.

For $a,\ b \in {\fukuso}^2$ then
\[
a\otimes b=
\left(
\begin{array}{c}
a_{1}b \\
a_{2}b
\end{array}
\right)
=
\left(
\begin{array}{c}
a_{1}b_{1} \\
a_{1}b_{2} \\
a_{2}b_{1} \\
a_{2}b_{2} 
\end{array}
\right),
\quad 
b\otimes a=
\left(
\begin{array}{c}
b_{1}a_{1} \\
b_{1}a_{2} \\
b_{2}a_{1} \\
b_{2}a_{2} 
\end{array}
\right)
=
\left(
\begin{array}{c}
a_{1}b_{1} \\
a_{2}b_{1} \\
a_{1}b_{2} \\
a_{2}b_{2} 
\end{array}
\right),
\]
so it is easy to see
\[
\left(
\begin{array}{cccc}
1& 0& 0& 0 \\
0& 0& 1& 0 \\
0& 1& 0& 0 \\
0& 0& 0& 1
\end{array}
\right)
\left(
\begin{array}{c}
a_{1}b_{1} \\
a_{1}b_{2} \\
a_{2}b_{1} \\
a_{2}b_{2} 
\end{array}
\right)
=
\left(
\begin{array}{c}
a_{1}b_{1} \\
a_{2}b_{1} \\
a_{1}b_{2} \\
a_{2}b_{2} 
\end{array}
\right).
\]
That is, the swap operator is 
\begin{equation}
U=
\left(
\begin{array}{cccc}
1& 0& 0& 0 \\
0& 0& 1& 0 \\
0& 1& 0& 0 \\
0& 0& 0& 1
\end{array}
\right). 
\end{equation}
This matrix can be written as follows by making use of three 
Controlled--NOT matrices (gates)
\begin{equation}
\left(
\begin{array}{cccc}
1& 0& 0& 0 \\
0& 0& 1& 0 \\
0& 1& 0& 0 \\
0& 0& 0& 1
\end{array}
\right) 
=
\left(
\begin{array}{cccc}
1& 0& 0& 0 \\
0& 1& 0& 0 \\
0& 0& 0& 1 \\
0& 0& 1& 0
\end{array}
\right) 
\left(
\begin{array}{cccc}
1& 0& 0& 0 \\
0& 0& 0& 1 \\
0& 0& 1& 0 \\
0& 1& 0& 0
\end{array}
\right) 
\left(
\begin{array}{cccc}
1& 0& 0& 0 \\
0& 1& 0& 0 \\
0& 0& 0& 1 \\
0& 0& 1& 0
\end{array}
\right),
\end{equation}
or graphically
\begin{center}
\setlength{\unitlength}{1mm}   %
\begin{picture}(150,40)
\put(15,30){\line(1,0){45}}   
\put(66,30){\line(1,0){45}}   
\put(15,10){\line(1,0){21}}   
\put(42,10){\line(1,0){42}}   
\put(90,10){\line(1,0){21}}   

\put(0,25){\makebox(9,10)[r]{$A$}} 
\put(0,5){\makebox(9,10)[r]{$B$}} 
\put(117,25){\makebox(25,10)[l]{$B$}} 
\put(117,5){\makebox(25,10)[l]{$A$}} 
\put(39,13){\line(0,1){17}}     
\put(63,10){\line(0,1){17}}     
\put(87,13){\line(0,1){17}}     
\put(36,25){\makebox(6,10){$\bullet$}} 
\put(84,25){\makebox(6,10){$\bullet$}} 
\put(60,5){\makebox(6,10){$\bullet$}} 
\put(60,25){\makebox(6,10){X}}         
\put(36,5){\makebox(6,10){X}}         
\put(84,5){\makebox(6,10){X}}         
\put(63,30){\circle{6}}               
\put(39,10){\circle{6}}               
\put(87,10){\circle{6}}               
\end{picture}
\end{center}
See for example \cite{KF1} and \cite{KF14}. 

\par \noindent
A comment is in order.\quad In this case we can happen to write $U$ as 
\begin{equation}
U=\frac{1}{2}\left({\bf 1}\otimes {\bf 1} + \sum_{j=1}^{3}
\sigma_{j}\otimes \sigma_{j} \right),
\end{equation}
where $\{\sigma_{1}, \sigma_{2}, \sigma_{}\}$ are Pauli matrices. 
But unfortunately we cannot extend this formula further. 

\par \noindent
It is not easy for us to conjecture its general form from this 
swap operator. Let us try for $n=3$. The result is 
\[
\left(
\begin{array}{ccccccccc}
1& 0& 0& 0& 0& 0& 0& 0& 0 \\
0& 0& 0& 1& 0& 0& 0& 0& 0 \\
0& 0& 0& 0& 0& 0& 1& 0& 0 \\
0& 1& 0& 0& 0& 0& 0& 0& 0 \\
0& 0& 0& 0& 1& 0& 0& 0& 0 \\
0& 0& 0& 0& 0& 0& 0& 1& 0 \\
0& 0& 1& 0& 0& 0& 0& 0& 0 \\
0& 0& 0& 0& 0& 1& 0& 0& 0 \\
0& 0& 0& 0& 0& 0& 0& 0& 1 
\end{array}
\right)
\left(
\begin{array}{c}
a_{1}b_{1} \\
a_{1}b_{2} \\
a_{1}b_{3} \\
a_{2}b_{1} \\
a_{2}b_{2} \\
a_{2}b_{3} \\
a_{3}b_{1} \\
a_{3}b_{2} \\
a_{3}b_{3} 
\end{array}
\right)
=
\left(
\begin{array}{c}
a_{1}b_{1} \\
a_{2}b_{1} \\
a_{3}b_{1} \\
a_{1}b_{2} \\
a_{2}b_{2} \\
a_{3}b_{2} \\ 
a_{1}b_{3} \\
a_{2}b_{3} \\
a_{3}b_{3} 
\end{array}
\right).
\]
Here we rewrite the swap operator above as follows.
\begin{equation}
U=
\left(
\begin{array}{ccc}
 \left(
  \begin{array}{ccc}
   1& 0& 0\\
   0& 0& 0\\
   0& 0& 0
  \end{array}
 \right) &
 \left(
  \begin{array}{ccc}
   0& 0& 0\\
   1& 0& 0\\
   0& 0& 0
  \end{array}
 \right) &
 \left(
  \begin{array}{ccc}
   0& 0& 0\\
   0& 0& 0\\
   1& 0& 0
  \end{array}
 \right) \\
  \left(
  \begin{array}{ccc}
   0& 1& 0\\
   0& 0& 0\\
   0& 0& 0
  \end{array}
 \right) &
 \left(
  \begin{array}{ccc}
   0& 0& 0\\
   0& 1& 0\\
   0& 0& 0
  \end{array}
 \right) &
 \left(
  \begin{array}{ccc}
   0& 0& 0\\
   0& 0& 0\\
   0& 1& 0
  \end{array}
 \right) \\
 \left(
  \begin{array}{ccc}
   0& 0& 1\\
   0& 0& 0\\
   0& 0& 0
  \end{array}
 \right) &
 \left(
  \begin{array}{ccc}
   0& 0& 0\\
   0& 0& 1\\
   0& 0& 0
  \end{array}
 \right) &
 \left(
  \begin{array}{ccc}
   0& 0& 0\\
   0& 0& 0\\
   0& 0& 1
  \end{array}
 \right) 
\end{array}
\right).
\end{equation}
Now, from the above form 
we can conjecture the general form of the swap operator.

\par \noindent
We note that 
\begin{equation}
({\bf 1}\otimes {\bf 1})_{ij,kl}=\delta_{ik}\delta_{jl},
\end{equation}
so after some trials we conclude 
\[
U : {\fukuso}^{n}\otimes {\fukuso}^{n} \longrightarrow 
    {\fukuso}^{n}\otimes {\fukuso}^{n}
\]
as 
\begin{equation}
U=(U_{ij,kl})\quad ;\quad U_{ij,kl}=\delta_{il}\delta_{jk},
\end{equation}
where $ij=11,12,\cdots,1n, 21,22,\cdots,2n,\ \cdots,\ n1,n2,\cdots,nn$.

\par \noindent
The proof is simple and as follows.\quad 
\begin{eqnarray}
(a\otimes b)_{ij}=a_{i}b_{j} \longrightarrow 
\left\{U(a\otimes b)\right\}_{ij}
&=&\sum_{kl=11}^{nn}U_{ij,kl}a_{k}b_{l}
=\sum_{kl=11}^{nn}\delta_{il}\delta_{jk}a_{k}b_{l} \nonumber \\
&=&\sum_{l=1}^{n}\delta_{il}b_{l} \sum_{k=1}^{n}\delta_{jk}a_{k}
=b_{i}a_{j}=(b\otimes a)_{ij}.  \nonumber 
\end{eqnarray}
At this stage there is no problem to take a limit $n \rightarrow \infty$. 

\par \noindent 
Let $\calh$ be a Hilbert space with a basis $\{e_{n}\}$ ($n \geq 1$). 
Then the universal swap operator is given by
\begin{equation}
U=(U_{ij,kl})\quad ;\quad U_{ij,kl}=\delta_{il}\delta_{jk},
\end{equation}
where $ij=11,12,\cdots, \cdots$.

\par \noindent
{\bf Problem} Is it possible to construct this universal swap operator 
by making use of Laser techniques ?

\par \vspace{10mm} \noindent
\begin{Large}
{\bf Appendix C\quad Calculation of Path Integral}
\end{Large}

\par \noindent
Let us calculate (\ref{eq:path-integral-formula}) explicitly.\quad 
Noting $z_{N}=z_{0}$ and rewriting 
\begin{eqnarray}
&&\sum_{j=1}^{N} \left\{ {\bar z_{j}}(z_{j}-z_{j-1}) +
i\omega{\Delta}t{\bar z_{j}}z_{j-1} \right\}  \nonumber \\
=&&
\sum_{j=1}^{N} \left\{ {\bar z_{j}}z_{j}-{\bar z_{j}}
(1-i\omega{\Delta}t)z_{j-1} \right\}
=
({\bar z_{N}},{\bar z_{N-1}}, \cdots, {\bar z_{1}}){\bf A}
\left(
\begin{array}{c}
z_{N} \\
z_{N-1} \\
\cdot \\
\cdot \\
z_{1}
\end{array}
\right) 
\equiv {\bf Z}^{\dagger}{\bf A}{\bf Z}, \nonumber
\end{eqnarray}
where 
\[
{\bf A}=
\left(
\begin{array}{cccccc}
1& -(1-i\omega{\Delta}t)& 0& \cdots& 0& 0 \\
0& 1& -(1-i\omega{\Delta}t)& \cdots& 0& 0 \\
\cdots& \cdots& \cdots& \cdots& \cdots& \cdots \\
\cdots& \cdots& \cdots& \cdots& \cdots& \cdots \\
0& 0& 0& \cdots& 1& -(1-i\omega{\Delta}t) \\
-(1-i\omega{\Delta}t)& 0& 0& 0& \cdots& 1
\end{array}
\right)
\]
we have 
\begin{eqnarray}
&&\int_{PBC} \prod_{j=1}^{N} \frac{[d^{2}z_{j}]}{\pi} 
\mbox{exp}\left\{ -\sum_{j=1}^{N} \left\{ {\bar z_{j}}(z_{j}-z_{j-1}) +
i\omega{\Delta}t{\bar z_{j}}z_{j-1} \right\} 
\right\} \nonumber \\
&=&
\int \prod_{j=1}^{N} \frac{[d^{2}z_{j}]}{\pi} 
\mbox{exp}\left(-{\bf Z}^{\dagger}{\bf A}{\bf Z}\right)
=
\frac{1}{\mbox{det}{\bf A}}, \nonumber 
\end{eqnarray}
where we have used that ${\bf A}$ is a normal matrix ($X^{\dagger}X=
XX^{\dagger}$). Since it is easy to see 
\[
\mbox{det}{\bf A}=1-(1-i\omega{\Delta}t)^{N},
\]
we obtain 
\begin{equation}
(\ref{eq:path-integral-formula})=\lim_{N \to \infty} 
\frac{1}{1-(1-i\omega{\Delta}t)^{N}}
=\lim_{N \to \infty}\frac{1}{1-(1-\frac{i\omega T}{N})^{N}}
=\frac{1}{1-\mbox{e}^{-i\omega T}}.
\end{equation}
This is just (\ref{eq:Abel sum}).

\par \vspace{10mm} \noindent
\begin{Large}
{\bf Appendix D\quad Representation from $SU(2)$ to $SO(3)$}
\end{Large}

\par \noindent
We in this appendix give a useful expression to the well--known 
representation from $SU(2)$ to $SO(3)$. This result is no direct relation 
to the text of this paper, but may become useful in the near future. 
Now let us define 
\[
\rho : SU(2)\ {\longrightarrow}\ SO(3).
\]
First of all we note a simple fact : 
\[
g=
\left(
 \begin{array}{cc}
  a+ib&  c+id \\
  -c+id& a-ib
  \end{array}
\right)
\quad a, b, c, d\ \in \real
\]
where 
\[
g \in SU(2) \Longleftrightarrow a^{2}+b^{2}+c^{2}+d^{2}=1. 
\]

Let us set $\{\sigma_{1}, \sigma_{2}, \sigma_{3}\}$ 
Pauli matrices
\[
\sigma_{1}=
\left(
  \begin{array}{cc}
    0& 1 \\
    1& 0 
  \end{array}
\right), \quad
\sigma_{2}=
\left(
  \begin{array}{cc}
    0& -i \\
    i& 0 
  \end{array}
\right), \quad
\sigma_{3}=
\left(
  \begin{array}{cc}
    1& 0 \\
    0& -1 
  \end{array}
\right) 
\]
and set 
\[
\tau_{j}=\frac{1}{2}\sigma_{j} \quad \mbox{for}\quad j=1,2,3.
\]
The representation $\rho$ is given as follows : it is easy to see 
\begin{eqnarray}
g^{-1}\tau_{1}g&=&(a^{2}-b^{2}-c^{2}+d^{2})\tau_{1}
+2(ab+cd)\tau_{2}-2(ac-bd)\tau_{3}, \nonumber \\
g^{-1}\tau_{2}g&=&-2(ab-cd)\tau_{1}
+(a^{2}-b^{2}+c^{2}-d^{2})\tau_{2}+2(ad+bc)\tau_{3}, \nonumber \\
g^{-1}\tau_{3}g&=&2(ac+bd)\tau_{1}
-2(ad-bc)\tau_{2}+(a^{2}+b^{2}-c^{2}-d^{2})\tau_{3}, \nonumber 
\end{eqnarray}
so we have 
\[
\left(g^{-1}\tau_{1}g, g^{-1}\tau_{2}g, g^{-1}\tau_{3}g \right)
=\left(\tau_{1}, \tau_{2}, \tau_{3} \right)\rho(g)
\]
where 
\begin{equation}
G\equiv \rho(g)=
\left(
 \begin{array}{ccc}
  a^{2}-b^{2}-c^{2}+d^{2}& -2(ab-cd)& 2(ac+bd) \\
  2(ab+cd)& a^{2}-b^{2}+c^{2}-d^{2}& -2(ad-bc) \\
  -2(ac-bd)& 2(ad+bc)& a^{2}+b^{2}-c^{2}-d^{2}
  \end{array}
\right).
\end{equation}
Here let us transform the above $G$. 
Noting that 
\begin{eqnarray}
a^{2}-b^{2}-c^{2}+d^{2}&=&1-2(b^{2}+c^{2}),  \nonumber \\
a^{2}-b^{2}+c^{2}-d^{2}&=&1-2(b^{2}+d^{2}),  \nonumber \\
a^{2}+b^{2}-c^{2}-d^{2}&=&1-2(c^{2}+d^{2}),  \nonumber 
\end{eqnarray}
from $a^{2}+b^{2}+c^{2}+d^{2}=1$, we have 
\begin{eqnarray}
G&=&
\left(
 \begin{array}{ccc}
  1-2(b^{2}+c^{2})& -2(ab-cd)& 2(ac+bd) \\
  2(ab+cd)& 1-2(b^{2}+d^{2})& -2(ad-bc) \\
  -2(ac-bd)& 2(ad+bc)& 1-2(c^{2}+d^{2})
  \end{array}
\right)  \nonumber \\
&=&
\left(
 \begin{array}{ccc}
  1& 0& 0 \\
  0& 1& 0 \\
  0& 0& 1
  \end{array}
\right)
+
\left(
 \begin{array}{ccc}
  -2(b^{2}+c^{2})& -2(ab-cd)& 2(ac+bd) \\
  2(ab+cd)& -2(b^{2}+d^{2})& -2(ad-bc) \\
  -2(ac-bd)& 2(ad+bc)& -2(c^{2}+d^{2})
  \end{array}
\right) \nonumber \\
&=&
\left(
 \begin{array}{ccc}
  1& 0& 0 \\
  0& 1& 0 \\
  0& 0& 1
  \end{array}
\right)
+
\left(
 \begin{array}{ccc}
  0& -2ab& 2ac \\
  2ab& 0& -2ad \\
  -2ac& 2ad& 0
  \end{array}
\right)
+
\left(
 \begin{array}{ccc}
  -2(b^{2}+c^{2})& 2cd& 2bd \\
  2cd& -2(b^{2}+d^{2})& 2bc\\
  2bd& 2bc& -2(c^{2}+d^{2})
  \end{array}
\right) \nonumber \\
&=&
\left(
 \begin{array}{ccc}
  1& 0& 0 \\
  0& 1& 0 \\
  0& 0& 1
  \end{array}
\right)
+ 2a
\left(
 \begin{array}{ccc}
  0& -b& c \\
  b& 0& -d \\
  -c& d& 0
  \end{array}
\right)
+ 2
\left(
 \begin{array}{ccc}
  -(b^{2}+c^{2})& cd& bd \\
  cd& -(b^{2}+d^{2})& bc\\
  bd& bc& -(c^{2}+d^{2})
  \end{array}
\right) \nonumber 
\end{eqnarray}
If we define $M$ as 
\begin{equation}
M=
\left(
 \begin{array}{ccc}
  0& -b& c \\
  b& 0& -d \\
  -c& d& 0
  \end{array}
\right)
\end{equation}
then easily 
\[
M^{2}=
\left(
 \begin{array}{ccc}
  -(b^{2}+c^{2})& cd& bd \\
  cd& -(b^{2}+d^{2})& bc\\
  bd& bc& -(c^{2}+d^{2})
  \end{array}
\right), 
\]
so we finally obtain 
\begin{equation}
G={\bf 1}+2aM+2M^{2}.
\end{equation}
This equation is very simple and interesting. 

\par \noindent 
The author could not find standard books (not papers) in representation 
theory which write this equation.

\vspace{20cm}
%

\end{document}